\documentclass[twocolumn]{aastex631}

\usepackage{amsmath}
\usepackage{url}
\usepackage{appendix}
\graphicspath{{./}{figures/}}

\begin{document}

\title{Multi-messenger Emission by Magnetically Arrested Disks and \\ Relativistic Jets of Black Hole X-ray Binaries}
\shortauthors{Kuze et al.}

\correspondingauthor{Riku Kuze}
\email{r.kuze@astr.tohoku.ac.jp}

\author[0000-0002-5916-788X]{Riku Kuze}
\affiliation{Astronomical Institute, Graduate School of Science, Tohoku University, Sendai 980-8578, Japan}
\affiliation{Center for Gravitational Physics, Yukawa Institute for Theoretical Physics, Kyoto University, Kyoto 606-8502, Japan}

\author[0000-0003-2579-7266]{Shigeo S. Kimura}
\affiliation{Frontier Research Institute for Interdisciplinary Sciences, Tohoku University, Sendai 980-8578, Japan}
\affiliation{Astronomical Institute, Graduate School of Science, Tohoku University, Sendai 980-8578, Japan}

\author[0000-0002-5387-8138]{Ke Fang}
\affiliation{Department of Physics, Wisconsin IceCube Particle Astrophysics Center, University of Wisconsin-Madison, Madison, WI, USA}



\begin{abstract}
Black hole X-ray binaries (BHXBs) are observed in various wavelengths from radio to GeV gamma-ray. Several BHXBs, including MAXI J1820+070 and Cygnus X-1, are also found to emit ultrahigh-energy (UHE; photon energy  $>$100~TeV) gamma rays. The origin and production mechanism of the multi-wavelength emission of BHXBs are under debate. We propose a scenario where relativistic particles from magnetically arrested disks (MADs), which could form when BHXBs are in quiescent or hard states, produce UHE gamma rays, while electrons in the jets produce GeV gamma-ray emission. Specifically, magnetic turbulence in MADs heats up and accelerates electrons and protons, while magnetic reconnection in jets accelerates electrons. Sub-PeV gamma rays and neutrinos are produced when relativistic protons interact with the thermal protons and the radiation by thermal electrons in the disk.  We discuss the perspectives of observing sub-PeV multi-messenger signals from individual BHXBs. Finally, we evaluate the integrated fluxes of the quiescent and hard-state BHXB population and find that BHXBs may contribute to the Galactic diffuse emission above $\sim 100$~TeV. 
\end{abstract}



\section{Introduction} \label{sec:intro}
Black hole X-ray binaries (BHXBs) consist of a stellar-mass black hole (BH) and a companion star. An accretion disk forms around the BH due to mass accretion from the companion star and other nearby material. This type of system is observed from radio to X-rays and sometimes in GeV gamma rays \citep[e.g.,][]{Motta+21}. Recently, Large High Altitude Air Shower Observatory (LHAASO) reported the detection of ultrahigh-energy (UHE; $>100$~TeV) gamma rays from several BHXBs. In particular, they find point-like gamma-ray sources associated with MAXI J1820+070 and Cygnus X-1 with spectra extending to $\sim 100$~TeV \citep{LHAASO2024_mq}. 

The origins of these ultrahigh-energy (UHE) gamma rays are unknown. Two general emission mechanisms exist--leptonic and hadronic. In a leptonic scenario, UHE electrons upscatter photons to the UHE regime via inverse Compton scattering \citep[e.g.,][]{Zdziarski+2009}. In a hadronic scenario, neutral pions produced by the inelastic $pp$  collisions ($p+p \rightarrow p+p+\pi$) and the photomeson production ($p+\gamma \rightarrow p+\pi$) \citep[e.g.,][]{Pepe+2015, Kantzas+2021} of PeV protons decay into UHE gamma rays. 

Galactic BHXBs exhibit multiple spectral states. In a hard state, the X-ray spectrum shows a hard power law with a cutoff around 30 - 100 keV \citep{Done+2007, Belloni2010}. Hard-state X-rays are typically explained as the thermal inverse Compton scattering by hot electrons in a corona \citep{McConnell+2002}. The nature of the corona, however, is still under debate. Studies \citep[e.g.,][]{Esin+1997, Wang+2024} have proposed that the corona may be identical to a hot accretion flow, also known as a radiatively inefficient accretion flow  \citep[RIAF;][]{Narayah&Yi1994, Yuan&Narayan2014}. Hot accretion flows may also occur in a quiescent state, where the mass accretion rate is low.

In a RIAF, high disk temperature yields high viscosity \citep{Shakura1973}, leading to high radial velocity and efficient magnetic flux accumulation onto the central BH \citep{Cao2011, KimuraSudoh2021}. As a result, magnetically arrested disks \citep[MADs;][]{Bisnovatyi-Kogan1974, Narayan2012} likely form, in which a magnetic flux threading in the BH is at a saturation level. 

In \citet{RK+2024}, we developed a two-zone emission model, called the `Jet-MAD model', to explain the multi-wavelength emission by nearby radio galaxies. This model consists of two emission components--MADs and jets. In MADs, we consider emission processes of thermal electrons, nonthermal electrons, nonthermal protons, and secondary particles. In jets, our model accounts for the emission processes of nonthermal electrons. Notable features of this model include the careful treatment of plasma injection into the jet, which is based on phenomena at the jet base exhibited by general relativistic magnetohydrodynamic (GRMHD) simulations \citep{Ripperda2022}, and a consistent evaluation of plasma parameters at the jet dissipation region, derived from those determined at the jet base.

In this paper, we examine whether such a two-zone emission model explains the multi-wavelength data of BHXBs. We also investigate the role of BHXBs in producing the diffuse gamma-ray and neutrino emission of the Galactic plane recently reported by air shower gamma-ray observatories \citep{TibetAsg2021, CaoDiffuse2023, HAWC:2023wdq} and IceCube  \citep{IceCubeDiffuse2023}. BHXBs that emit 10-100 TeV gamma rays and neutrinos should produce PeV cosmic rays (CRs). This motivates us to evaluate the contribution of BHXBs to the Galactic CR spectrum, especially at PeV energies \citep[cf.][]{Cooper+2020, KimuraSudoh2021, Kantzas+2023}.

The paper is organized as follows. We describe the Jet-MAD model in Section \ref{sec:JetMAD}. In Section \ref{sec:results}, we show the results of applying our model to selected BHXBs. In Section \ref{sec: diffuse}, we discuss the contribution of the BHXBs to the Galactic diffuse emission. We present our conclusion in Section \ref{sec:summary}.

\section{Jet-MAD model} \label{sec:JetMAD}
Studying the multi-wavelength emission from jets and MADs is well motivated as jets are efficiently produced when the accretion disk is in the MAD regime \citep{Tchekhovskoy2011}. Here, we briefly describe the Jet-MAD model developed by \citet{RK+2024}. We define a MAD as an accretion flow with $\beta \lesssim 1$, where $\beta$ is plasma beta treated as a free parameter. The MAD with $\beta\lesssim 1$ may be realized in the inner region within 10 gravitational radii \citep[see Figure 10 of ][]{White2019}. We consider the MAD as a corona with the Thomson optical depth lower than unity. In the MAD, magnetohydrodynamic (MHD) turbulence is induced by the plasma instability, such as magnetic Rayleigh-Taylor instability \citep[e.g.,][]{Mckinney2012, Marshall+2018, XieZdziarski2019}. The MHD turbulence heats up thermal plasma and accelerates nonthermal particles. Recent particle-in-cell \citep[e.g.,][]{Galishnikova+2023, Vos+2024} and GRMHD simulations \citep{Ripperda2020, Ripperda2022} show that magnetic reconnection occurs near the BH. Based on these results, our model assumes magnetic reconnection in that region. This magnetic reconnection is expected to accelerate the electrons emitting high-energy gamma rays \citep{Hakobyan2023}, which interact with each other, resulting in copious pair loading into the jet \citep{Kimura+2022, Chen+2022}. The loaded pairs move outward, and at some distance from the BH, the jets are unstable due to some plasma instability \citep{Nishikawa_2016, Borse+2021, Sironi+2021}, triggering magnetic reconnection inside the jets with the entrainment of ambient gas. In our model, we assume that magnetic reconnection accelerates electron-positron pairs in the jets, leading to multi-wavelength photon emission. In our model, jets and MADs are physically connected through the plasma loading and energy extraction processes.

In Sections \ref{subsec:MAD} and \ref{subsec:Jet}, we briefly explain the emission and particle acceleration inside the MADs and the jets, respectively. Very-high-energy gamma rays are absorbed through the Breit-Wheeler process ($\gamma+\gamma \rightarrow e^+ + e^-$) by the radiation field of the companion star. We estimate the optical depth of this process in Section \ref{atn_comp}. We discuss the formation and properties of a MAD in the appendices. In particular, we show the condition required to form a MAD and estimate the corresponding critical mass accretion rate, $\dot{m}_{\rm crit}$, in Appendix \ref{MAD_condition}. We evaluate whether the MAD is in a collisionless system in Appendix \ref{relax_MAD}. We show the method to calculate the particle energy distribution for the MAD model in Appendix \ref{MAD_pair}.

\subsection{MAD model}\label{subsec:MAD}
Below, we summarize the emission model inside the MADs constructed by \cite{Kimura2020} and \cite{RK+2024}. MHD turbulence in the MADs heats up the plasma and accelerates nonthermal particles \citep{Lynn+2014, Kimura2016, Zhdankin2018, KimuraTomida2019, SunBai2021}, which emit multi-wavelength photons via synchrotron radiation.

In the MAD model, we normalize the mass accretion rate, $\dot{M}$, and the size of the emission region, $R_d$, by the Eddington rate and the gravitational radius, respectively, i.e., $\dot{M}c^2 \equiv \dot{m}L_{\rm Edd}$, $R_d \equiv r R_g = r GM_{\rm BH}/c^2$, where $L_{\rm Edd}$ is the Eddington luminosity, $c$ is the speed of light, and $G$ is the gravitational constant. We use the notation $Q_X = Q/10^X$ in cgs unit unless otherwise noted. For the BH mass, we use the notation $M_1 = M_{\rm BH}/(10M_\odot)$. We treat $\dot{m}$ as a free parameter and determine it to fit the multi-wavelength data of the BHXBs. To limit the number of free parameters in our model, we set $r=10$.\footnote{
MAD is defined as a disk where the magnetic flux threading the BH is at a saturated level \citep{Narayan2012}. The size of a MAD is determined by the magnetic flux at the BH horizon or the inner edge of the accretion flows. Recent GRMHD simulations \citep{Ripperda2020, Ripperda2022, Liska+2022} find a hot, corona-like accretion flow within $20 R_g$. Thus, $r\sim 10$ is a reasonable assumption.}

The radial velocity, number density, and magnetic field inside the MAD are analytically estimated to be \citep[see][for parameters for active galactic nuclei]{KimuraMurase2019, Kimura2020}
\begin{eqnarray}
    V_R &\approx& \frac{1}{2}\alpha V_K \simeq 1.5\times10^{9} r_{1}^{-1/2}\alpha_{-0.5}~ \rm cm ~s^{-1}, \\
    n_{p, \rm mad} &\approx& \frac{\dot{M}}{4\pi R_d H V_R m_p} \\ 
    &\simeq& 4.1\times10^{16} \dot{m}_{-1}M^{-1}_1 r_1^{-3/2}\alpha_{-0.5}^{-1}~ \rm cm^{-3}, \nonumber \\
    B_d &\approx & \sqrt{\frac{8\pi n_{p,\rm mad}m_p C_s^2}{\beta}} \\
    &\simeq& 2.0\times10^{7} \dot{m}_{-1}^{1/2} M_{1}^{-1/2}r_{1}^{-5/4} \alpha_{-0.5}^{-1/2}\beta_{-1}^{-1/2}~ \rm G, \nonumber    
\end{eqnarray}
where $\alpha$ is the viscous parameter \citep{Shakura1973}, which we leave it as a free, $V_K = \sqrt{GM_{\rm BH}/R_d} $ is the Keplerian velocity,  $H\approx (C_s/V_K)R_d\approx R_d/2$ is the scale height, $C_s \approx V_K/2$ is the sound velocity, and $m_p$ is the proton mass.

In the MAD model, we consider four particle species: thermal electrons, primary protons, secondary electron-positron pairs produced by the Breit-Wheeler process, and those by the Bethe-Heitler process ($p+\gamma \rightarrow p+e^+ + e^-$). The electrons are thermalized through the Coulomb collisions inside the MAD when $\dot{m}\gtrsim 10^{-3}$. Thus, we do not consider the primary electrons in the paper (see Appendix \ref{relax_MAD}).

Thermal electrons heated by the MHD turbulence emit multi-wavelength photons via synchrotron and Comptonization processes. We determine the electron temperature by balancing the heating and the cooling rates. As the electron heating process inside the MAD is still unknown \citep[e.g.,][]{Kawazura2020}, we set the electron-proton-heating ratio, $Q_i/Q_e$, as a parameter and set it to be 1. We calculate the cooling rates of the thermal electrons in the same manner as \citet{Kimura2015}.

The MHD turbulence accelerates primary protons, emitting nonthermal synchrotron radiation. To obtain the energy distribution of nonthermal particles, we solve the energy transport equation with one-zone and steady-state approximations (see Appendix \ref{MAD_pair} for a more detailed calculation method). We assume the injection term to be a power-law with an index $s_{\rm inj}$ and an exponential cutoff at a cutoff energy. We determine $s_{\rm inj}$ from the power-law solution of the transport equation in Section 2 of \citet{Stawarz2008}, where a hard-sphere-like momentum diffusion coefficient, $D(p)\propto p^2$, is assumed. We estimate the cutoff energy in two ways. In one approach, it is obtained by balancing the acceleration and energy loss timescales, referred to as $E_{\rm cut, cl}$. In the other approach, it is evaluated as the Hillas energy, $E_{\rm Hillas} \approx eB_dR_d(V_A/c) \simeq 3.0\times10^{15} ~ B_{d, 6}R_{d, 7}(V_A/c) ~ \rm eV$, where $e$ is the electric charge, $B_d$ is the magnetic field strength in the MAD, $V_A/c \approx 0.71 r_1^{-1} \beta_{-1}^{-1/2}$ is the Alfv\'en velocity. We set the cutoff energy as the lower one of these, $E_{\rm cut}={\rm min}(E_{\rm cut, cl},~E_{\rm Hillas})$.

MHD turbulence accelerates the particles stochastically, and we set the effective mean free path for interacting with the turbulence as $\eta_{\rm turb} H$, where $\eta_{\rm turb}$ is the numerical factor. We treat $\eta_{\rm turb}$ as a free not to have a higher value than unity. We account for the following energy loss processes: diffusive escape, infall to the BH, and relevant cooling processes, including synchrotron radiation, Bethe-Heitler process, $pp$ inelastic collision, and photomeson production. 

Nonthermal protons interact with thermal protons through the inelastic $pp$ collisions and optical-to-X-ray photons through the Bethe-Heitler process and photomeson production. These processes cause the production of secondary electron-positron pairs and pions. Produced secondary electron-positron pairs and pions emit high-energy gamma rays. Since the production of the secondary electron-positron pairs and pions depends on the photon spectrum, we calculate the photon and the pair spectrum iteratively until they converge (see Appendix \ref{MAD_pair}). The pion decay also emits high-energy neutrinos, and we calculate the neutrino spectrum based on Section 2 of \citet{Kimura2020}.

In BHXBs, a standard geometrically thin disk may exist outside the MAD. Such a geometry is consistent with the observation of Cygnus X-1 by Imaging X-ray Polarimetry Explorer \citep[IXPE;][]{IXPE2022}, though other geometries have also been proposed (for example, \citealp{Moscibrodzka2024}). At an even larger distance, the companion star provides a radiation field. Photons from the standard disk and the companion star can be targets of the hadronic and thermal Comptonization processes. We evaluate their number densities assuming a truncation radius of the standard disk as $R_{\rm disk} \sim 10^2 R_g$. We find that the densities of these photons are low due to their large distances from the gamma-ray emission region in the MAD. Therefore, we ignore both components in our calculation. In the hard state of Cygnus X-1, the truncation radius is estimated to be $\sim15 R_g$ \citep[e.g.,][]{Suzaku2008}, but the contribution by the disk photons is still sub-dominant to the MAD photons.

\begin{table*}[t]
\caption{Physical quantities for each BHXB.}\label{tab1}
\begin{tabular}{r|lllll}
\hline \hline
 \ & Cygnus X-1 & MAXI J1820+070 &GRO J1655-40 & GX 339-4 & XTE J1118+480 \\
\hline 
$M_{\rm BH}(M_{\odot})$ & 21 & 7.0 & 6.3 & 9.0 & 7.0 \\
$\Gamma_j$ & 2.55 & 2.1 & 3.88 & 2.55 & 1.01 \\
$\theta_j$ (deg) & 27.5 & 70 & 70 & 40 & 68 \\
$L_{\rm star}~\rm (erg ~ s^{-1})$ & $1.6 \times10^{39}$ & $7.1 \times10^{32}$ & $1.4 \times10^{35}$ & $2.5 \times10^{34}$ & $2.7\times10^{31}$ \\
$T_{\rm star}$ (K)&$3.1\times10^4$ & $4.7\times10^3$ & $6.3\times10^3$ &$3.9\times10^3$ & $2.8\times10^3$ \\
$a_{\rm binary}$ (cm) &$3.7\times10^{12}$ & $4.5\times10^{11}$ & $1.2\times10^{12}$ & $8.1\times10^{11}$ & $1.8\times10^{11}$ \\
$d_L$ (kpc) & 2.2 & 3.0 & 1.7 & 8.0 & 1.8 \\
$\delta$ (deg) & 35 & 7.2 & -39 & -48 & 48 \\
\hline
\end{tabular}
\tablecomments{$M_{\rm BH}$, $\Gamma_j$, $\theta_j$, $L_{\rm star}$, $T_{\rm star}$, $a_{\rm binary}$, $d_L$, and $\delta$ are the mass of the black hole, jet Lorentz factor, jet inclination angle, the luminosity of the companion star, the temperature of the companion star, the separation distance, luminosity distance from Earth to each BHXB, and declination of each BHXB, respectively. The references for each physical quantity of individual BHXBs are \citet{Wood+2021}, \citet{Yoshitake+2024}, \citet{Koljonen+2023} for MAXI J1820+070, \citet{Tetarenko+2019}, \citet{MillerJones+2021} for Cygnus X-1, \citet{Saikia+2019}, \citet{Migliari+2007}, \cite{Shidatsu+2016}, \cite{Liu+2007_GRO} for GRO J1655-40, \citet{Heida2017}, \citet{Kosenkov+2020}, \citet{Malzac+2018}, \citet{Buxton+2012} for GX 339-4, and \citet{Chatterjee+2019}, \citet{Saikia+2019}, \citet{Hernandez+2012}, \citet{McClintock+2001} for XTE J1118+480.}
\end{table*}

\begin{table*}[t]
\centering
\caption{The list of the parameters and the physical quantities of the Jet-MAD model for various BHXBs.} \label{MAD_parameter_BHXB}

Parameters of the Jet-MAD model for each BHXB
\begin{tabular}{rllllll}
\hline \hline 
& Cygnus X-1 & MAXI J1820+070 & GRO J1655-40 & GX 339-4 & XTE J1118+480 &\\ \hline
$\dot{m}$ & 0.1 & $5\times10^{-3}$ & 0.03 & 0.1 & 0.1 &\\
$r^{\dagger}$ & 10 & 10 & 10 &10 & 10&\\
$Q_i/Q_e^{\dagger}$ & 1.0 & 1.0& 1.0& 1.0& 1.0&\\
$\eta_{\rm turb}$ & 1.0 & 0.44 & 0.33 & 0.66 & 0.47 &\\
$\alpha$ & 0.3 & 0.1 & 0.11 & 0.25 & 0.3& \\
$\beta$ & 0.135 & 0.8 & 0.7 & 0.18 & 0.5& \\
$\epsilon_{\rm NT}$& $5.0\times10^{-3}$& 0.33 & $3.6\times10^{-3}$& $3.3\times10^{-3}$& $3.3\times10^{-3}$& \\
$\epsilon_{\rm dis}$ & 0.1 & 0.15 & 0.14 & 0.15 & 0.15 &\\ \hline 
$\sigma_{\rm ent}^{\dagger}$ & $10^{-3}$ & $10^{-3}$& $10^{-3}$& $10^{-3}$& $10^{-3}$&\\
$p^{\dagger}$ & 2.1 & 2.1& 2.1& 2.1& 2.1 &\\
$\xi^{\dagger}$ & $10^3$ & $10^3$& $10^3$& $10^3$& $10^3$&\\
$\delta B/B^{\dagger}$& 0.35 & 0.35& 0.35& 0.35& 0.35&\\
$R_j^{\dagger}~ (R_g)$ & $10^3$ & $10^3$& $10^3$& $10^3$& $10^3$& \\
\hline  
\end{tabular}   
\begin{tabular}{c}
Parameters with $\dagger$ are fixed, and the others are free.
\\ \\
\end{tabular}

Derived physical quantities for each BHXB
    \begin{tabular}{r|llllll}
    \hline \hline
        \ & Cygnus X-1 & MAXI J1820+070 & GRO J1655-40 & GX 339-4 & XTE J1118+480 \\ \hline
        $s_{\rm inj}$ & 1.21 & 1.24 & 1.12 & 1.13 & 1.17 \\
        $\theta_e$ & 0.25 & 0.81 & 0.34 & 0.22 & 0.66 \\
        $B_d$ (G) & $1.2\times10^7$ & $3.3\times10^6$ & $8.7\times10^6$ & $1.7\times10^7$ & $5.9\times10^6$ \\
        $\dot{m}_{\rm crit}$ & 0.22 & 0.15 & 0.049 & 0.13 & 0.96 \\
        $E_{p,\rm max}$ (GeV) & $1.6\times10^5$ & $2.7\times10^5$ & $3.3\times10^6$ & $2.3\times10^5$& $3.1\times10^5$ \\ 
        $E_{\pi, \rm cut}$ (GeV) &$7.9\times10^3$ & $1.8\times10^4$ &$1.0\times10^4$ & $5.2\times10^3$& $1.3\times10^4$& \\
        \hline
        $\sigma_\pm$ & $1.1\times10^2$ & $2.4\times10^2$ & $1.9\times10^2$ & $9.3\times10^1$ & $1.3\times10^2$ \\
        $B_{\rm dis}~\rm(G)$ & $6.0\times10^2$ &  $2.3\times10^2$ & $8.4\times10^2$ & $2.2\times10^3$ & $1.4\times10^3$ \\
        $n_e ~ \rm (cm^{-3})$ & $1.9\times10^{10}$ & $4.3\times10^{9}$ & $2.6\times10^{9}$ & $4.4\times10^{10}$ & $8.9\times10^{10}$ \\
        $L_j ~ \rm (erg~s^{-1})$ & $1.6\times10^{38}$& $2.6\times10^{36}$ & $1.4\times10^{37}$& $6.7\times10^{37}$& $1.6\times10^{37}$ \\ 
        $L_e~ \rm (erg~s^{-1})$ & $9.7\times10^{34}$ & $2.6\times10^{33}$ & $3.6\times10^{33}$ & $4.1\times10^{34}$ & $5.6\times10^{34}$ \\
        $\delta_D$ & 2.3 & 0.68 & 0.38 & 1.3 & 1.02 \\
        \hline
    \end{tabular}
    \tablecomments{$\theta_e= k_BT_e/(m_ec^2)$ is the normalized electron temperature, $k_B$ is the Boltzmann constant, $L_j$ is the jet luminosity where the pairs are injected at the jet base, and $L_e$ is the electron luminosity at the jet dissipation region.}
    
\end{table*}

\subsection{Jet model}\label{subsec:Jet}
In this subsection, we briefly explain the pair loading into the jet, particle acceleration, and the emission from the jet. A more detailed model description is shown in \citet{RK+2024}.

In the MAD state, magnetic reconnection periodically occurs in the vicinity of the BH, where magnetic energy density is much higher than the plasma energy density \citep{Ripperda2022}. Magnetic reconnection in such a magnetic-energy dominant region accelerates the surrounding plasma particles, and the accelerated electrons emit high-energy photons. These photons interact with each other and produce electron-positron pairs, forming a blob. We determine the magnetization parameter of the blob formed at the jet base, $\sigma_\pm$, by equating the pair production rate with the pair escape rate \citep{Chen+2022}. We find that $\sigma_\pm \simeq 10^2 \gg 1$ is satisfied for typical BHXB parameters, showing that the jet base is highly magnetized.

The blob created at the jet base travels outward, and we assume that the blob dissipates its energy at $\sim3\times10^4R_g$ away from the BH. At the outer part of the jet, the toroidal magnetic field is dominant, where we can write $B\propto R^{-1}$ owing to the magnetic flux conservation, where $R$ is the cylindrical radial distance from the jet axis. If the number flux of the jet plasma is conserved, i.e., $n_e\propto R^{-2}$, then the magnetization parameter conserves along the jet. However, the blob expands along the magnetic field lines due to their velocity dispersion, causing a more rapid decline in the number density. Thus, by introducing a numerical factor $\xi \simeq 10^3$, the magnetization parameter in the jet dissipation region can be written as $\xi \sigma_\pm$. We fix $\xi\simeq 10^3$ since the interval time of the reconnection flare at the BH magnetosphere is $\sim 10^3 R_g/c$ \citep{Ripperda2022}.

Since $\xi \sigma_\pm \gg 1$, magnetic energy is dominant as the internal energy inside the jet, and the magnetic energy dissipation accelerates the pairs. Due to the velocity difference between the jet and the ambient gas, we consider that Kelvin-Helmholtz instability (KHI) is triggered at the jet edge \citep{Sironi+2021}. The KHI generates turbulence inside the jet, which induces magnetic reconnection \citep[e.g.,][]{Yang+2024}. Owing to KHI, electron-proton plasma can be entrained into the jet from the ambient gas. Thus, the magnetization parameter decreases due to the plasma entrainment and the magnetic reconnection. Turbulent magnetic reconnection accelerates the nonthermal particles \citep[e.g.,][]{HoshinoPhyLV2012, Xu2023}. We define the magnetization parameter after plasma entrainment, $\sigma_{\rm ent}$, given as a free parameter. This parameter determines the electron number density after plasma entrainment. 

We solve the energy transport equation with the one-zone and the steady-state approximations to obtain the electron energy distribution (see Equation (7) of \citealp{RK+2024}). We assume that the injection term follows a power-law with an index  $p$ given as a free parameter (see Equation (8) of \citealp{RK+2024}). We determine the maximum energy as $\gamma_{\rm max} = (\delta B /B)^2 \xi \sigma_\pm$, where $\delta B$ is the amplitude of the turbulent magnetic field, which is free and tuned to fit the multi-wavelength data.

Nonthermal electrons lose their energy through synchrotron radiation, synchrotron self-Compton scattering (SSC), adiabatic cooling, and diffusive escape by resonant scattering with the KHI-induced turbulence (see Appendix D of \citealp{RK+2024}). Since the jet emission region is far from the BH, the photons emitted from the MADs and the companion star cannot be the dominant target photons of the inverse Compton scattering, and we ignore these components in our calculation of the jet emission.

We calculate the jet emission as $\nu_{\rm obs}L_{\nu,\rm obs} = \delta_D^4 \nu_{\rm src}L_{\nu, \rm src}$, where $\nu_{\rm obs}L_{\nu, \rm obs}$ is the observed luminosity, $\delta_D$ is the Doppler factor, and $\nu_{\rm src}L_{\nu, \rm src}$ is the intrinsic photon spectrum obtained by the jet model. We estimate the Doppler factor of each BHXB using the estimated jet Lorenz factor, $\Gamma_j$, and the jet inclination angle, $\theta_j$ (see Table \ref{tab1}).

\subsection{Photo-attenuation by the companion star}\label{atn_comp}
\begin{figure}
    \centering
    \includegraphics[width=\linewidth]{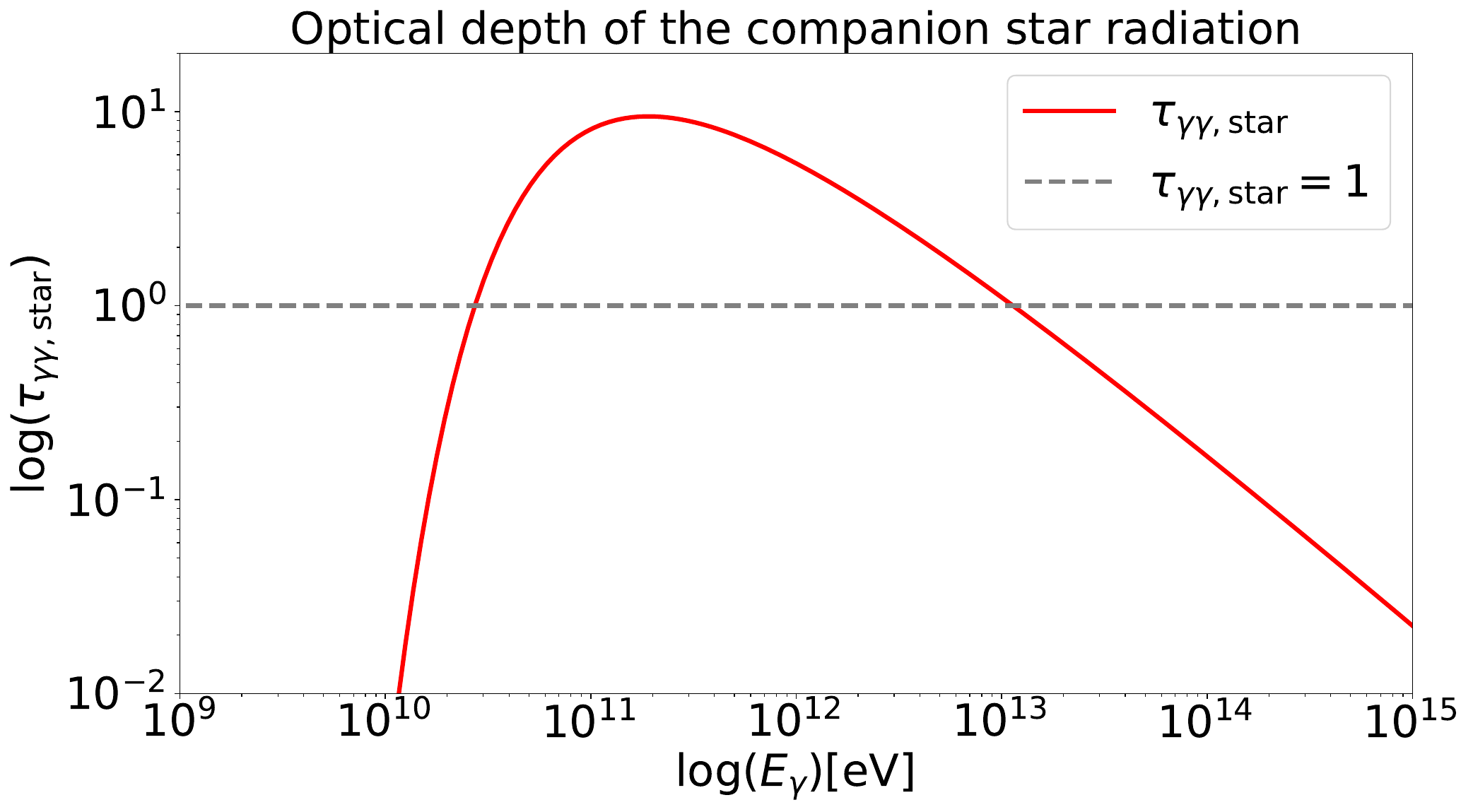}
    \caption{Optical depth of the Breit-Wheeler process by the companion star radiation for Cygnus X-1 as a function of the target photon energy (red line). The gray dashed line represents $\tau_{\gamma\gamma, \rm star} = 1$.}
    \label{opt-depth}
\end{figure}

Very-high-energy gamma rays produced in the MADs and jets may be absorbed by the radiation field of the companion star when they leave the source. We calculate the optical depth of the Breit-Wheeler process by the companion star radiation for Cygnus X-1, $\tau_{\gamma\gamma, \rm star}$, following the method in \citet{Coppi1990}. Figure \ref{opt-depth} shows the phase-averaged $\tau_{\gamma\gamma, \rm star}$ as a function of the gamma-ray photon energy. In general, the optical depth is less than unity at $E_\gamma \gtrsim 10^{13}~\rm eV$. Since the magnetic field around the companion star is not strong,  cascade pairs cool by the inverse Compton scattering process. We calculate the attenuation factor as $F_{\rm atn, star} \sim 1/\tau_{\gamma\gamma, \rm star}$. As $F_{\rm atn, star} \lesssim 1$ at $E_{\gamma} \gtrsim 10^{13}$ eV, photons above 10~TeV are not absorbed by the companion star of Cygnus X-1. 

Using the physical quantities tabulated in Table \ref{tab1}, we find  $\tau_{\gamma\gamma, \rm star} \lesssim 0.1$ in other BHXBs. Thus, their companion stars do not absorb very-high-energy gamma rays. 

\section{Photon and neutrino emissions from individual BHXBs} \label{sec:results}
We apply our Jet-MAD model to five BHXBs: Cygnus X-1, MAXI J1820+070, GRO J1655-40, GX 339-4, and XTE J1118+480. We selected these BHXBs because they have better-quality observational data than other BHXBs. We list the physical quantities for each BHXB in Table \ref{tab1} and the model parameter sets and derived physical quantities in Table \ref{MAD_parameter_BHXB}.

We note that UHE gamma rays have also been observed from SS 433 \citep{HAWC:2018gwz}, V4641 Sgr \citep{Alfaro:2024cjd}, and GRS 1915+105 \citep{LHAASO2024_mq}. These gamma-ray sources are extended and likely originate from structures around the binaries, such as the large-scale jets of the microquasars. Thus, we do not discuss these three objects in this paper.

Below, we first analytically estimate the typical photon energies, gamma-ray luminosity, efficiencies of each physical process, and the internal optical depth of MADs in Section \ref{typ}. We discuss the results of the photon spectra in Sections \ref{Cygnus}, \ref{MAXI}, and \ref{others:ph} for Cygnus X-1, MAXI J1820+070, and the other BHXBs, respectively. We show the results of the neutrino spectra for five BHXBs in Section \ref{neutrino}.

\subsection{Typical features of photon and neutrino spectra}\label{typ}
\begin{figure}[tbp]
    \centering
    \includegraphics[width=\linewidth]{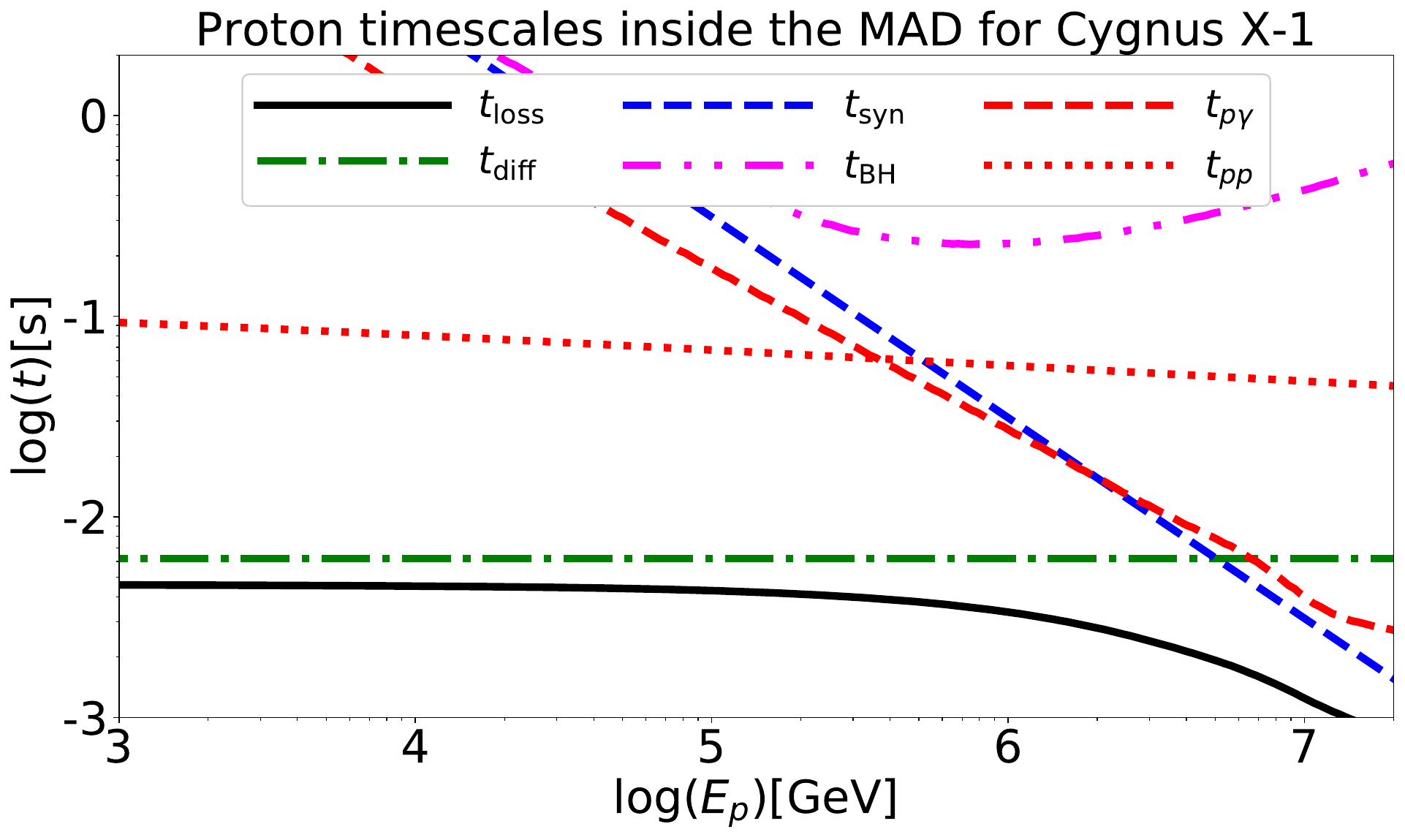}
    \includegraphics[width=\linewidth]{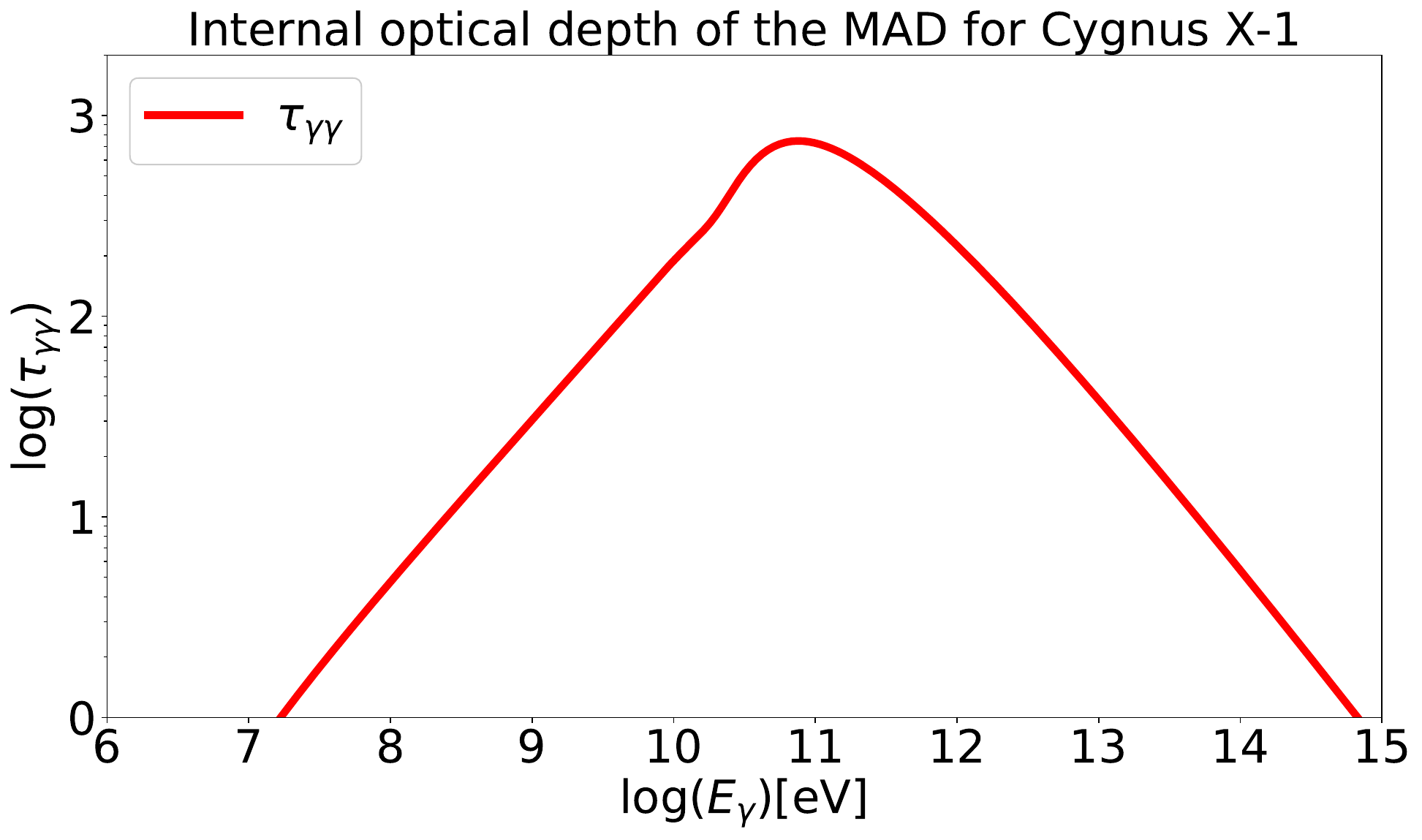}
    \includegraphics[width=\linewidth]{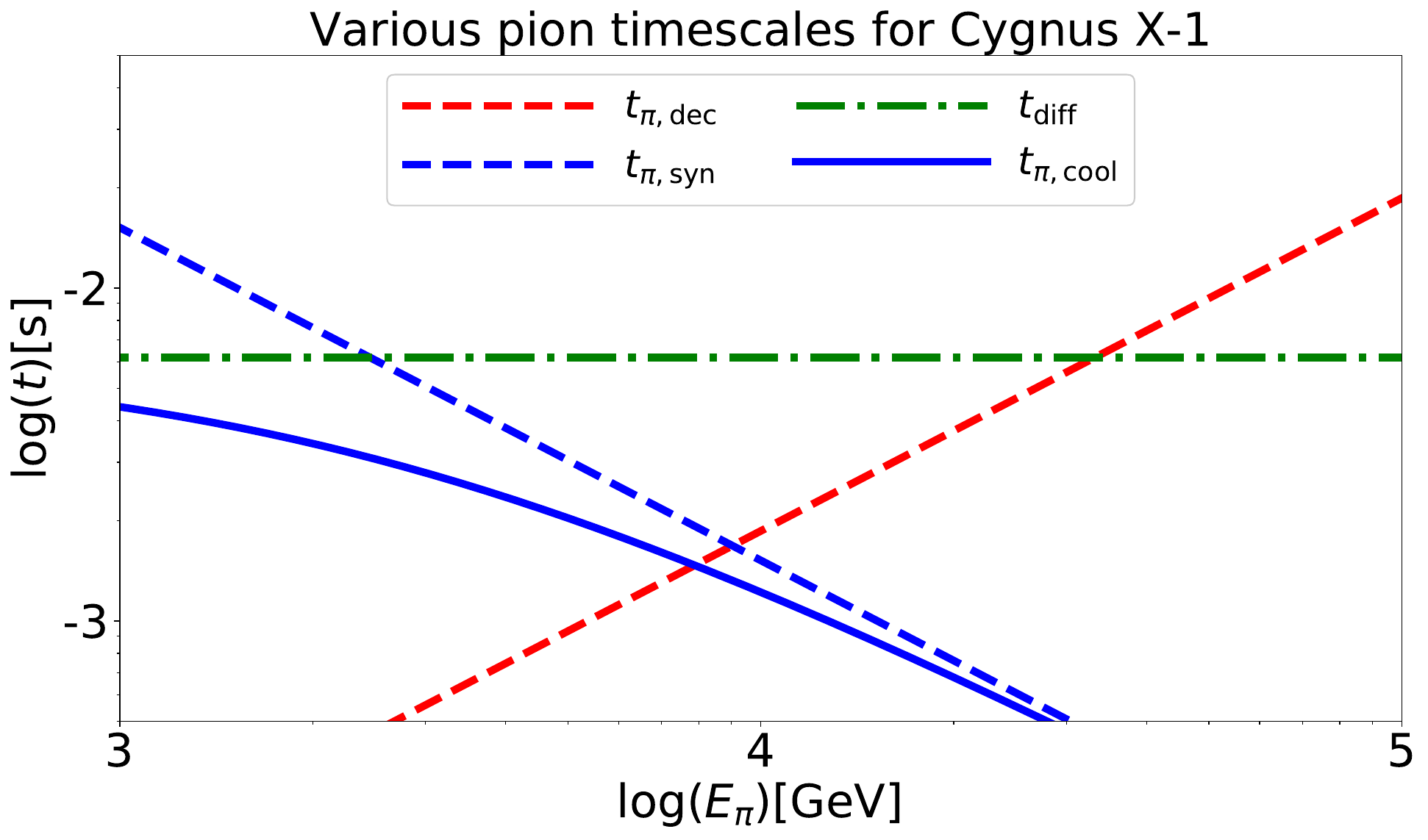}
    \caption{{\it Top}: Proton timescales in the MAD for Cygnus X-1 as a function of the proton energy. Here $t_{\rm loss}$, $t_{\rm fall}$, $t_{\rm diff}$, $t_{\rm syn}$, $t_{\rm BH}$, $t_{pp}$, and $t_{p\gamma}$ are the energy loss (black solid; $t_{\rm loss}^{-1} = t_{\rm cool}^{-1} + t_{\rm esc}^{-1}$, where  $t_{\rm cool}$ is the cooling time due to various interaction processes and $t_{\rm esc}$ is the escape time defined as $t_{\rm esc}^{-1} = t_{\rm diff}^{-1} + t_{\rm fall}^{-1}$), diffusion (green dot-dashed), synchrotron (blue dashed), Bethe–Heitler (magenta dot-dot-dashed), $pp$ inelastic collision (red dotted), and photomeson production (red dashed) timescales, respectively. {\it Middle}: Optical depth of the Breit-Wheeler process inside the MAD for Cygnus X-1 as a function of the photon energy. Low-energy photons emitted by the thermal electrons absorb the GeV-TeV gamma rays. {\it Bottom}: Various pion timescales inside the MAD for Cygnus X-1 as a function of the pion energy. Here $t_{\pi, \rm dec}$, $t_{\pi, \rm syn}$, $t_{\pi, \rm cool}$ are the pion decay (red dashed), pion synchrotron (blue dashed), and pion cooling (blue solid), respectively.}
    \label{timescale-fraction}
\end{figure}

\begin{figure*}
    \centering
    \includegraphics[width=0.9\linewidth]{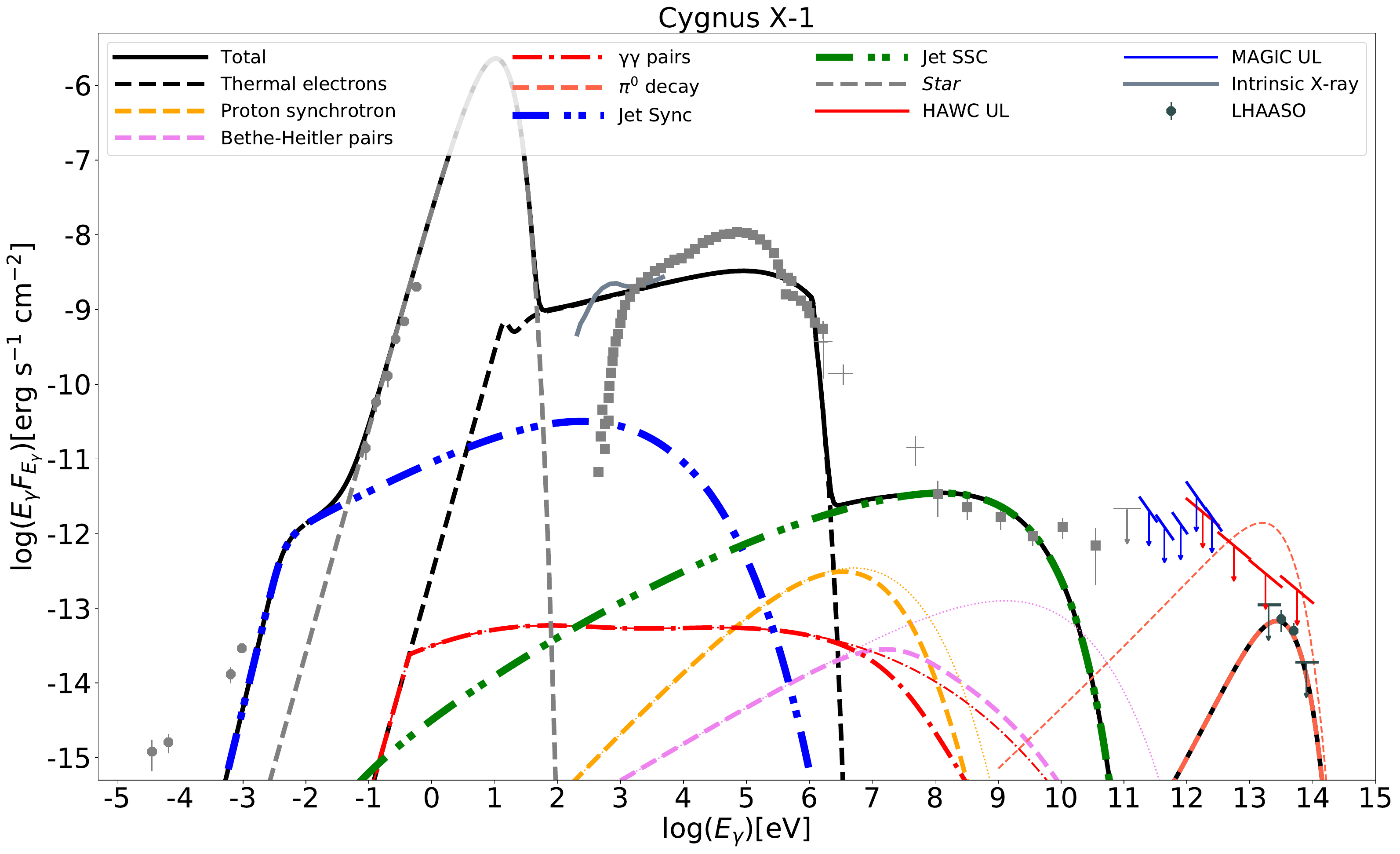}
    \caption{Photon spectra obtained by our model for Cygnus X-1. The thick and thin lines are the photon spectra after and before internal attenuation by the Breit-Wheeler process, respectively. Data points are taken from \citet{Zdziarski+2017, LHAASO2024_mq}. The gray solid line in the X-ray band is the soft X-ray spectrum before absorption by the interstellar medium. The sensitivity line for CTA is taken from \citet{CTA2011}. The upper limit lines for HAWC and MAGIC are taken from \citet{HAWC2021} and \citet{MAGIC2017}, respectively.}
    \label{CygX-1}
\end{figure*}
\begin{figure*}
    \centering
    \includegraphics[width=0.9\linewidth]{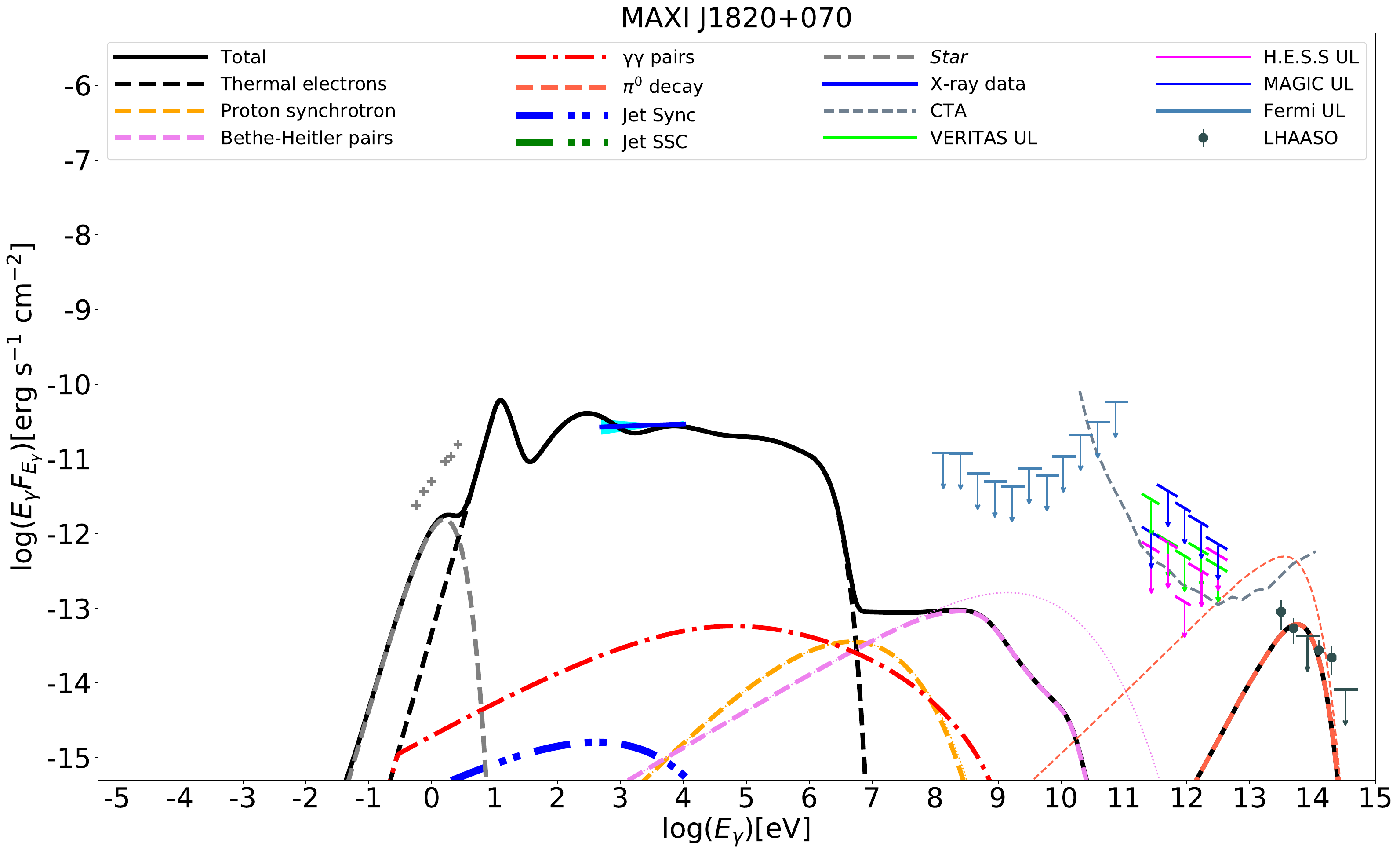}
    \caption{Same as Figure \ref{CygX-1}, but for MAXI J1820+070. We take the data points (gray) when MAXI J1820+070 is in the hard state from \citet{Yoshitake+2024}. The cyan bow-tie and blue solid line are the faintest X-ray data after 400 days of the outburst in 2015 \citep{Arabaci+2022}. The upper limit lines of Fermi, VERITAS, H.E.S.S., and MAGIC are taken from \citet{Abe+2022}. The sub-PeV data points are taken from \citet{LHAASO2024_mq}}
    \label{MAXIJ1820}
\end{figure*}
\begin{figure*}
    \centering
    \includegraphics[width=0.6\linewidth]{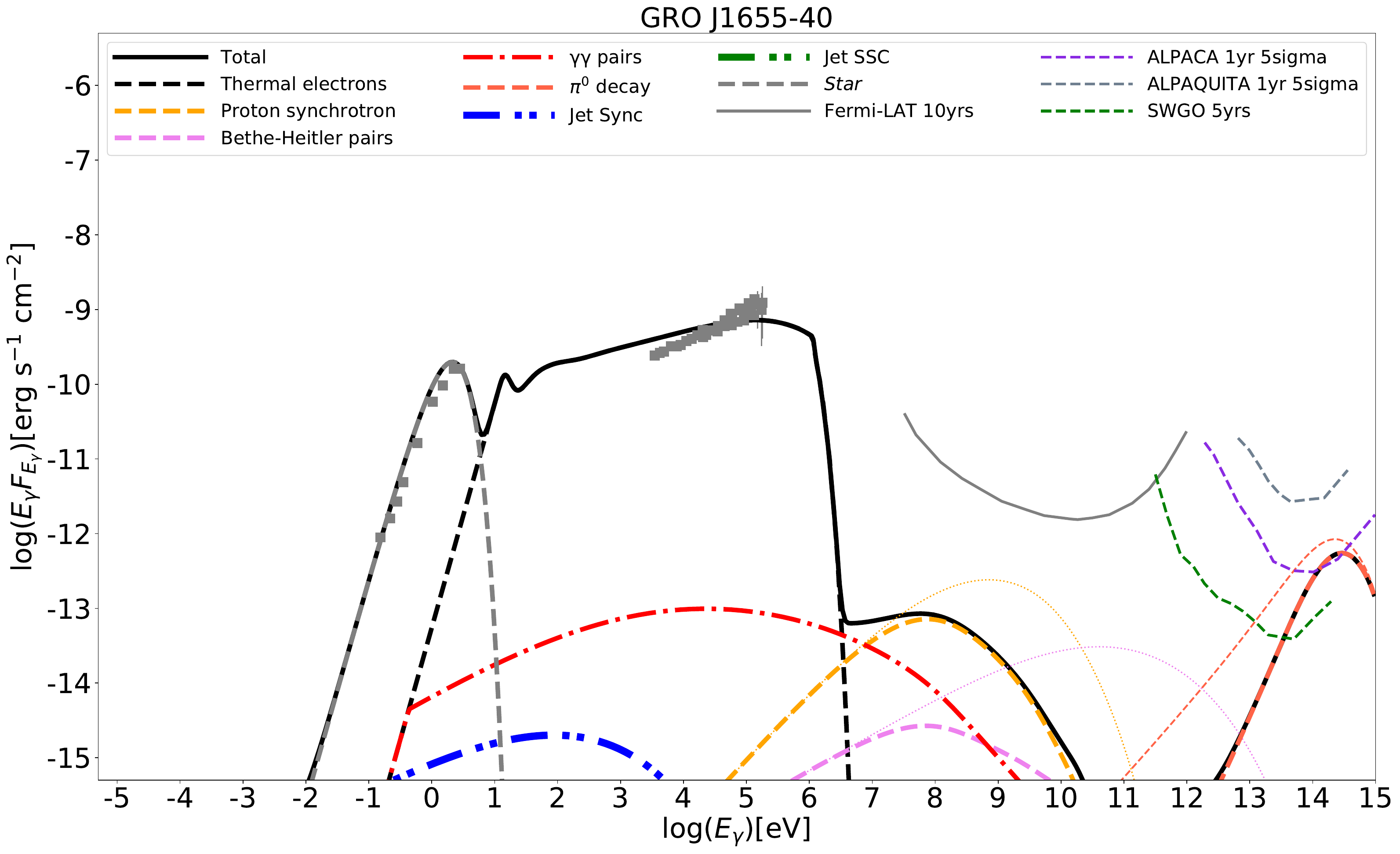}
     \includegraphics[width=0.6\linewidth]{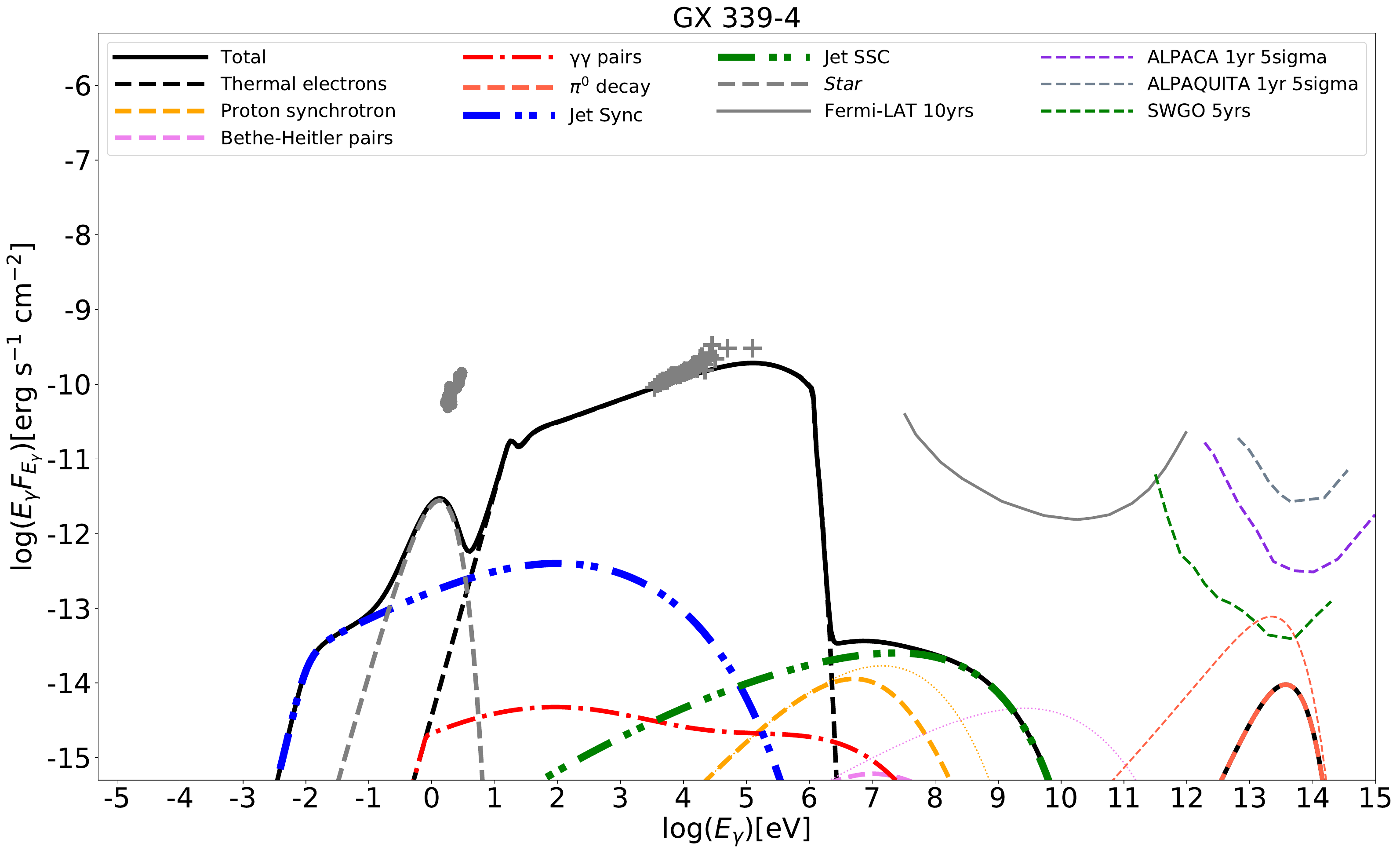}
    \includegraphics[width=0.6\linewidth]{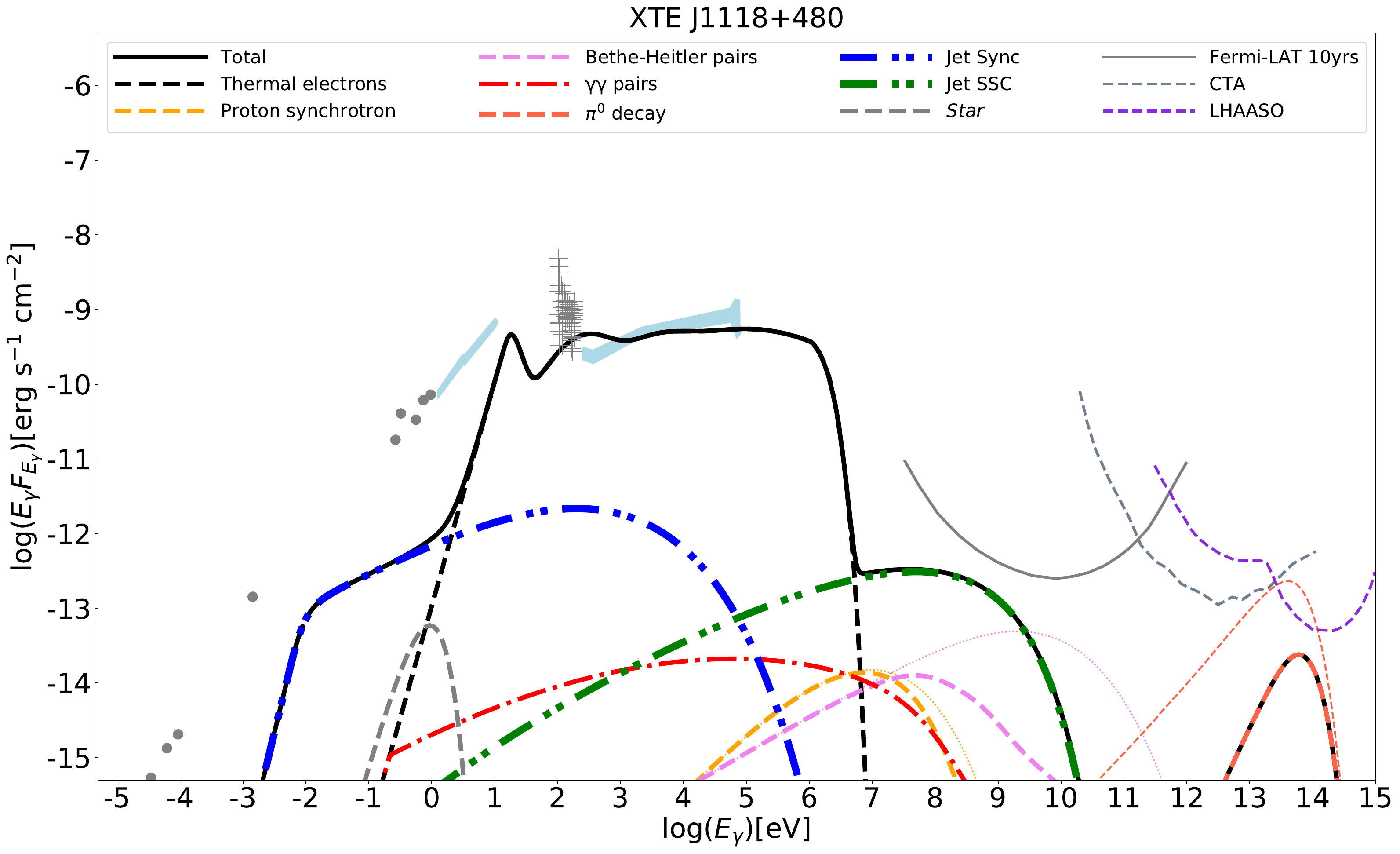}
    \caption{Same as Figure \ref{CygX-1}, but for GRO J1655-40 ({\it top}), GX 339-4 ({\it middle}), and XTE J1118+480 ({\it bottom}). Data points are taken from \citet{Migliari+2007} for GRO J1655-40, \citet{Gandhi+2010} for GX 339-4, and \citet{Chaty+2003} for XTE J1118+480. The Fermi sensitivity line is taken from \url{https://www.slac.stanford.edu/exp/glast/groups/canda/lat_Performance.htm}. The ALPACA, LHAASO, and SWGO sensitivity lines are taken from \citet{ALPACA2022}, \citet{LHAASO2019}, and \citet{SWGO2019}, respectively.}
    \label{GROJ1655-40}
\end{figure*}

The jets emit the radio signals and GeV gamma rays by the synchrotron radiation and the SSC, respectively, and MADs emit very-high-energy gamma rays by the pion decay produced by the $pp$ inelastic collisions.

The electron maximum Lorentz factor in the jets is $ \gamma_{\rm max} = (\delta B/B)^2 \xi \sigma_\pm \simeq 10^4~(\delta B/B)_{-0.5}^2 \xi_3 \sigma_{\pm,2} $, corresponding to a synchrotron photon energy  
\begin{align}
E_{\gamma, \rm max, syn} &= \gamma_{\rm max}^2 \frac{heB_{\rm dis}}{2\pi m_ec} \nonumber \\ 
&\simeq 1.2\times10^3 \left(\frac{\delta B/B}{10^{-0.5}}\right)^4 \xi_3^2 \sigma_{\pm, 2}^2 B_{\rm dis, 3} ~ \rm eV,
\end{align}
where $h$ is the Planck constant, $B_{\rm dis}$ is the magnetic field at the jet dissipation region, which we estimate by the equation of the energy conservation at the jet dissipation region (see Equation (10) of \citet{RK+2024}), and $m_e$ is the electron mass. Since the jet is in a slow cooling regime and $p<3$, the synchrotron photon flux peaks at $E_\gamma \simeq E_{\gamma, \rm max, syn}$ (see Table \ref{MAD_parameter_BHXB}). Due to the Klein-Nishina effect, the SSC photon spectrum peaks when the electron Lorentz factor is $\gamma_e \simeq 10^3$. The corresponding SSC peak photon energy is $E_{\gamma, \rm pk, SSC} \approx \gamma_e^2 E_{\gamma, \rm max, syn} \simeq 1.2\times 10^9 $ eV.

The cutoff energy of the nonthermal protons in the MADs is $E_{p, \rm cut}\sim 1\times10^{14}$ eV. The luminosity of nonthermal protons  peaks at $E_p \sim E_{p,\rm cut}$ since $s_{\rm inj} \lesssim 2$ (see Table \ref{MAD_parameter_BHXB}). Photons from the pion decay carry an energy $E_\gamma \simeq 0.1E_p\sim 1\times 10^{13}$~eV. The nonthermal proton luminosity is estimated to be $L_p \approx \epsilon_{\rm dis}\epsilon_{\rm NT}\dot{M}c^2 \simeq 4.0\times10^{34} ~\epsilon_{\rm dis, -1}\epsilon_{\rm NT, -2.5}\dot{m}_{-1}M_{1} ~ \rm erg ~ s^{-1}$, where $\epsilon_{\rm dis}$ is the ratio of the dissipation power to the accretion power and $\epsilon_{\rm NT}$ is the ratio of the nonthermal particle energy to the dissipation energy. We tune these plasma parameters to fit the multi-wavelength data but restrict them not to taking unphysical values based on the plasma simulations \citep[][and see Appendix \ref{MAD_pair}]{Zhdankin+2019}. By denoting the fraction of the hadronic process cooling timescale to the energy loss timescale as $f_\pi \sim 0.1$, the gamma-ray luminosity emitted by the hadronic processes is estimated to be 
\begin{align}
L_{\gamma, \rm had} &= \frac13 f_\pi L_p \label{Lgamma,had}\\
&\simeq 1.4\times10^{33}~ f_{\pi, -1}\epsilon_{\rm dis, -1}\epsilon_{\rm NT, -2.5}\dot{m}_{-1}M_{1} ~ \rm erg ~ s^{-1}.\nonumber 
\end{align}
The prefactor $1/3$ in the above expression arises because in inelastic $pp$ collisions, neutral pions are produced with roughly 1/3 probability. These UHE photons are attenuated by the lower-energy photons in MADs by the Breit-Wheeler process. The pion decay process also produces neutrinos with a similar luminosity to that given in Equation (\ref{Lgamma,had}), but neutrinos are not attenuated. 

We show various interaction timescales as a function of the proton energy for Cygnus X-1 in the top panel of Figure \ref{timescale-fraction}. For $E_p \lesssim 10^{14}$ eV, the accelerated protons will escape from the MADs diffusively, as shown in Figure \ref{timescale-fraction}. For $E_p \sim 10^{14}$ eV, the $pp$ inelastic collision cooling timescale and energy loss timescale are $t_{pp} \simeq 6\times10^{-2}$ s and $t_{\rm loss} \simeq 4\times10^{-3}$ s, respectively, and thus, the assumption of $f_\pi \sim 0.1$ used in Equation (\ref{Lgamma,had}) is reasonable.

We also show the optical depth of the Breit-Wheeler process inside the MAD for Cygnus X-1 in the middle panel of Figure \ref{timescale-fraction}. Photons in the optical to X-ray bands emitted by the thermal synchrotron radiation and the thermal Comptonization process absorb the TeV and GeV gamma rays emitted by the decay of the neutral pions and synchrotron radiation by the secondary pairs produced via the Bethe-Heitler process, respectively.

High-energy neutrinos are emitted by the decay of charged pions produced by the hadronic processes. Since the magnetic field strength inside the MAD is high, the charged pions cool through the synchrotron radiation efficiently. If the pion cooling timescale, $t_{\pi, \rm cool}$, is shorter than the charged pion decay timescale, $t_{\pi, \rm dec}$, neutrino spectra are suppressed by a factor of $f_{\pi, \rm sup}\approx 1-\exp(t_{\pi, \rm dec}^{-1}/t_{\pi, \rm cool}^{-1})$, where $t_{\pi, \rm cool}^{-1} =t_{\pi, \rm syn}^{-1} + t_{\rm diff}^{-1}$ and $t_{\pi, \rm syn}$ is the pion synchrotron cooling timescale. We estimate $t_{\pi, \rm dec}$ and $t_{\pi,\rm syn}$ as $t_{\pi, \rm dec} = (E_\pi/m_\pi c^2)t_{\pi,0}$ and $t_{\pi, \rm syn} = (6\pi m_\pi^4 c^3)/(m_e^2 \sigma_T B_d^2E_\pi)$, where $E_\pi$ is the pion energy, $m_\pi$ is the charged pion mass, and $t_{\pi,0}$ is the decay time for charged pion, respectively. The bottom panel of Figure \ref{timescale-fraction} shows the pion cooling timescales in Cygnus X-1. Above $E_{\pi, \rm cut} $, charged pions lose their energy before decaying into neutrinos. This cutoff energy may be estimated by equating $t_{\pi, \rm dec}$ with $t_{\pi, \rm cool}$ assuming that synchrotron radiation is the dominant cooling process, 
\begin{align}
E_{\pi, \rm cut} &\approx \sqrt{\frac{6\pi m_\pi^5 c^5}{m_e^2 \sigma_T B_d^2 t_{\pi, 0}}} \\ 
&\simeq 5.6\times10^{3} \dot{m}_{-1}^{-1/2}m_1^{1/2}r_1^{5/4}\alpha_{-0.5}^{1/2} \beta_{-1}^{1/2}~\rm GeV. \nonumber
\label{Epicut}
\end{align}
We tabulate $E_{\pi, \rm cut}$ of each BHXB in Table \ref{MAD_parameter_BHXB}. A corresponding suppression of the neutrino spectrum is then expected at $E_\nu \approx 0.25E_{\pi, \rm cut}$.

\subsection{Cygnus X-1}\label{Cygnus}
Figure \ref{CygX-1} presents our model for the multi-wavelength emission from Cygnus X-1. The MAD pion decay component shows up around $\sim 10$~TeV, explaining the LHAASO data. GeV-to-TeV gamma rays are also produced in the MAD but attenuated by the optical and X-ray photons through the Breit-Wheeler process, which explains the non-detection of this source at 0.1-10~TeV. Due to the Klein-Nishina effect, attenuation of the sub-PeV gamma rays is moderate (see the middle panel of Figure \ref{timescale-fraction}). 

As $\sigma_{\rm ent}\ll 1$ in the jet model, the magnetic energy density is comparable to the energy density of the emitted photons, leading to efficient SSC. Moreover, the jet has a large emission power due to the high $\delta_D$ of Cygnus X-1. As a result, the jet SSC component explains the GeV gamma-ray data.

Our model does not account for the X-ray data at 3-100 keV. This is because the MAD formation condition (as discussed in  Section~\ref{MAD_condition}) restricts $\dot{m}$ and, subsequently, a harder Comptonization spectrum. A recent X-ray polarization study suggests that the Faraday effect would depolarize X-ray photons if magnetic fields are strong in the  X-ray emitting region \citep{Barner24}. Consequently, as is the case here, the MAD is not expected to be the source of 2-10 keV X-rays of Cygnus X-1, meaning that the MAD will not work as corona for Cygnus X-1. To reproduce the X-ray data, other components, such as outer hot accretion flow or standard disks inside the MAD supplying the seed photons, are necessary. 

The missing 3-100 keV component, however, does not impact the sub-PeV gamma-ray flux. This is because it does not attenuate UHE gamma rays yet barely contributes to the photomeson production (see top panel of Figure \ref{timescale-fraction}). 

\subsection{MAXI J1820+070}\label{MAXI}

Figure \ref{MAXIJ1820} presents our result for MAXI J1820+070. MAXI J1820+070 has been in a quiescent state with a few weak outbursts during the LHAASO operation time \citep{Yoshitake+2024}. We show the X-ray data taken during the decay phase of the small outbursts of MAXI J1820+070, which is slightly higher than the quiescent state (cyan bow-tie and blue solid line in Figure \ref{MAXIJ1820}). This is motivated since the average X-ray flux of this source is likely in between the quiescent flux and the peak of the outbursts during the LHAASO observation time. 

Our model explains both the X-ray and sub-PeV gamma-ray data. A higher  $\epsilon_{\rm NT}$, is used, comparing to that of Cygnus X-1 (see the next subsection and Section \ref{sec:summary} for discussion on $\epsilon_{\rm NT}$). High electron temperature and number density are needed to reproduce the X-ray data by Comptonization in the MAD. To achieve these, we use high $\beta$ and low $\alpha$ compared to the models of other BHXBs (see Table \ref{MAD_parameter_BHXB}).

MAXI J1820+070 has a low mass accretion rate. This yields low photon and proton number densities and consequently a long hadronic interaction time. While, owing to high $\beta$ and low $\dot{m}$, the magnetic field inside the MAD is low, leading to a long acceleration timescale. Thus, the proton maximum energy is determined by $E_{p, \rm max} = E_{\rm cut, cl}\simeq 3\times10^{14} ~\rm eV$, corresponding to a peak of the pion decay photons at $E_\gamma \simeq 3\times10^{13}~\rm eV$.

As the jet power scales to $\delta_D^4$ (see Section \ref{subsec:Jet}) and $\delta_D \lesssim 1$ in MAXI J1820+070, the jet component of this source is rather weak and does not appear in Figure \ref{MAXIJ1820}. The jet parameters are not constrained by the multi-wavelength observation, and we have adopted the parameter values of the jet component of Cygnus X-1.

\subsection{Other BHXBs}\label{others:ph}
Figure \ref{GROJ1655-40} shows the results for GRO J1655-40, GX 339-4, and XTE J1118+480. Since $\dot{m}$ of these three BHXBs are close to Cygnus X-1, we use the same plasma parameters as those for Cygnus X-1. Our model explains the X-ray data of these three BHXBs. This suggests that MAD may work as a coronal plasma in most cases. Since the magnetic fields are strong in the MAD, this result suggests that the X-ray polarization would be low in the soft X-ray band. An observation of such a feature by IXPE may test our model. 

XTE J1118+480 is not detected by LHAASO \citep{LHAASO2024_mq}, and our model is consistent with the non-detection. GX 339-4 and GRO J1655-40 are in the southern sky, outside the field of view of the current-generation air shower gamma-ray observatories. Figure \ref{GROJ1655-40} suggests that GRO J1655-40 may be a favored target for future southern hemisphere observatories such as Andes Large area PArticle detector for Cosmic ray physics and Astronomy (ALPACA) \citep{ALPACA2022} and Southern Wide-Field Gamma-Ray Observatory (SWGO) \citep{SWGO2019}.

The jet luminosities of GRO J1655-40, GX 339-4, and XTE J1118+480 are boosted by $\delta_D^4\simeq0.02$, 2.9, and 1.08, respectively. This explains why the jet component is faint in GRO J1655-40 but presents in the other two sources. We again adopt the parameter values of Cygnus X-1 for the jet components in these three sources as the models are not constrained by observations. 

Our model does not reproduce the optical data of GX 339-4 and the radio-to-optical data of XTE J1118+480. In both systems, the companion star is not massive, and additional optical emitting components may be present. In GX 339-4, optical photons may come from the standard disk located outside the MAD. In XTE J1118+480, the radio-to-optical spectrum almost follows a single power-law. These photons may come from nonthermal electrons in the jet and RIAFs \citep[e.g.,][]{Markoff+2001, Yuan+2003}. In both jets and RIAFs, the inner parts are optically thick for the synchrotron self-absorption (SSA) process, whereas the outer parts are SSA thin. Such a feature creates a power-law spectrum as the SSA peak frequency depends on the radius. We speculate that the RIAF and jet could explain optical-IR and radio data, respectively.

In both GX 339-4 and XTE J1118+480, the low-mass companion star is fainter than the accretion disk. As a result, the photometric data does not constrain the luminosity of the companion star. We estimate the star luminosity using the mass-luminosity relationship. The uncertainty of this subdominant component does not affect our results. 

\subsection{Neutrino emissions}\label{neutrino}
\begin{figure}[tbp]
    \centering
    \includegraphics[width=\linewidth]{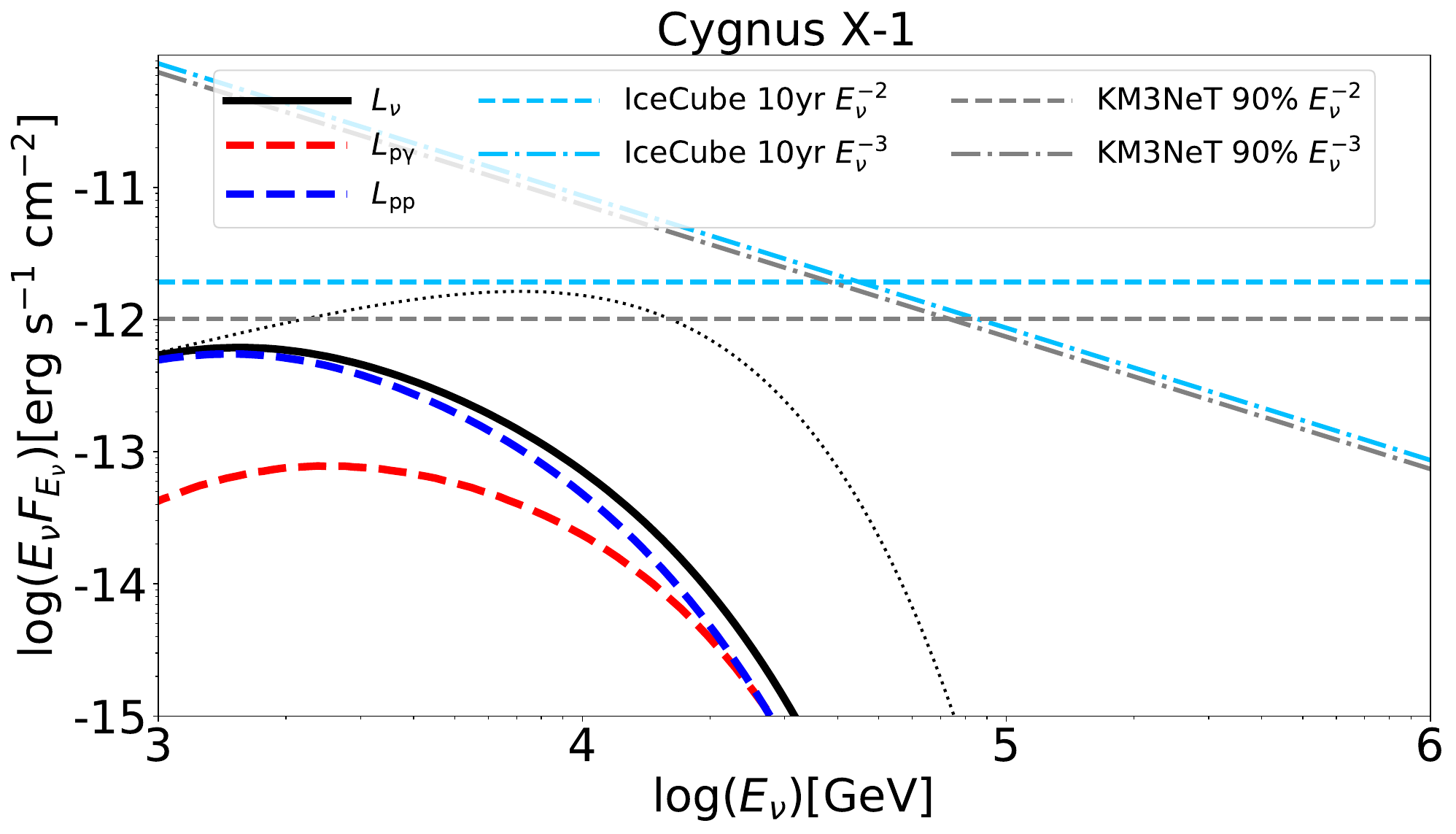}
    \includegraphics[width = \linewidth]{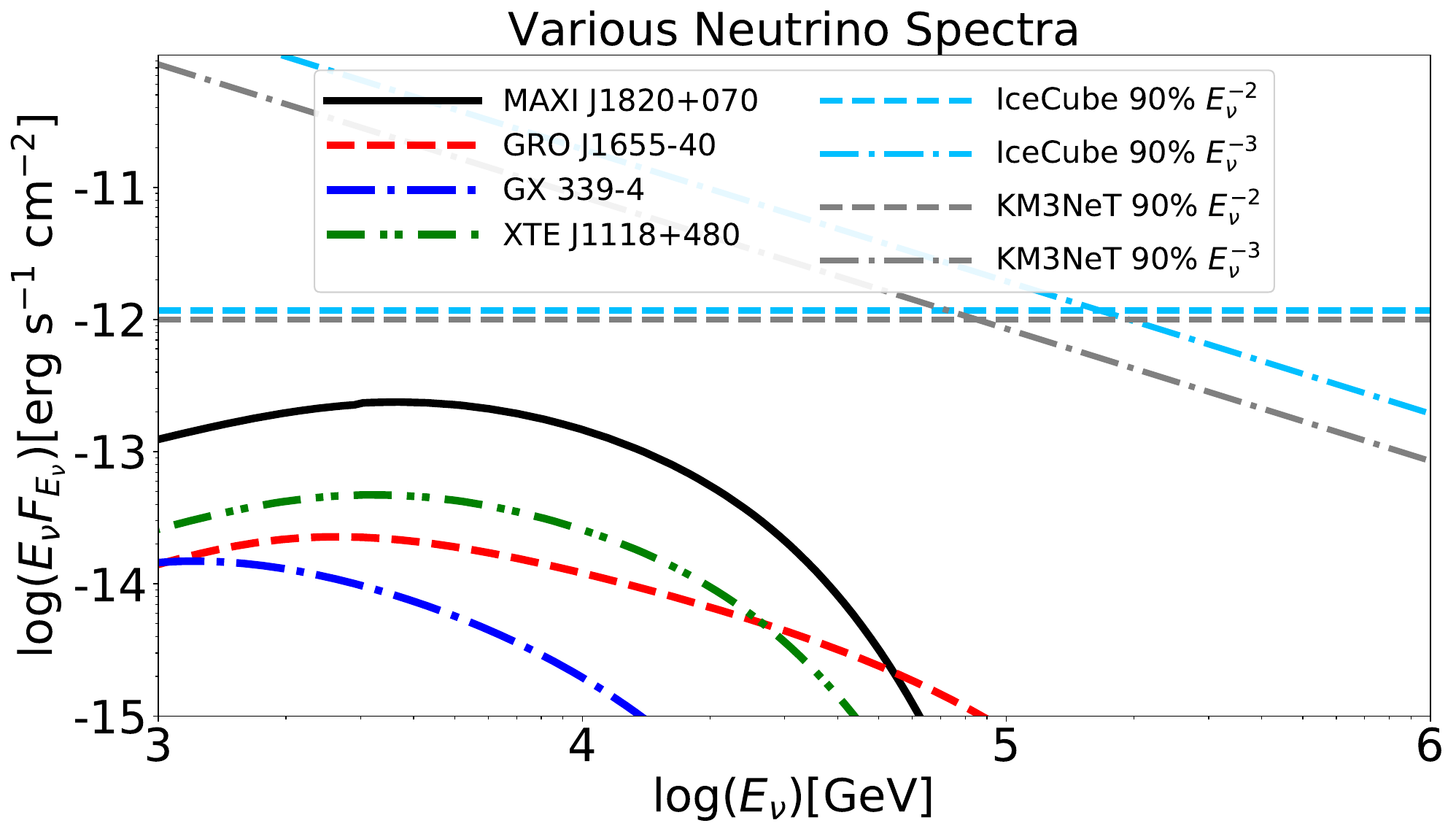}
    \caption{{\it Top}: Neutrino spectrum obtained by the MAD model for Cygnus X-1. Black, red dashed, and blue dashed lines are the total neutrino emission from MAD, neutrino emission by the photomeson production, and neutrino emission by the $pp$ inelastic collisions, respectively. The thin black dotted line is the total neutrino emission before pion suppression. IceCube sensitivities obtained by 10-year data (light-blue) are taken from \citet{IceCubeUL2020}. KM3NeT sensitivity curves (gray) are taken from \citet{TRIDENT2022}. The dashed and dot-dashed lines represent the neutrino spectra with $E_\nu^{-2}$ and $E_\nu^{-3}$, respectively. {\it Bottom}: Same as the top panel but for the other BHXBs. Black solid, red dashed, blue dot-dashed, and green dot-dot-dashed curves correspond to neutrino spectra of MAXI J1820+070, GRO J1655-40, GX 339-4, and XTE J1118+480, respectively. For comparison, gray and light-blue lines are the KM3NeT and IceCube sensitivities, respectively, assuming $E^{-2}$ (dashed) and $E^{-3}$ (dot-dashed) spectra, for a source at declination $\delta = 0.0$. IceCube sensitivities are taken from \citet{IceCubeUL2020}.}
    \label{neutrino_BHXB1}
\end{figure}

Figure \ref{neutrino_BHXB1} shows the neutrino spectra from the MADs. For Cygnus X-1, we calibrate $\epsilon_{\rm NT}$ using LHAASO data. The predicted neutrino flux is below the sensitivities obtained by IceCube 10-year data \citep{IceCubeUL2020}. Our model shows that the neutrino flux from Cygnus X-1 is $\approx 5\times 10^{-13}~\rm erg~s^{-1}~cm^{-2}$ at $E_\nu \simeq 3\times10^{12}$ eV. This obtained neutrino flux is slightly lower than the sensitivity of the future neutrino experiment, KM3NeT \citep{KM3NeT2016}, and KM3NeT might be able to detect the neutrino emission from MAD of Cygnus X-1 by taking a long operation time (see top panel of Figure \ref{neutrino_BHXB1}). Future neutrino detectors that improve the point-source sensitivities at TeV-PeV energies, such as IceCube-Gen2 \citep{IceCubeGen22021}, TRIDENT \citep{TRIDENT2022}, P-ONE \citep{Twagirayezu:2023Sd}, and HUNT \citep{HUNT2023}, have the potential to detect the neutrinos from Cygnus X-1. 

In the bottom panel of Figure \ref{neutrino_BHXB1}, we present the neutrino spectra of the other BHXBs. For comparison, we show the point-source sensitivities of IceCube and KM3NeT with track-like events for a source at $\delta = 0$. We note that these sensitivities depend on the source declination. The neutrino flux of MAXI J1820+070 is about half that of Cygnus X-1 and could be detected by future neutrino detectors. For GRO J1655-40, $\dot{m}$ is lower, whereas $\beta$ is higher compared to the other hard state BHXBs (see Table \ref{MAD_parameter_BHXB}). Consequently, the magnetic field inside the MAD is weaker, causing inefficient proton synchrotron cooling. This results in higher pion production efficiency and higher $E_{p,\rm max}$. On the other hand, the neutrino spectrum at $E_\nu \gtrsim 10^4$ GeV is strongly suppressed due to the pion suppression, making the detection of high-energy neutrinos more difficult. The other two BHXBs are not observable even with the next-generation detectors. These high-energy neutrino signals are crucial for testing our model.

In the jet, magnetic reconnection also accelerates the entrained protons, and the accelerated protons can emit neutrinos through the hadronic processes. However, due to low proton and photon number density, hadronic processes inside the jet are inefficient. Thus, neutrino luminosity from the jets is much lower than that from the MADs.

\section{Diffuse Emission from the quiescent and hard states BHXBs}\label{sec: diffuse}
We evaluate the contributions by MADs in BHXBs to PeV CRs, diffuse gamma-rays, and diffuse galactic neutrinos. We discuss the diffuse gamma-ray and neutrino emission in Section \ref{diffuse} and CR spectra in Section \ref{CR_BHXB}.

\subsection{Galactic neutrino and TeV gamma-rays}\label{diffuse}
To estimate the diffuse gamma-ray and neutrino fluxes from BHXBs, we first need to estimate the number of BHXBs in our galaxy. This number is largely uncertain and is suggested to be $\sim 10^4$ based on a binary population synthesis \citep{Yungelson+2006} and $\sim 10^3$ based on the event rate of the BH X-ray transients \citep{Corral+2016} and the recent binary population synthesis \citep{Shao&Li2020}. Thus we adopt a range of  $10^3-10^4$ for the total BHXB population. 

We then evaluate the luminosity of BHXBs in the Galaxy. For a simple estimation, we assume that all quiescent BHXBs have $\dot{m} = 10^{-4}$ and $M_{\rm BH}=10M_\odot$ and hard-state BHXBs have $\dot{m} = 0.05$ and $M_{\rm BH}=10M_\odot$, respectively. The other model parameters are set to be the same as those of MAXI J1820+070 for quiescent BHXBs and Cygnus X-1 for hard BHXBs, except with $\beta = 0.1$ in the latter case.   

We assume that  BHXBs are uniformly distributed in our Galaxy within 15 kpc from the Galactic center. We randomly position $10^3 - 10^4$ BHXBs and evaluate their distances to the Earth. Using these distances, we calculate the total flux of the gamma rays and the neutrinos as
\begin{equation}
F_{\rm tot} = \sum^{N_{\rm BHXB}}_{i=1} \frac{L_{\rm fiducial}}{4\pi D_i^2} \cdot f_{\rm duty},
\label{dif_emi}
\end{equation}
where $F_{\rm tot}$ is the total flux, $L_{\rm fiducial}$ is the luminosity of the fiducial BHXB, $D_i$ is the distance between Earth to each BHXB, $i$ is the $i$-th BHXB, and $f_{\rm duty}$ is the duty cycle of the quiescent and hard states.  Based on 19 years of X-ray observation of all known BHXBs, \citet{Tetarenko+2016} finds a median duty cycle of 2.7\% for the hard state (see Figure 9 of \citealp{Tetarenko+2016}). BHXBs in quiescent states are usually undetectable unless they are nearby. We set $f_{\rm duty} \simeq 0.027$ and  $1$ for the hard and quiescent states, respectively. 

\begin{figure}
    \centering
    \includegraphics[width = \linewidth]{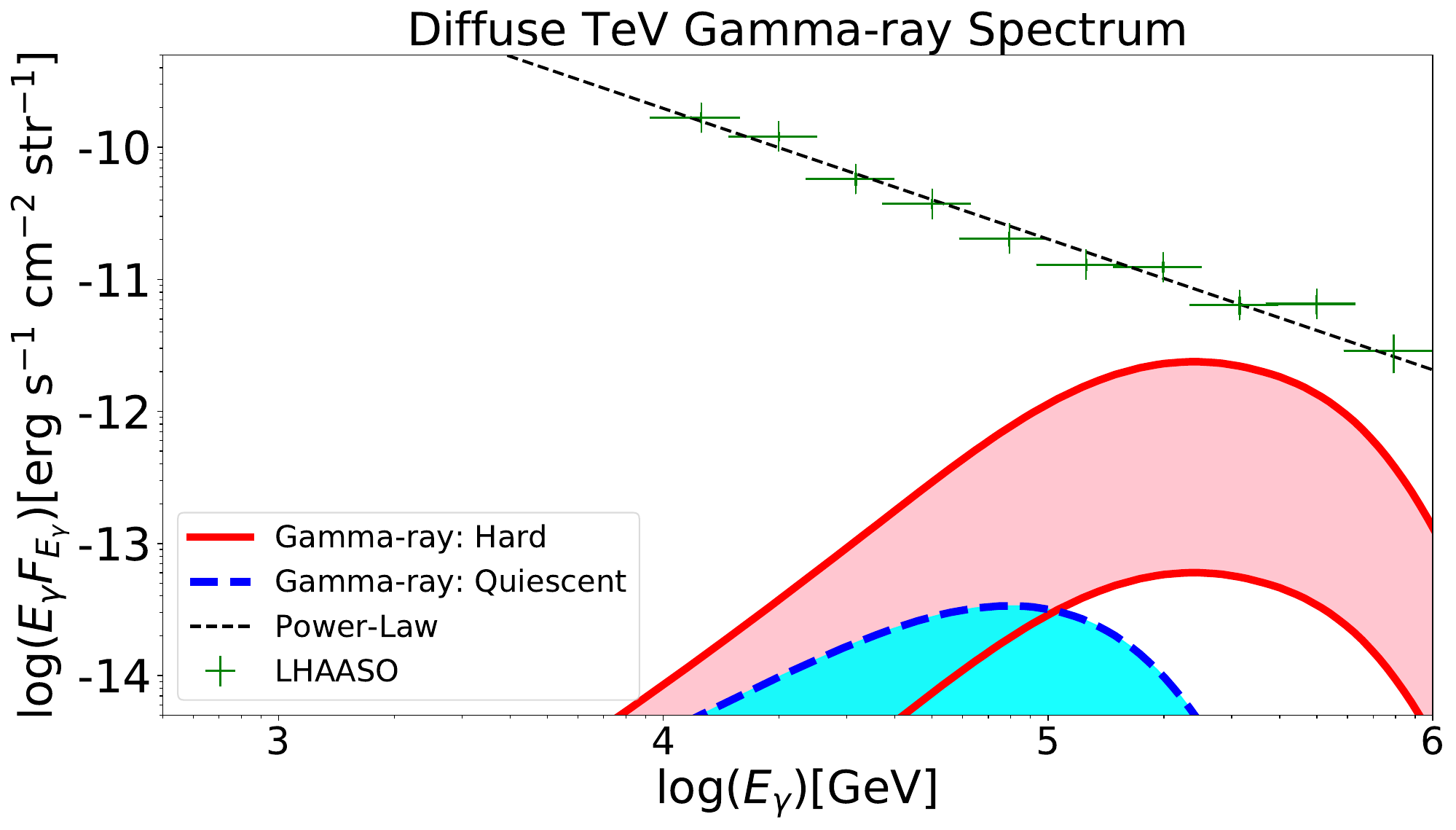}
    \includegraphics[width = \linewidth]{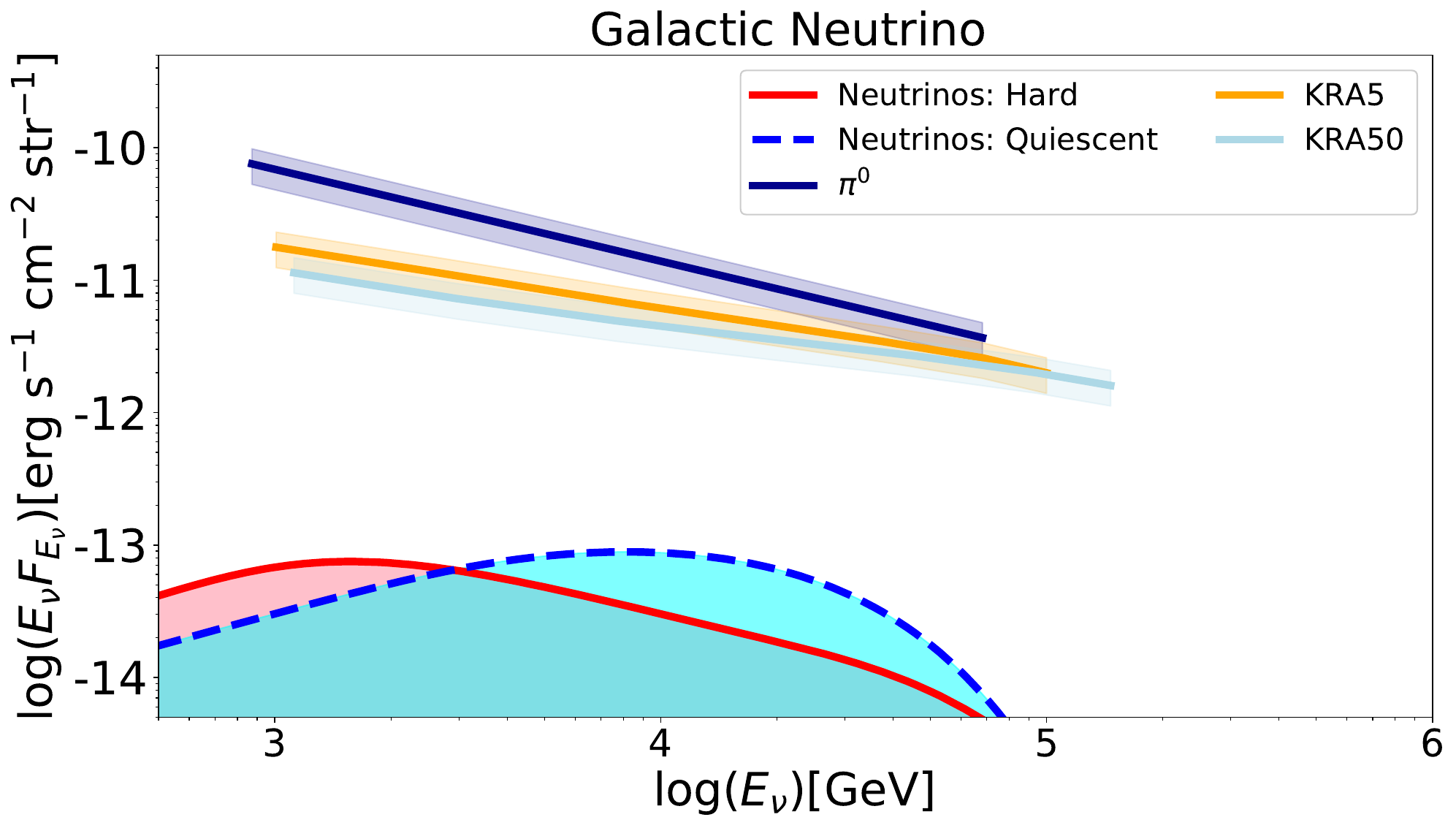}
    \caption{Contribution to the diffuse gamma rays and neutrinos from the BHXBs. The blue dashed and red solid lines show the gamma-ray and neutrino emission from the quiescent state BHXBs and the hard state BHXBs, respectively. {\it Top}: Diffuse TeV gamma-ray spectra from the BHXBs. Data points are taken from \citet{CaoDiffuse2023}. {\it Bottom}: Galactic neutrino spectra from the BHXBs. Data points are taken from \citet{IceCubeDiffuse2023}.}
    \label{Diff_contribution}
\end{figure}

Figure \ref{Diff_contribution} shows that hard-state BHXBs may contribute to 30\% of the diffuse gamma-ray emission observed by LHAASO at $E_\gamma \sim 100$ TeV, whereas quiescent BHXBs barely contribute due to their inefficient pion production. Due to the pion suppression, BHXBs do not contribute to the  Galactic diffuse neutrino emission. 

The Galactic diffuse emission observed by air shower gamma-ray observatories may largely arise from the cosmic-ray sea interacting with the interstellar medium (ISM) \citep{Fang&Muraase2023, Fang:2023azx, Lipari:2024pzo}, although it may also include a contribution from unresolved sources \citep{Vecchiotti_2022}. Pulsar halos have been suggested as such sources \citep{Dekker:2023six, Yan+2024}. Our results imply that hard-state BHXBs may also be a promising contributor to the diffuse gamma-ray emission from the Galactic plane. 

\subsection{CRs from the quiescent and hard states BHXBs}\label{CR_BHXB}
Inside the MAD, the MHD turbulence accelerates the protons, and the protons escape from the MAD diffusively. We estimate the escaped CR luminosity and evaluate the contribution of the BHXBs to the CR spectra. We calculate the CR injection rate to the ISM as
\begin{equation}
E_pL_{E_p} = \frac{t_{\rm loss}}{t_{\rm diff}} E_p^2 \dot{N}_{E_{p}, \rm inj} \, N_{\rm BHXB} \, f_{\rm duty},
\label{CR_lumi}
\end{equation}
where $t_{\rm diff}$ is the diffusive escape timescale, $\dot{N}_{E_{p}, \rm inj}$ is the primary proton injection term of the MAD model. The escaped CRs propagate in the ISM and arrive on Earth. The confinement time inside the ISM, $t_{\rm conf}$, is provided by the grammage, $X_{\rm esc} = n_{\rm ISM}\mu m_p c t_{\rm conf}$, where $n_{\rm ISM}$ is the ISM number density and $\mu$ is the ISM mean atomic mass. Based on the boron-to-carbon ratio measurements, the grammage of CRs is estimated to be $X_{\rm esc} \simeq 2.0 (E_p/250 ~ {\rm GeV})^{-s} ~ \rm g ~ cm^{-2}$, where $s = 0.46$ for $E_p < 250 ~ \rm GeV$ and $s = 0.33$ for $E_p > 250 ~ \rm GeV$ \citep{Aguilar+2016, MuraseFukugita2018, Adriani+2022}. Then, the CR escape rate from the ISM is estimated to be $E_pU_{E_p}V_{\rm gal}/t_{\rm conf} \approx E_pU_{E_p}cM_{\rm gas}/X_{\rm esc}$, where $U_{E_p}$ is the differential energy density of the CR and $M_{\rm gas} \simeq 8\times10^9 M_\odot$ is the total gas mass in our galaxy \citep{Nakanishi&Sofue2016}. CR escape rate should balance with the CR injection rate, and we estimate the CR intensity, $\Phi_p = cU_{E_p}/(4\pi E_p)$, as \citep[e.g.,][]{KimuraMurase2018}
\begin{equation}
E_p^2\Phi_p \approx \frac{E_pL_{E_p}X_{\rm esc}}{4\pi M_{\rm gas}}.
\label{CR_intensity}
\end{equation}

\begin{figure}
    \centering
    \includegraphics[width = \linewidth]{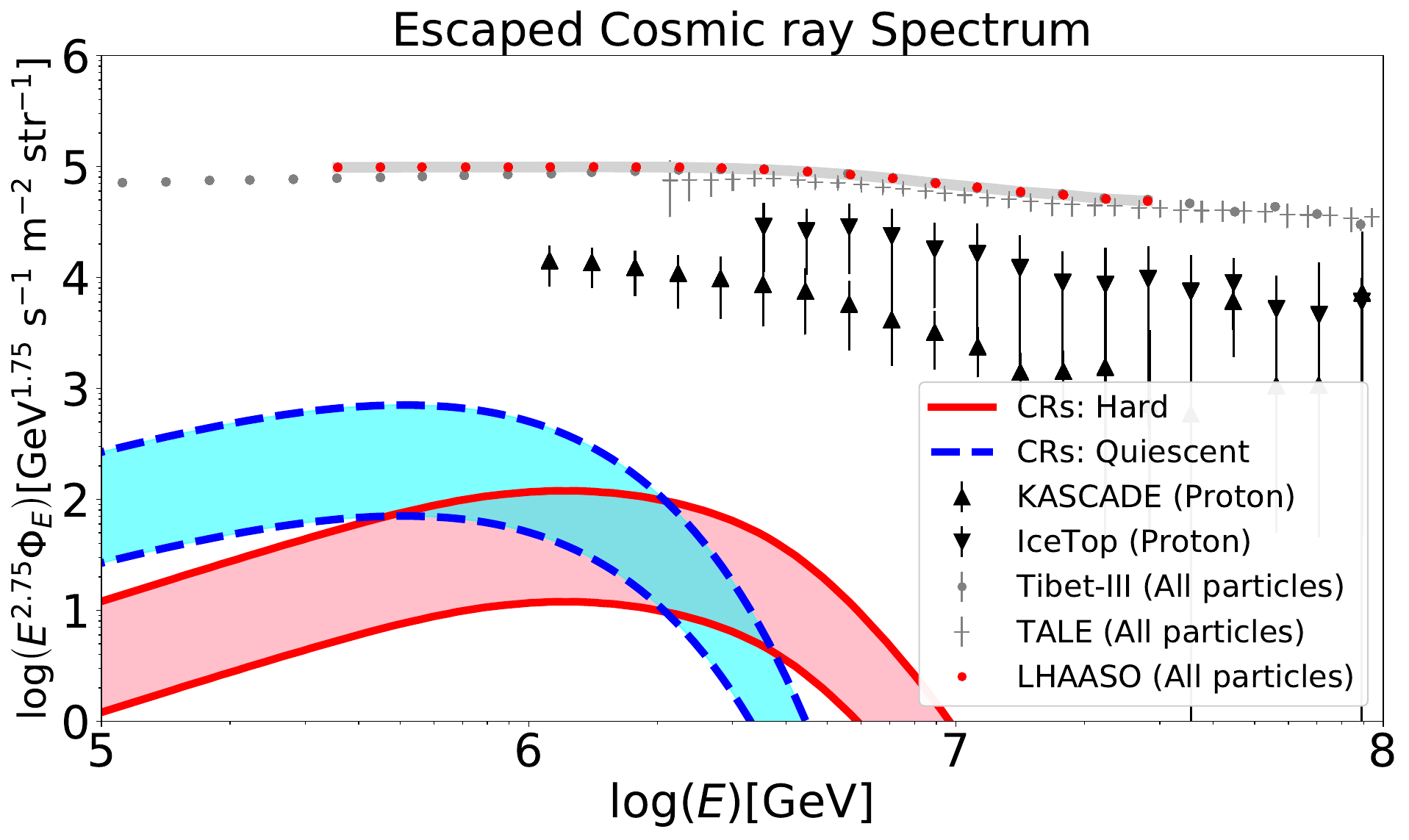}
    \caption{CR spectrum predicted by our Jet-MAD model for the quiescent state (blue dashed line) and the hard state (red solid line). The experimental data for protons and all-particle CR energy spectra are taken from \citet{Apel+2013}, \citet{Aartsen+2019_IceTop} and \citet{Amenomori+2008_TibetIII}, \citet{Abbasi+2018_TALE}, and \citet{LHAASO_CRspe_2024}, respectively.}
    \label{CR_spectrum}
\end{figure}

We show the CR spectra from the quiescent and hard states BHXBs in Figure \ref{CR_spectrum}. BHXBs cannot contribute significantly to the observed CR spectra in both states. This comes from the low mass accretion rate for the quiescent state BHXBs and low $\epsilon_{\rm NT}$ and duty cycle for the hard state BHXBs. \citet{KimuraSudoh2021} have estimated the CR spectrum from the MAD of the quiescent state of BHXBs (QBHXBs) and showed that the CRs from QBHXBs can explain the PeV CRs (see their Figure 5). In their scenario, the magnetic reconnection and the turbulence accelerates the protons inside the MAD. \citet{KimuraSudoh2021} considers the flat $\dot{m}$ distribution in the range of $\dot{m} = 10^{-5} -10^{-2}$, and they involve the contribution by the BHXBs with a higher $\dot{m}$ than that we assume in this paper. Owing to this, the QBHXBs can reproduce the observed CR spectra. Without specifying the nature of the corona, \citet{Fang:2024wmf} considers a general scenario where protons are accelerated by a turbulent magnetic field and magnetic reconnection in a BHXB corona in hard and soft states, respectively. The work suggests that CRs from BHXB coronae may significantly contribute to the Galactic CR spectra. 

The observed CRs may come from other source classes. In particular, CRs with $E_p \lesssim 10^6$ GeV could come from the Galactic supernova remnants (SNRs; e.g., \citealt{Helder+2012, Ackermann:2013wqa, Fang:2022uge}). CRs with $E_p \gtrsim 3\times10^7$ GeV could come from binary neutron-star merger remnants \citep{KimuraMurase2018, RBB18a} and Sgr A*'s past activities \citep{FujitaMuraseKimura2017, Kuze+2022}.

Magnetic reconnection in the jets of BHXBs also accelerates the entrained protons. At the initial phase of the plasma entrainment, the entrained protons are subdominant of the internal energy. Then, the proton magnetization parameter is the same as $\sigma_{\pm}$, leading to proton maximum Lorentz factor as $\gamma_{p, \rm max}\approx (\delta B/B)^2 \xi \sigma_\pm$. With our BHXB parameter sets (see Table \ref{MAD_parameter_BHXB}), the proton maximum energy is estimated to be $E_p \sim10^4 ~ \rm GeV$. Thus, the jets of hard-state BHXBs do not significantly contribute to the CR spectra above the knee.

Our estimation does not account for the luminosity function of BHXBs. Based on a relationship between radio and jet kinetic luminosities, \citet{Heinz&Grimm2005} suggests that the mean jet kinetic luminosity of BHXBs is $\left<L_j\right> \approx 2.5\times10^{37} ~\rm erg~s^{-1}$. In our Jet-MAD model, the jet luminosity at the jet base is calculated as $L_j = \eta \dot{M}c^2$, where $\eta\approx 0.58$ is the fraction of accretion luminosity to the jet kinetic luminosity (see Section 2.2 of \citealp{RK+2024}). With our fiducial values $M = 10M_\odot$ and $\dot{m} = 0.05$,  $L_j\simeq 4.0 \times10^{37} ~ \rm erg~s^{-1}$, which is consistent with the findings of \citet{Heinz&Grimm2005}. 
 
\section{Conclusion} \label{sec:summary}
We have investigated the high-energy radiation of BHXBs by applying our Jet-MAD model to five BHXBs. We find that hadronic processes in the MAD may explain the UHE gamma-ray sources associated with Cygnus X-1 and MAXI J1820+070 recently detected by LHAASO (see Figures \ref{CygX-1} and \ref{MAXIJ1820}). Among southern-sky sources, future gamma-ray detectors such as ALPACA and SWGO may detect GRO J1655-40 (see Figure \ref{GROJ1655-40}).
Comptonization by thermal electrons roughly explains the X-ray spectra of the BHXBs, except Cygnus X-1, where additional components are needed to explain the 3-100 keV data. 

Inside the jet, relativistic electrons accelerated by magnetic reconnection produce multi-wavelength emission via synchrotron and SSC radiation. The jet emission explains the GeV gamma-ray observation of Cygnus X-1 and is too faint to be detected in other sources.  

Our model predicts high-energy neutrino emission from the MADs. Future TeV-PeV neutrino detectors such as IceCube-Gen2, KM3NeT, P-ONE, TRIDENT, and HUNT have the potential to detect Cygnus X-1 and MAXI J1820+070 (see Figure \ref{neutrino_BHXB1}). The MADs of these objects have weak magnetic fields, allowing efficient pion production before synchrotron cooling.  

We evaluate the contribution of BHXBs to the Galactic diffuse gamma-ray and neutrino fluxes. Our results suggest that hard-state BHXBs may contribute up to a few 10\% of the diffuse gamma rays. The BHXB population, which is largely unresolved by air shower gamma-ray observatories, may plausibly explain the Galactic diffuse emission together with the diffuse cosmic-ray sea. This population, however, barely contributes to the neutrino emission of the Galactic plane as a result of the pion suppression. 

MHD turbulence accelerates the CR protons in the MAD up to PeV, and they escape diffusively from the MAD to the ISM. We evaluate the contribution of the quiescent and the hard BHXBs to the CR spectra observed on Earth. We find that BHXBs barely contribute to the observed CR spectra due to low $\dot{m}$ for the quiescent state and low $\epsilon_{\rm NT}$ and low duty cycle for the hard state. 

To explain the LHAASO data, relatively low and high values of $\epsilon_{\rm NT}$ are chosen for Cygnus X-1 and MAXI J1820+070, respectively. This may be understood by considering the formation process of a MAD. Cygnus X-1 has a high $\dot{m}$, and then, the standard disk will be located close to the BH. In the standard thin disk, the magnetorotational instability (MRI) and subsequent disk dynamo will create a large-scale poloidal field, and these large-scale fields are accumulated onto the BH \citep[e.g.,][]{Salvesen2016, Hayashi2024}. MRI dynamo changes the polarity of the large-scale fields in $10-20$ dynamical timescales, and thus, the magnetic fields accumulated on the BH would disappear by magnetic reconnection within several dynamical timescales of the inner radius of the standard accretion disk. Thus, the MAD formation is rather unlikely for BHXBs with high $\dot{m}$. Nevertheless, we speculate that the MAD can form even for the high $\dot{m}$ BHXBs, although the MAD state does not continue for a long time. This may cause a lower value of $\epsilon_{\rm NT}$, as nonthermal particles are efficiently produced only in the MAD state. On the other hand, MAXI J1820+070 has a low $\dot{m}$, and then, the standard disk would be located far from the BH, leading to the formation of a persistent MAD state. This might cause a higher value of $\epsilon_{\rm NT}$, which is consistent with the cases with radio galaxies and quiescent X-ray binaries \citep{Kuze+2022, KimuraSudoh2021}. Therefore, it may be natural to use a higher $\epsilon_{\rm NT}$ for MAXI J1820+070 than that of the other BHXBs. Future modeling of $\epsilon_{\rm NT}$ as a function of mass accretion rate remains as a future work.

A feature of our model is that photons at different wavelengths originate from distinct emission regions. In particular, the MAD emits both X-rays and sub-PeV gamma rays, whereas the jet produces radio signals and GeV gamma rays. As a result, one should expect time variability correlations between X-ray and TeV gamma-ray bands, as well as between radio and GeV gamma-ray bands. The sizes of the emission regions correspond to a variability timescale of $\sim 10^{-3}$~s in the MAD, and $\sim 10^{-1}$~s in the jet. Although current-generation gamma-ray detectors cannot resolve such rapid variability in BHXBs, long-term monitoring and a multi-wavelength campaign combined with variability analyses could identify the high-energy radiation our model predicts to occur during the hard state, which typically lasts for a few to several months.

Our model suggests that individual BHXBs, including Cygnus X-1 and MAXI J1820+070, are promising multi-messenger targets for future UHE gamma-ray and high-energy neutrino detectors. These future observations can test our model and are crucial to understanding the nonthermal phenomena near BHs.

\section*{acknowledgments}
We would like to thank Hirofumi Noda, Megumi Shidatsu, and Chris Done for their useful comments and discussions. Discussions for this paper were performed many times at Science Lounge of FRIS CoRE. This work is partly supported by JSPS KAKENHI grant Nos. 22K14028, 21H04487, and 23H04899 (S.S.K.). This work is also supported by JST SPRING, grant No. JPMJSP2114 (R.K.). R.K. acknowledges the support of the JSPS Overseas Challenge Program for Young Researchers. S.S.K. acknowledges support from the Tohoku Initiative for Fostering Global Researchers for Interdisciplinary Sciences (TI-FRIS) of MEXTs Strategic Professional Development Program for Young Researchers. K.F. acknowledges support from the National Science Foundation (PHY-2238916) the Sloan Research Fellowship. This work was supported by a grant from the Simons Foundation (00001470, KF).

\appendix
\section{Condition for MAD}\label{MAD_condition}
This section explains the condition for forming the MADs. Thermal protons in the MAD cannot cool through the radiative processes, and their temperature is comparable to the virial temperature. Thermal electrons receive a significant fraction of the dissipation energy and become relativistic. Unlike protons, electrons may be efficiently cooled by radiative processes, and their temperature is mildly relativistic. Thermal electrons and protons interact through the Coulomb collisions and exchange their energy. When the electron-proton relaxation timescale, $t_{p-e}$, is shorter than the infall timescale, $t_{\rm fall}$, protons cool by the Coulomb collision. This results in a low proton temperature, with which the accretion disk is no longer in the RIAF regime. Therefore, we assume that the MAD is not formed when such a condition is met. 

We estimate the electron-proton relaxation timescale inside the MADs following Equation (13) of \citet{Takahara&Kusunose1985}:
\begin{equation}
t_{p-e} = \frac{1}{n_{p, \rm mad} \sigma_T c}\sqrt{\frac{\pi}{2}} \frac{m_p}{m_e }\frac{1}{\ln \Lambda} \left( \theta_e + \theta_p \right)^{3/2},
\label{tpe}
\end{equation}
where $\sigma_T$ is the Thomson cross section, $\ln \Lambda$ is the Coulomb logarithm, and $\theta_p = k_BT_p/(m_pc^2)$ is the normalized proton temperature. 

We estimate the infall timescale as
\begin{equation}
    t_{\rm fall} \approx \frac{R_d}{V_R} \simeq 9.8\times10^{-3} r_1^{3/2}M_1\alpha_{-0.5}^{-1}~\rm s.
    \label{tfall}
\end{equation}
From Equations (\ref{tpe}) and (\ref{tfall}), we compare $t_{p-e}$ and $t_{\rm fall}$ and determine whether the disk is in the MAD state. By equating $t_{p-e}$ with $t_{\rm fall}$, we derive the critical mass accretion rate, $\dot{m}_{\rm crit}$. Assuming $\theta_e > \theta_p$, we estimate $\dot{m}_{\rm crit}$ as
\begin{equation}
    \dot{m}_{\rm crit} \simeq 0.35\alpha_{-0.5}^{2} \theta_{e, -0.5}^{3/2}.
    \label{mdotcrit}
\end{equation}
If $\dot{m} > \dot{m}_{\rm crit}$, the relaxation timescale is shorter than the infall timescale, and the accretion disk is not in the MAD state. The assumption of $\theta_e > \theta_p$ is always satisfied for the MADs as long as electrons are relativistic, which is reasonable at $R_d < (m_p/m_e)R_g$.

Observations and numerical simulations both imply the existence of MADs in hard-state BHXBs. \citet{You+2023} interprets the time lag of MAXI J1820+070 between the radio and X-ray bands as the MAD formation time. Using general relativistic radiation magnetohydrodynamic simulations, \citet{Moscibrodzka2024} shows that MADs can reproduce the X-ray polarization data of Cygnus X-1 observed by IXPE \citep{IXPE2022}. 

BHXBs are in the quiescent state if the mass accretion rate is highly sub-Eddington, e.g., $\dot{m}<0.01$. MADs may also form in such a state due to inefficient matter cooling (see \citealp{KimuraSudoh2021}).

\section{Relaxation timescale for the MAD}\label{relax_MAD}
In this section, we evaluate whether the MADs are in a collisionless system. Based on Equations (12) and (14) of \citet{Takahara&Kusunose1985}, we estimate the proton-proton relaxation timescale, $t_{p-p}$, and the electron-electron relaxation timescale, $t_{e-e}$, as
\begin{align}
    t_{p-p} &= \frac{1}{n_{p, \rm mad} \sigma_T c}\frac{4\sqrt{\pi}}{\ln \Lambda} \left(\frac{m_p}{m_e}\right)^2 \theta_p^{3/2}, \\
     t_{e-e} &= \frac{1}{n_{p, \rm mad} \sigma_T c}\frac{4\sqrt{\pi}}{\ln \Lambda}\theta_e^{3/2}.
\label{tpp-tee}
\end{align}
As $\theta_p \approx m_p C_s^2/(m_pc^2) \simeq 2.5\times10^{-2}r_1^{-1}$ and $\theta_e > \theta_p$, each relaxation timescale for the proton-electron relaxation timescale is estimated to be
\begin{align}
\frac{t_{p-p}}{t_{p-e}} &\approx 4\sqrt{2}\frac{m_p}{m_e} \left( \frac{\theta_p}{\theta_e}\right)^{3/2} \simeq 2.3\times10^2r_1^{-3/2}\theta_{e,-0.5}^{-3/2}, \label{tpp_tpe} \\
 \frac{t_{e-e}}{t_{p-e}} &\approx 4\sqrt{2}\frac{m_e}{m_p}\simeq 3.1\times10^{-3}.
\label{tee_tpe}
\end{align}
From Equation (\ref{tpp_tpe}), the proton-proton relaxation timescale is longer than the infall timescale, and this means that the protons are not thermalized by the Coulomb collisions inside the MAD, allowing the production of the nonthermal protons. 

In contrast, from Equation (\ref{tee_tpe}), the electron-electron relaxation timescale is shorter than the proton-electron relaxation timescale; hence, electrons may be thermalized by Coulomb collisions if the mass accretion rate has a higher value. Since the dependence of electron temperature for these timescales are the same, the critical mass accretion rate at which electrons are thermalized can be obtained by multiplying Equation (\ref{tee_tpe}) to Equation (\ref{mdotcrit}), indicating that when $\dot{m}\gtrsim10^{-3}$, electrons will be thermalized by the Coulomb collisions.

\section{Particle energy distribution for the MAD model}\label{MAD_pair}
This section shows how to calculate the particle energy distribution in the MAD model. We obtain the energy distribution of nonthermal particles by solving the energy transport equation with one-zone and steady-state approximations:
\begin{equation}
    -\frac{d}{dE_i}\left(\frac{E_iN_{E_i}}{t_{i,\rm cool}}\right) = -\frac{N_{E_i}}{t_{\rm esc}}+\dot{N}_{E_i, \rm inj},
    \label{trans_eq}
\end{equation}
where $i$ is the particle species, $E_i$ is the particle energy, $N_{E_i}$ is the differential number spectrum, $\dot{N}_{E_i, \rm inj}$ is the injection term, and $t_{i,\rm cool}$ and $t_{\rm esc}$ is cooling and escape timescales, respectively. As expressed in Section \ref{subsec:MAD}, the injection term is assumed to be a power-law with the index $s_{\rm inj}$ and the exponential cutoff at $E_{i, \rm cut}$. For the normalization of the injection term, we calculate the primary proton luminosity as $L_p \approx (Q_i/(Q_i+Q_e))\epsilon_{\rm NT}\epsilon_{\rm dis}\dot{M}c^2 = \int E_p \dot{N}_{p, \rm inj}dE_p$. While it has not been established yet in what proportion nonthermal particles have energy by particle acceleration, \citet{Zhdankin+2019} has shown that nonthermal particles have around 30\% of the total particle energy. Thus, in the paper, we set the value of $\epsilon_{\rm NT} \lesssim 0.33$. For $\epsilon_{\rm dis}$, since the central BH will have a high spin parameter due to the strong jet ejection, the dissipation efficiency will be high, and we set it to be about 0.15.

We approximate the injection term of the secondary electron-positron pairs produced by the Bethe-Heitler process as
\begin{equation}
    E^2_{\rm BH,\pm}\dot{N}_{\rm BH, \pm, \rm inj} \approx E_p^2N_{E_p}t_{\rm BH}^{-1},
    \label{Bethe-Heitler}
\end{equation}
where $E_{\rm BH,\pm} \approx (m_e/m_p)E_p$ is the energy of the produced pairs and $t_{\rm BH}$ is the Bethe-Heitler process timescale. These pairs and pions produce high-energy gamma rays, which interact with lower-energy photons in MADs, leading to electromagnetic cascades. We estimate the optical depth of the Breit-Wheeler process, $\tau_{\gamma \gamma}$, following \citet{Coppi1990}, and the gamma rays are attenuated by a factor of $F_{\rm atn} \approx (1-\exp [-\tau_{\gamma\gamma}])/\tau_{\gamma\gamma}$ (see middle panel of Figure \ref{timescale-fraction}). The injection term of the secondary electron-positron pairs produced by the Breit-Wheeler process is approximated to be
\begin{equation}
    E^2_{\gamma\gamma,\pm}\dot{N}_{\gamma\gamma, \pm, \rm inj} \approx (1-F_{\rm atn})E_\gamma L_{E_\gamma},
    \label{breit-wheeler}
\end{equation}
where $E_{\gamma\gamma,\pm}\approx E_\gamma/2$ is the pair energy and $L_{E_\gamma}$ is the intrinsic differential photon luminosity of all components. Since the injection term in Equation (\ref{breit-wheeler}) depends on the gamma-ray spectrum, we iteratively calculate the photon and the electron-positron pair spectra until they converge.

\bibliography{sample631}{}

\begin{thebibliography}{}
\expandafter\ifx\csname natexlab\endcsname\relax\def\natexlab#1{#1}\fi
\providecommand{\url}[1]{\href{#1}{#1}}
\providecommand{\dodoi}[1]{doi:~\href{http://doi.org/#1}{\nolinkurl{#1}}}
\providecommand{\doeprint}[1]{\href{http://ascl.net/#1}{\nolinkurl{http://ascl.net/#1}}}
\providecommand{\doarXiv}[1]{\href{https://arxiv.org/abs/#1}{\nolinkurl{https://arxiv.org/abs/#1}}}

\bibitem[{{Aartsen} {et~al.}(2019){Aartsen}, {Ackermann}, {Adams}, {Aguilar}, {Ahlers}, {Ahrens}, {Alispach}, {Andeen}, {Anderson}, {Ansseau}, {Anton}, {Arg{\"u}elles}, {Auffenberg}, {Axani}, {Backes}, {Bagherpour}, {Bai}, {Barbano}, {Barwick}, {Baum}, {Baur}, {Bay}, {Beatty}, {Becker}, {Becker Tjus}, {BenZvi}, {Berley}, {Bernardini}, {Besson}, {Binder}, {Bindig}, {Blaufuss}, {Blot}, {Bohm}, {B{\"o}rner}, {B{\"o}ser}, {Botner}, {B{\"o}ttcher}, {Bourbeau}, {Bourbeau}, {Bradascio}, {Braun}, {Bretz}, {Bron}, {Brostean-Kaiser}, {Burgman}, {Buscher}, {Busse}, {Carver}, {Chen}, {Cheung}, {Chirkin}, {Clark}, {Classen}, {Collin}, {Conrad}, {Coppin}, {Correa}, {Cowen}, {Cross}, {Dave}, {de Andr{\'e}}, {De Clercq}, {DeLaunay}, {Dembinski}, {Deoskar}, {De Ridder}, {Desiati}, {de Vries}, {de Wasseige}, {de With}, {DeYoung}, {Diaz}, {D{\'\i}az-V{\'e}lez}, {Dujmovic}, {Dunkman}, {Dvorak}, {Eberhardt}, {Ehrhardt}, {Eller}, {Evenson}, {Fahey}, {Fazely}, {Felde}, {Feusels}, {Filimonov}, {Finley}, {Franckowiak}, {Friedman},
  {Fritz}, {Gaisser}, {Gallagher}, {Ganster}, {Garrappa}, {Gerhardt}, {Ghorbani}, {Glauch}, {Gl{\"u}senkamp}, {Goldschmidt}, {Gonzalez}, {Grant}, {Griffith}, {G{\"u}nder}, {G{\"u}nd{\"u}z}, {Haack}, {Hallgren}, {Halve}, {Halzen}, {Hanson}, {Hebecker}, {Heereman}, {Heix}, {Helbing}, {Hellauer}, {Henningsen}, {Hickford}, {Hignight}, {Hill}, {Hoffman}, {Hoffmann}, {Hoinka}, {Hokanson-Fasig}, {Hoshina}, {Huang}, {Huber}, {Hultqvist}, {H{\"u}nnefeld}, {Hussain}, {In}, {Iovine}, {Ishihara}, {Jacobi}, {Japaridze}, {Jeong}, {Jero}, {Jones}, {Jonske}, {Joppe}, {Kang}, {Kappes}, {Kappesser}, {Karg}, {Karl}, {Karle}, {Katz}, {Kauer}, {Kelley}, {Kheirandish}, {Kim}, {Kintscher}, {Kiryluk}, {Kittler}, {Klein}, {Koirala}, {Kolanoski}, {K{\"o}pke}, {Kopper}, {Kopper}, {Koskinen}, {Kowalski}, {Krings}, {Kr{\"u}ckl}, {Kulacz}, {Kunwar}, {Kurahashi}, {Kyriacou}, {Labare}, {Lanfranchi}, {Larson}, {Lauber}, {Lazar}, {Leonard}, {Leuermann}, {Liu}, {Lohfink}, {Lozano Mariscal}, {Lu}, {Lucarelli}, {L{\"u}nemann}, {Luszczak},
  {Madsen}, {Maggi}, {Mahn}, {Makino}, {Mallik}, {Mallot}, {Mancina}, {Mari{\c{s}}}, {Maruyama}, {Mase}, {Maunu}, {Meagher}, {Medici}, {Medina}, {Meier}, {Meighen-Berger}, {Menne}, {Merino}, {Meures}, {Miarecki}, {Micallef}, {Moment{\'e}}, {Montaruli}, {Moore}, {Morse}, {Moulai}, {Muth}, {Nagai}, {Nahnhauer}, {Nakarmi}, {Naumann}, {Neer}, {Niederhausen}, {Nowicki}, {Nygren}, {Obertacke Pollmann}, {Olivas}, {O'Murchadha}, {O'Sullivan}, {Palczewski}, {Pandya}, {Pankova}, {Park}, {Peiffer}, {P{\'e}rez de los Heros}, {Philippen}, {Pieloth}, {Pinat}, {Pizzuto}, {Plum}, {Porcelli}, {Price}, {Przybylski}, {Raab}, {Raissi}, {Rameez}, {Rauch}, {Rawlins}, {Rea}, {Reimann}, {Relethford}, {Renzi}, {Resconi}, {Rhode}, {Richman}, {Robertson}, {Rongen}, {Rott}, {Ruhe}, {Ryckbosch}, {Rysewyk}, {Safa}, {Sanchez Herrera}, {Sandrock}, {Sandroos}, {Santander}, {Sarkar}, {Sarkar}, {Satalecka}, {Schaufel}, {Schlunder}, {Schmidt}, {Schneider}, {Schneider}, {Schumacher}, {Sclafani}, {Seckel}, {Seunarine}, {Shefali}, {Silva},
  {Snihur}, {Soedingrekso}, {Soldin}, {Song}, {Spiczak}, {Spiering}, {Stachurska}, {Stamatikos}, {Stanev}, {Stasik}, {Stein}, {Stettner}, {Steuer}, {Stezelberger}, {Stokstad}, {St{\"o}{\ss}l}, {Strotjohann}, {St{\"u}rwald}, {Stuttard}, {Sullivan}, {Sutherland}, {Taboada}, {Tenholt}, {Ter-Antonyan}, {Terliuk}, {Tilav}, {Tomankova}, {T{\"o}nnis}, {Toscano}, {Tosi}, {Tselengidou}, {Tung}, {Turcati}, {Turcotte}, {Turley}, {Ty}, {Unger}, {Unland Elorrieta}, {Usner}, {Vandenbroucke}, {Van Driessche}, {van Eijk}, {van Eijndhoven}, {Vanheule}, {van Santen}, {Vraeghe}, {Walck}, {Wallace}, {Wallraff}, {Wandkowsky}, {Watson}, {Weaver}, {Weiss}, {Weldert}, {Wendt}, {Werthebach}, {Westerhoff}, {Whelan}, {Whitehorn}, {Wiebe}, {Wiebusch}, {Wille}, {Williams}, {Wills}, {Wolf}, {Wood}, {Wood}, {Woschnagg}, {Wrede}, {Xu}, {Xu}, {Xu}, {Yanez}, {Yodh}, {Yoshida}, {Yuan}, {Z{\"o}cklein}, \& {IceCube Collaboration}}]{Aartsen+2019_IceTop}
{Aartsen}, M.~G., {Ackermann}, M., {Adams}, J., {et~al.} 2019, \prd, 100, 082002, \dodoi{10.1103/PhysRevD.100.082002}

\bibitem[{{Aartsen} {et~al.}(2020){Aartsen}, {Ackermann}, {Adams}, {Aguilar}, {Ahlers}, {Ahrens}, {Alispach}, {Andeen}, {Anderson}, {Ansseau}, {Anton}, {Arg{\"u}elles}, {Auffenberg}, {Axani}, {Backes}, {Bagherpour}, {Bai}, {Balagopal}, {Barbano}, {Barwick}, {Bastian}, {Baum}, {Baur}, {Bay}, {Beatty}, {Becker}, {Becker Tjus}, {BenZvi}, {Berley}, {Bernardini}, {Besson}, {Binder}, {Bindig}, {Blaufuss}, {Blot}, {Bohm}, {B{\"o}rner}, {B{\"o}ser}, {Botner}, {B{\"o}ttcher}, {Bourbeau}, {Bourbeau}, {Bradascio}, {Braun}, {Bron}, {Brostean-Kaiser}, {Burgman}, {Buscher}, {Busse}, {Carver}, {Chen}, {Cheung}, {Chirkin}, {Choi}, {Clark}, {Classen}, {Coleman}, {Collin}, {Conrad}, {Coppin}, {Correa}, {Cowen}, {Cross}, {Dave}, {De Clercq}, {DeLaunay}, {Dembinski}, {Deoskar}, {De Ridder}, {Desiati}, {de Vries}, {de Wasseige}, {de With}, {DeYoung}, {Diaz}, {D{\'\i}az-V{\'e}lez}, {Dujmovic}, {Dunkman}, {Dvorak}, {Eberhardt}, {Ehrhardt}, {Eller}, {Engel}, {Evenson}, {Fahey}, {Fazely}, {Felde}, {Filimonov}, {Finley}, {Fox},
  {Franckowiak}, {Friedman}, {Fritz}, {Gaisser}, {Gallagher}, {Ganster}, {Garrappa}, {Gerhardt}, {Ghorbani}, {Glauch}, {Gl{\"u}senkamp}, {Goldschmidt}, {Gonzalez}, {Grant}, {Griffith}, {Griswold}, {G{\"u}nder}, {G{\"u}nd{\"u}z}, {Haack}, {Hallgren}, {Halliday}, {Halve}, {Halzen}, {Hanson}, {Haungs}, {Hebecker}, {Heereman}, {Heix}, {Helbing}, {Hellauer}, {Henningsen}, {Hickford}, {Hignight}, {Hill}, {Hoffman}, {Hoffmann}, {Hoinka}, {Hokanson-Fasig}, {Hoshina}, {Huang}, {Huber}, {Huber}, {Hultqvist}, {H{\"u}nnefeld}, {Hussain}, {In}, {Iovine}, {Ishihara}, {Japaridze}, {Jeong}, {Jero}, {Jones}, {Jonske}, {Joppe}, {Kang}, {Kang}, {Kappes}, {Kappesser}, {Karg}, {Karl}, {Karle}, {Katz}, {Kauer}, {Kelley}, {Kheirandish}, {Kim}, {Kintscher}, {Kiryluk}, {Kittler}, {Klein}, {Koirala}, {Kolanoski}, {K{\"o}pke}, {Kopper}, {Kopper}, {Koskinen}, {Kowalski}, {Krings}, {Kr{\"u}ckl}, {Kulacz}, {Kurahashi}, {Kyriacou}, {Labare}, {Lanfranchi}, {Larson}, {Lauber}, {Lazar}, {Leonard}, {Leszczy{\'n}ska}, {Leuermann}, {Liu},
  {Lohfink}, {Lozano Mariscal}, {Lu}, {Lucarelli}, {L{\"u}nemann}, {Luszczak}, {Lyu}, {Ma}, {Madsen}, {Maggi}, {Mahn}, {Makino}, {Mallik}, {Mallot}, {Mancina}, {Mari{\c{s}}}, {Maruyama}, {Mase}, {Matis}, {Maunu}, {McNally}, {Meagher}, {Medici}, {Medina}, {Meier}, {Meighen-Berger}, {Menne}, {Merino}, {Meures}, {Micallef}, {Mockler}, {Moment{\'e}}, {Montaruli}, {Moore}, {Morse}, {Moulai}, {Muth}, {Nagai}, {Naumann}, {Neer}, {Niederhausen}, {Nisa}, {Nowicki}, {Nygren}, {Obertacke Pollmann}, {Oehler}, {Olivas}, {O'Murchadha}, {O'Sullivan}, {Palczewski}, {Pandya}, {Pankova}, {Park}, {Peiffer}, {P{\'e}rez de los Heros}, {Philippen}, {Pieloth}, {Pinat}, {Pizzuto}, {Plum}, {Porcelli}, {Price}, {Przybylski}, {Raab}, {Raissi}, {Rameez}, {Rauch}, {Rawlins}, {Rea}, {Reimann}, {Relethford}, {Renschler}, {Renzi}, {Resconi}, {Rhode}, {Richman}, {Robertson}, {Rongen}, {Rott}, {Ruhe}, {Ryckbosch}, {Rysewyk}, {Safa}, {Sanchez Herrera}, {Sandrock}, {Sandroos}, {Santander}, {Sarkar}, {Sarkar}, {Satalecka}, {Schaufel},
  {Schieler}, {Schlunder}, {Schmidt}, {Schneider}, {Schneider}, {Schr{\"o}der}, {Schumacher}, {Sclafani}, {Seckel}, {Seunarine}, {Shefali}, {Silva}, {Snihur}, {Soedingrekso}, {Soldin}, {Song}, {Spiczak}, {Spiering}, {Stachurska}, {Stamatikos}, {Stanev}, {Stein}, {Steinm{\"u}ller}, {Stettner}, {Steuer}, {Stezelberger}, {Stokstad}, {St{\"o}{\ss}l}, {Strotjohann}, {St{\"u}rwald}, {Stuttard}, {Sullivan}, {Taboada}, {Tenholt}, {Ter-Antonyan}, {Terliuk}, {Tilav}, {Tollefson}, {Tomankova}, {T{\"o}nnis}, {Toscano}, {Tosi}, {Trettin}, {Tselengidou}, {Tung}, {Turcati}, {Turcotte}, {Turley}, {Ty}, {Unger}, {Unland Elorrieta}, {Usner}, {Vandenbroucke}, {Van Driessche}, {van Eijk}, {van Eijndhoven}, {Vanheule}, {van Santen}, {Vraeghe}, {Walck}, {Wallace}, {Wallraff}, {Wandkowsky}, {Watson}, {Weaver}, {Weindl}, {Weiss}, {Weldert}, {Wendt}, {Werthebach}, {Whelan}, {Whitehorn}, {Wiebe}, {Wiebusch}, {Wille}, {Williams}, {Wills}, {Wolf}, {Wood}, {Wood}, {Woschnagg}, {Wrede}, {Xu}, {Xu}, {Xu}, {Yanez}, {Yodh}, {Yoshida},
  {Yuan}, \& {Z{\"o}cklein}}]{IceCubeUL2020}
---. 2020, \prl, 124, 051103, \dodoi{10.1103/PhysRevLett.124.051103}

\bibitem[{{Aartsen} {et~al.}(2021){Aartsen}, {Abbasi}, {Ackermann}, {Adams}, {Aguilar}, {Ahlers}, {Ahrens}, {Alispach}, {Allison}, {Amin}, {Andeen}, {Anderson}, {Ansseau}, {Anton}, {Arg{\"u}elles}, {Arlen}, {Auffenberg}, {Axani}, {Bagherpour}, {Bai}, {Balagopal V}, {Barbano}, {Bartos}, {Bastian}, {Basu}, {Baum}, {Baur}, {Bay}, {Beatty}, {Becker}, {Tjus}, {BenZvi}, {Berley}, {Bernardini}, {Besson}, {Binder}, {Bindig}, {Blaufuss}, {Blot}, {Bohm}, {Bohmer}, {B{\"o}ser}, {Botner}, {B{\"o}ttcher}, {Bourbeau}, {Bourbeau}, {Bradascio}, {Braun}, {Bron}, {Brostean-Kaiser}, {Burgman}, {Burley}, {Buscher}, {Busse}, {Bustamante}, {Campana}, {Carnie-Bronca}, {Carver}, {Chen}, {Chen}, {Cheung}, {Chirkin}, {Choi}, {Clark}, {Clark}, {Classen}, {Coleman}, {Collin}, {Connolly}, {Conrad}, {Coppin}, {Correa}, {Cowen}, {Cross}, {Dave}, {Deaconu}, {De Clercq}, {DeLaunay}, {De Kockere}, {Dembinski}, {Deoskar}, {De Ridder}, {Desai}, {Desiati}, {de Vries}, {de Wasseige}, {de With}, {DeYoung}, {Dharani}, {Diaz}, {D{\'\i}az-V{\'e}lez},
  {Dujmovic}, {Dunkman}, {DuVernois}, {Dvorak}, {Ehrhardt}, {Eller}, {Engel}, {Evans}, {Evenson}, {Fahey}, {Farrag}, {Fazely}, {Felde}, {Fienberg}, {Filimonov}, {Finley}, {Fischer}, {Fox}, {Franckowiak}, {Friedman}, {Fritz}, {Gaisser}, {Gallagher}, {Ganster}, {Garcia-Fernandez}, {Garrappa}, {Gartner}, {Gerhard}, {Gernhaeuser}, {Ghadimi}, {Glaser}, {Glauch}, {Gl{\"u}senkamp}, {Goldschmidt}, {Gonzalez}, {Goswami}, {Grant}, {Gr{\'e}goire}, {Griffith}, {Griswold}, {G{\"u}nd{\"u}z}, {Haack}, {Hallgren}, {Halliday}, {Halve}, {Halzen}, {Hanson}, {Hanson}, {Hardin}, {Haugen}, {Haungs}, {Hauser}, {Hebecker}, {Heinen}, {Heix}, {Helbing}, {Hellauer}, {Henningsen}, {Hickford}, {Hignight}, {Hill}, {Hill}, {Hoffman}, {Hoffmann}, {Hoffmann}, {Hoinka}, {Hokanson-Fasig}, {Holzapfel}, {Hoshina}, {Huang}, {Huber}, {Huber}, {Huege}, {Hughes}, {Hultqvist}, {H{\"u}nnefeld}, {Hussain}, {In}, {Iovine}, {Ishihara}, {Jansson}, {Japaridze}, {Jeong}, {Jones}, {Jonske}, {Joppe}, {Kalekin}, {Kang}, {Kang}, {Kang}, {Kappes}, {Kappesser},
  {Karg}, {Karl}, {Karle}, {Katori}, {Katz}, {Kauer}, {Keivani}, {Kellermann}, {Kelley}, {Kheirandish}, {Kim}, {Kin}, {Kintscher}, {Kiryluk}, {Kittler}, {Kleifges}, {Klein}, {Koirala}, {Kolanoski}, {K{\"o}pke}, {Kopper}, {Kopper}, {Koskinen}, {Koundal}, {Kovacevich}, {Kowalski}, {Krauss}, {Krings}, {Kr{\"u}ckl}, {Kulacz}, {Kurahashi}, {Gualda}, {Lahmann}, {Lanfranchi}, {Larson}, {Latif}, {Lauber}, {Lazar}, {Leonard}, {Leszczy{\'n}ska}, {Li}, {Liu}, {Lohfink}, {LoSecco}, {Mariscal}, {Lu}, {Lucarelli}, {Ludwig}, {L{\"u}nemann}, {Luszczak}, {Lyu}, {Ma}, {Madsen}, {Maggi}, {Mahn}, {Makino}, {Mallik}, {Mancina}, {Mandalia}, {Mari{\c{s}}}, {Marka}, {Marka}, {Maruyama}, {Mase}, {Maunu}, {McNally}, {Meagher}, {Medina}, {Meier}, {Meighen-Berger}, {Merz}, {Meyers}, {Micallef}, {Mockler}, {Moment{\'e}}, {Montaruli}, {Moore}, {Morse}, {Moulai}, {Muth}, {Naab}, {Nagai}, {Nam}, {Nauman}, {Necker}, {Neer}, {Nelles}, {Nguyn}, {Niederhausen}, {Nisa}, {Nowicki}, {Nygren}, {Oberla}, {Pollmann}, {Oehler}, {Olivas}, {O'Sullivan},
  {Pan}, {Pandya}, {Pankova}, {Papp}, {Park}, {Parker}, {Paudel}, {Peiffer}, {P{\'e}rez de los Heros}, {Petersen}, {Philippen}, {Pieloth}, {Pieper}, {Pinfold}, {Pizzuto}, {Plaisier}, {Plum}, {Popovych}, {Porcelli}, {Rodriguez}, {Price}, {Przybylski}, {Raab}, {Raissi}, {Rameez}, {Rauch}, {Rawlins}, {Rea}, {Rehman}, {Reimann}, {Renschler}, {Renzi}, {Resconi}, {Reusch}, {Rhode}, {Richman}, {Riedel}, {Riegel}, {Roberts}, {Robertson}, {Roellinghoff}, {Rongen}, {Rott}, {Ruhe}, {Ryckbosch}, {Cantu}, {Safa}, {Herrera}, {Sandrock}, {Sandroos}, {Sandstrom}, {Santander}, {Sarkar}, {Sarkar}, {Satalecka}, {Scharf}, {Schaufel}, {Schieler}, {Schlunder}, {Schmidt}, {Schneider}, {Schneider}, {Schr{\"o}der}, {Schumacher}, {Sclafani}, {Seckel}, {Seunarine}, {Shaevitz}, {Sharma}, {Shefali}, {Silva}, {Smith}, {Smithers}, {Snihur}, {Soedingrekso}, {Soldin}, {S{\"o}ldner-Rembold}, {Song}, {Southall}, {Spiczak}, {Spiering}, {Stachurska}, {Stamatikos}, {Stanev}, {Stein}, {Stettner}, {Steuer}, {Stezelberger}, {Stokstad},
  {Strotjohann}, {St{\"u}rwald}, {Stuttard}, {Sullivan}, {Taboada}, {Taketa}, {Tanaka}, {Tenholt}, {Ter-Antonyan}, {Terliuk}, {Tilav}, {Tollefson}, {Tomankova}, {T{\"o}nnis}, {Torres}, {Toscano}, {Tosi}, {Trettin}, {Tselengidou}, {Tung}, {Turcati}, {Turcotte}, {Turley}, {Twagirayezu}, {Ty}, {Unger}, {Elorrieta}, {Vandenbroucke}, {van Eijk}, {van Eijndhoven}, {Vannerom}, {van Santen}, {Veberic}, {Verpoest}, {Vieregg}, {Vraeghe}, {Walck}, {Watson}, {Weaver}, {Weindl}, {Weinstock}, {Weiss}, {Weldert}, {Welling}, {Wendt}, {Werthebach}, {Whitehorn}, {Wiebe}, {Wiebusch}, {Williams}, {Wissel}, {Wolf}, {Wood}, {Woschnagg}, {Wrede}, {Wren}, {Wulff}, {Xu}, {Xu}, {Yanez}, {Yoshida}, {Yuan}, {Zhang}, {Zierke}, \& {Z{\"o}cklein}}]{IceCubeGen22021}
{Aartsen}, M.~G., {Abbasi}, R., {Ackermann}, M., {et~al.} 2021, Journal of Physics G Nuclear Physics, 48, 060501, \dodoi{10.1088/1361-6471/abbd48}

\bibitem[{{Abbasi} {et~al.}(2018){Abbasi}, {Abe}, {Abu-Zayyad}, {Allen}, {Azuma}, {Barcikowski}, {Belz}, {Bergman}, {Blake}, {Cady}, {Cheon}, {Chiba}, {Chikawa}, {Di Matteo}, {Fujii}, {Fujita}, {Fukushima}, {Furlich}, {Goto}, {Hanlon}, {Hayashi}, {Hayashi}, {Hayashida}, {Hibino}, {Honda}, {Ikeda}, {Inoue}, {Ishii}, {Ishimori}, {Ito}, {Ivanov}, {Jeong}, {Jeong}, {Jui}, {Kadota}, {Kakimoto}, {Kalashev}, {Kasahara}, {Kawai}, {Kawakami}, {Kawana}, {Kawata}, {Kido}, {Kim}, {Kim}, {Kim}, {Kishigami}, {Kitamura}, {Kitamura}, {Kuzmin}, {Kuznetsov}, {Kwon}, {Lee}, {Lubsandorzhiev}, {Lundquist}, {Machida}, {Martens}, {Matsuyama}, {Matthews}, {Mayta}, {Minamino}, {Mukai}, {Myers}, {Nagasawa}, {Nagataki}, {Nakamura}, {Nakamura}, {Nonaka}, {Nozato}, {Oda}, {Ogio}, {Ogura}, {Ohnishi}, {Ohoka}, {Okuda}, {Omura}, {Ono}, {Onogi}, {Oshima}, {Ozawa}, {Park}, {Pshirkov}, {Rodriguez}, {Rubtsov}, {Ryu}, {Sagawa}, {Sahara}, {Saito}, {Saito}, {Sakaki}, {Sakurai}, {Scott}, {Seki}, {Sekino}, {Shah}, {Shibata}, {Shibata}, {Shimodaira},
  {Shin}, {Shin}, {Smith}, {Sokolsky}, {Stokes}, {Stratton}, {Stroman}, {Suzawa}, {Takagi}, {Takahashi}, {Takamura}, {Takeda}, {Takeishi}, {Taketa}, {Takita}, {Tameda}, {Tanaka}, {Tanaka}, {Tanaka}, {Thomas}, {Thomson}, {Tinyakov}, {Tkachev}, {Tokuno}, {Tomida}, {Troitsky}, {Tsunesada}, {Tsutsumi}, {Uchihori}, {Udo}, {Urban}, {Wong}, {Yamamoto}, {Yamane}, {Yamaoka}, {Yamazaki}, {Yang}, {Yashiro}, {Yoneda}, {Yoshida}, {Yoshii}, {Zhezher}, {Zundel}, \& {Telescope Array Collaboration}}]{Abbasi+2018_TALE}
{Abbasi}, R.~U., {Abe}, M., {Abu-Zayyad}, T., {et~al.} 2018, \apj, 865, 74, \dodoi{10.3847/1538-4357/aada05}

\bibitem[{{Abe} {et~al.}(2022){Abe}, {Abe}, {Acciari}, {Aniello}, {Ansoldi}, {Antonelli}, {Arbet Engels}, {Arcaro}, {Artero}, {Asano}, {Baack}, {Babi{\'c}}, {Baquero}, {Barres de Almeida}, {Barrio}, {Batkovi{\'c}}, {Baxter}, {Becerra Gonz{\'a}lez}, {Bednarek}, {Bernardini}, {Bernardos}, {Berti}, {Besenrieder}, {Bhattacharyya}, {Bigongiari}, {Biland}, {Blanch}, {Bonnoli}, {Bo{\v{s}}njak}, {Burelli}, {Busetto}, {Carosi}, {Carretero-Castrillo}, {Ceribella}, {Chai}, {Chilingarian}, {Cikota}, {Colombo}, {Contreras}, {Cortina}, {Covino}, {D'Amico}, {D'Elia}, {da Vela}, {Dazzi}, {de Angelis}, {de Lotto}, {Del Popolo}, {Delfino}, {Delgado}, {Delgado Mendez}, {Depaoli}, {di Pierro}, {di Venere}, {Dominis Prester}, {Donini}, {Dorner}, {Doro}, {Elsaesser}, {Emery}, {Fallah Ramazani}, {Fari{\~n}a}, {Fattorini}, {Font}, {Fruck}, {Fukami}, {Fukazawa}, {Garc{\'\i}a L{\'o}pez}, {Garczarczyk}, {Gasparyan}, {Gaug}, {Giesbrecht Paiva}, {Giglietto}, {Giordano}, {Gliwny}, {Godinovi{\'c}}, {Grau}, {Green}, {Green}, {Hadasch},
  {Hahn}, {Hassan}, {Heckmann}, {Herrera}, {Hoang}, {Hrupec}, {H{\"u}tten}, {Imazawa}, {Inada}, {Iotov}, {Ishio}, {Jim{\'e}nez Mart{\'\i}nez}, {Jormanainen}, {Kerszberg}, {Kobayashi}, {Kubo}, {Kushida}, {Lamastra}, {Lelas}, {Leone}, {Lindfors}, {Linhoff}, {Lombardi}, {Longo}, {L{\'o}pez-Coto}, {L{\'o}pez-Moya}, {L{\'o}pez-Oramas}, {Loporchio}, {Lorini}, {Lyard}, {Fraga}, {Majumdar}, {Makariev}, {Maneva}, {Mang}, {Manganaro}, {Mangano}, {Mannheim}, {Mariotti}, {Mart{\'\i}nez}, {Mas Aguilar}, {Mazin}, {Menchiari}, {Mender}, {Mi{\'c}anovi{\'c}}, {Miceli}, {Miener}, {Miranda}, {Mirzoyan}, {Molina}, {Mondal}, {Moralejo}, {Morcuende}, {Moreno}, {Nakamori}, {Nanci}, {Nava}, {Neustroev}, {Nievas Rosillo}, {Nigro}, {Nilsson}, {Nishijima}, {Njoh Ekoume}, {Noda}, {Nozaki}, {Ohtani}, {Oka}, {Okumura}, {Otero-Santos}, {Paiano}, {Palatiello}, {Paneque}, {Paoletti}, {Paredes}, {Pavleti{\'c}}, {Persic}, {Pihet}, {Pirola}, {Podobnik}, {Prada Moroni}, {Prandini}, {Principe}, {Priyadarshi}, {Puljak}, {Rhode}, {Rib{\'o}},
  {Rico}, {Righi}, {Rugliancich}, {Sahakyan}, {Saito}, {Sakurai}, {Satalecka}, {Saturni}, {Schleicher}, {Schmidt}, {Schmuckermaier}, {Schubert}, {Schweizer}, {Sitarek}, {Sliusar}, {Sobczynska}, {Spolon}, {Stamerra}, {Stri{\v{s}}kovi{\'c}}, {Strom}, {Strzys}, {Suda}, {Suri{\'c}}, {Tajima}, {Takahashi}, {Takeishi}, {Tavecchio}, {Temnikov}, {Terauchi}, {Terzi{\'c}}, {Teshima}, {Tosti}, {Truzzi}, {Tutone}, {Ubach}, {van Scherpenberg}, {Vazquez Acosta}, {Ventura}, {Verguilov}, {Viale}, {Vigorito}, {Vitale}, {Vovk}, {Walter}, {Will}, {Wunderlich}, {Yamamoto}, {Zari{\'c}}, {Abdalla}, {Aharonian}, {Ait Benkhali}, {Ang{\"u}ner}, {Ashkar}, {Backes}, {Baghmanyan}, {Barbosa Martins}, {Batzofin}, {Becherini}, {Berge}, {Bernl{\"o}hr}, {B{\"o}ttcher}, {Boisson}, {Bolmont}, {de Bony de Lavergne}, {Bradascio}, {Breuhaus}, {Brose}, {Brun}, {Bulik}, {Bylund}, {Cangemi}, {Caroff}, {Casanova}, {Cerruti}, {Chand}, {Chandra}, {Chen}, {Chibueze}, {Cotter}, {Cristofari}, {Damascene Mbarubucyeye}, {Devin}, {Djannati-Ata{\"\i}},
  {Dmytriiev}, {Egberts}, {Ernenwein}, {Fiasson}, {Fichet de Clairfontaine}, {Fontaine}, {F{\"u}{\ss}ling}, {Funk}, {Gabici}, {Ghafourizadeh}, {Giavitto}, {Glawion}, {Glicenstein}, {Goswami}, {Grolleron}, {Hinton}, {H{\"o}rbe}, {Hoischen}, {Holch}, {Holler}, {Horns}, {Huang}, {Jamrozy}, {Jankowsky}, {Joshi}, {Jung-Richardt}, {Kasai}, {Katarzy{\'n}ski}, {Katz}, {Kh{\'e}lifi}, {Klu{\'z}niak}, {Komin}, {Kosack}, {Kostunin}, {Lang}, {Le Stum}, {Lemi{\`e}re}, {Lemoine-Goumard}, {Lenain}, {Leuschner}, {Lohse}, {Luashvili}, {Lypova}, {Mackey}, {Majumdar}, {Malyshev}, {Malyshev}, {Marandon}, {Marchegiani}, {Mart{\'\i}-Devesa}, {Marx}, {Maurin}, {Meyer}, {Mitchell}, {Moderski}, {Mohrmann}, {Montanari}, {Moulin}, {Muller}, {Murach}, {Nakashima}, {de Naurois}, {Nayerhoda}, {Niemiec}, {Priyana Noel}, {O'Brien}, {Ohm}, {Olivera-Nieto}, {de Ona Wilhelmi}, {Ostrowski}, {Panny}, {Panter}, {Parsons}, {Poireau}, {Prokhorov}, {Prokoph}, {P{\"u}hlhofer}, {Punch}, {Quirrenbach}, {Reichherzer}, {Reimer}, {Reimer}, {Renaud},
  {Rieger}, {Rowell}, {Rudak}, {Rueda Ricarte}, {Ruiz-Velasco}, {Sahakian}, {Salzmann}, {Santangelo}, {Sasaki}, {Sch{\"a}fer}, {Sch{\"u}ssler}, {Schutte}, {Schwanke}, {Shapopi}, {Sol}, {Specovius}, {Spencer}, {Stawarz}, {Steenkamp}, {Steinmassl}, {Steppa}, {Sushch}, {Suzuki}, {Takahashi}, {Tanaka}, {Thorpe-Morgan}, {Tluczykont}, {Tomankova}, {Tsuji}, {Uchiyama}, {van Eldik}, {van Soelen}, {Vecchi}, {Veh}, {Venter}, {Vink}, {Wagner}, {White}, {Wierzcholska}, {Wong}, {Yusafzai}, {Zacharias}, {Zanin}, {Zargaryan}, {Zdziarski}, {Zech}, {Zhu}, {Zouari}, {{\.Z}ywucka}, {Acharyya}, {Adams}, {Batista}, {Benbow}, {Capasso}, {Christiansen}, {Chromey}, {Errando}, {Falcone}, {Feng}, {Finley}, {Foote}, {Fortson}, {Furniss}, {Gent}, {Hanlon}, {Hervet}, {Holder}, {Hona}, {Humensky}, {Jin}, {Kaaret}, {Kertzman}, {Kherlakian}, {Kleiner}, {Kumar}, {Lang}, {Lundy}, {Maier}, {McGrath}, {Millis}, {Moriarty}, {Mukherjee}, {O'Brien}, {Ong}, {Park}, {Patel}, {Pfrang}, {Pohl}, {Pueschel}, {Quinn}, {Ragan}, {Reynolds}, {Ribeiro},
  {Roache}, {Ryan}, {Sadeh}, {Saha}, {Santander}, {Sembroski}, {Shang}, {Splettstoesser}, {Tak}, {Tucci}, {Weinstein}, {Williams}, {Williamson}, {Bosch-Ramon}, {Celma}, {Linares}, {Russell}, {Sala}, \& {The VERITAS Collaboration}}]{Abe+2022}
{Abe}, H., {Abe}, S., {Acciari}, V.~A., {et~al.} 2022, \mnras, 517, 4736, \dodoi{10.1093/mnras/stac2686}

\bibitem[{Abeysekara {et~al.}(2018)}]{HAWC:2018gwz}
Abeysekara, A.~U., {et~al.} 2018, Nature, 562, 82, \dodoi{10.1038/s41586-018-0565-5}

\bibitem[{Ackermann {et~al.}(2013)}]{Ackermann:2013wqa}
Ackermann, M., {et~al.} 2013, Science, 339, 807, \dodoi{10.1126/science.1231160}

\bibitem[{{Actis} {et~al.}(2011){Actis}, {Agnetta}, {Aharonian}, {Akhperjanian}, {Aleksi{\'c}}, {Aliu}, {Allan}, {Allekotte}, {Antico}, {Antonelli}, {Antoranz}, {Aravantinos}, {Arlen}, {Arnaldi}, {Artmann}, {Asano}, {Asorey}, {B{\"a}hr}, {Bais}, {Baixeras}, {Bajtlik}, {Balis}, {Bamba}, {Barbier}, {Barcel{\'o}}, {Barnacka}, {Barnstedt}, {Barres de Almeida}, {Barrio}, {Basso}, {Bastieri}, {Bauer}, {Becerra}, {Becherini}, {Bechtol}, {Becker}, {Beckmann}, {Bednarek}, {Behera}, {Beilicke}, {Belluso}, {Benallou}, {Benbow}, {Berdugo}, {Berger}, {Bernardino}, {Bernl{\"o}hr}, {Biland}, {Billotta}, {Bird}, {Birsin}, {Bissaldi}, {Blake}, {Blanch}, {Bobkov}, {Bogacz}, {Bogdan}, {Boisson}, {Boix}, {Bolmont}, {Bonanno}, {Bonardi}, {Bonev}, {Borkowski}, {Botner}, {Bottani}, {Bourgeat}, {Boutonnet}, {Bouvier}, {Brau-Nogu{\'e}}, {Braun}, {Bretz}, {Briggs}, {Brun}, {Brunetti}, {Buckley}, {Bugaev}, {B{\"u}hler}, {Bulik}, {Busetto}, {Buson}, {Byrum}, {Cailles}, {Cameron}, {Canestrari}, {Cantu}, {Carmona}, {Carosi}, {Carr},
  {Carton}, {Casiraghi}, {Castarede}, {Catalano}, {Cavazzani}, {Cazaux}, {Cerruti}, {Cerruti}, {Chadwick}, {Chiang}, {Chikawa}, {Cie{\'s}lar}, {Ciesielska}, {Cillis}, {Clerc}, {Colin}, {Colom{\'e}}, {Compin}, {Conconi}, {Connaughton}, {Conrad}, {Contreras}, {Coppi}, {Corlier}, {Corona}, {Corpace}, {Corti}, {Cortina}, {Costantini}, {Cotter}, {Courty}, {Couturier}, {Covino}, {Croston}, {Cusumano}, {Daniel}, {Dazzi}, {de Angelis}, {de Cea Del Pozo}, {de Gouveia Dal Pino}, {de Jager}, {de La Calle P{\'e}rez}, {de La Vega}, {de Lotto}, {de Naurois}, {de O{\~n}a Wilhelmi}, {de Souza}, {Decerprit}, {Deil}, {Delagnes}, {Deleglise}, {Delgado}, {Dettlaff}, {di Paolo}, {di Pierro}, {D{\'\i}az}, {Dick}, {Dickinson}, {Digel}, {Dimitrov}, {Disset}, {Djannati-Ata{\"\i}}, {Doert}, {Domainko}, {Dorner}, {Doro}, {Dournaux}, {Dravins}, {Drury}, {Dubois}, {Dubois}, {Dubus}, {Dufour}, {Durand}, {Dyks}, {Dyrda}, {Edy}, {Egberts}, {Eleftheriadis}, {Elles}, {Emmanoulopoulos}, {Enomoto}, {Ernenwein}, {Errando}, {Etchegoyen},
  {Falcone}, {Farakos}, {Farnier}, {Federici}, {Feinstein}, {Ferenc}, {Fillin-Martino}, {Fink}, {Finley}, {Finley}, {Firpo}, {Florin}, {F{\"o}hr}, {Fokitis}, {Font}, {Fontaine}, {Fontana}, {F{\"o}rster}, {Fortson}, {Fouque}, {Fransson}, {Fraser}, {Fresnillo}, {Fruck}, {Fujita}, {Fukazawa}, {Funk}, {G{\"a}bele}, {Gabici}, {Gadola}, {Galante}, {Gallant}, {Garc{\'\i}a}, {Garc{\'\i}a L{\'o}pez}, {Garrido}, {Garrido}, {Gasc{\'o}n}, {Gasq}, {Gaug}, {Gaweda}, {Geffroy}, {Ghag}, {Ghedina}, {Ghigo}, {Gianakaki}, {Giarrusso}, {Giavitto}, {Giebels}, {Giro}, {Giubilato}, {Glanzman}, {Glicenstein}, {Gochna}, {Golev}, {G{\'o}mez Berisso}, {Gonz{\'a}lez}, {Gonz{\'a}lez}, {Gra{\~n}ena}, {Graciani}, {Granot}, {Gredig}, {Green}, {Greenshaw}, {Grimm}, {Grube}, {Grudzi{\'n}ska}, {Grygorczuk}, {Guarino}, {Guglielmi}, {Guilloux}, {Gunji}, {Gyuk}, {Hadasch}, {Haefner}, {Hagiwara}, {Hahn}, {Hallgren}, {Hara}, {Hardcastle}, {Hassan}, {Haubold}, {Hauser}, {Hayashida}, {Heller}, {Henri}, {Hermann}, {Herrero}, {Hinton}, {Hoffmann},
  {Hofmann}, {Hofverberg}, {Horns}, {Hrupec}, {Huan}, {Huber}, {Huet}, {Hughes}, {Hultquist}, {Humensky}, {Huppert}, {Ibarra}, {Illa}, {Ingjald}, {Inoue}, {Inoue}, {Ioka}, {Jablonski}, {Jacholkowska}, {Janiak}, {Jean}, {Jensen}, {Jogler}, {Jung}, {Kaaret}, {Kabuki}, {Kakuwa}, {Kalkuhl}, {Kankanyan}, {Kapala}, {Karastergiou}, {Karczewski}, {Karkar}, {Karlsson}, {Kasperek}, {Katagiri}, {Katarzy{\'n}ski}, {Kawanaka}, {K{\c{e}}dziora}, {Kendziorra}, {Kh{\'e}lifi}, {Kieda}, {Kifune}, {Kihm}, {Klepser}, {Klu{\'z}niak}, {Knapp}, {Knappy}, {Kneiske}, {Kn{\"o}dlseder}, {K{\"o}ck}, {Kodani}, {Kohri}, {Kokkotas}, {Komin}, {Konopelko}, {Kosack}, {Kossakowski}, {Kostka}, {Kotu{\l}a}, {Kowal}, {Kozio{\l}}, {Kr{\"a}henb{\"u}hl}, {Krause}, {Krawczynski}, {Krennrich}, {Kretzschmann}, {Kubo}, {Kudryavtsev}, {Kushida}, {La Barbera}, {La Parola}, {La Rosa}, {L{\'o}pez}, {Lamanna}, {Laporte}, {Lavalley}, {Le Flour}, {Le Padellec}, {Lenain}, {Lessio}, {Lieunard}, {Lindfors}, {Liolios}, {Lohse}, {Lombardi}, {Lopatin}, {Lorenz},
  {Lubi{\'n}ski}, {Luz}, {Lyard}, {Maccarone}, {Maccarone}, {Maier}, {Majumdar}, {Maltezos}, {Ma{\l}kiewicz}, {Ma{\~n}{\'a}}, {Manalaysay}, {Maneva}, {Mangano}, {Manigot}, {Mar{\'\i}n}, {Mariotti}, {Markoff}, {Mart{\'\i}nez}, {Mart{\'\i}nez}, {Mastichiadis}, {Matsumoto}, {Mattiazzo}, {Mazin}, {McComb}, {McCubbin}, {McHardy}, {Medina}, {Melkumyan}, {Mendes}, {Mertsch}, {Meucci}, {Micha{\l}owski}, {Micolon}, {Mineo}, {Mirabal}, {Mirabel}, {Miranda}, {Mirzoyan}, {Mizuno}, {Moal}, {Moderski}, {Molinari}, {Monteiro}, {Moralejo}, {Morello}, {Mori}, {Motta}, {Mottez}, {Moulin}, {Mukherjee}, {Munar}, {Muraishi}, {Murase}, {Murphy}, {Nagataki}, {Naito}, {Nakamori}, {Nakayama}, {Naumann}, {Naumann}, {Nayman}, {Nedbal}, {Nied{\'z}wiecki}, {Niemiec}, {Nikolaidis}, {Nishijima}, {Nolan}, {Nowak}, {O'Brien}, {Ochoa}, {Ohira}, {Ohishi}, {Ohka}, {Okumura}, {Olivetto}, {Ong}, {Orito}, {Orr}, {Osborne}, {Ostrowski}, {Otero}, {Otte}, {Ovcharov}, {Oya}, {Ozi{\c{e}}b{\l}o}, {Paiano}, {Pallota}, {Panazol}, {Paneque}, {Panter},
  {Paoletti}, {Papyan}, {Paredes}, {Pareschi}, {Parsons}, {Paz Arribas}, {Pedaletti}, {Pepato}, {Persic}, {Petrucci}, {Peyaud}, {Piechocki}, {Pita}, {Pivato}, {P{\l}atos}, {Platzer}, {Pogosyan}, {Pohl}, {Pojma{\'n}ski}, {Ponz}, {Potter}, {Prandini}, {Preece}, {Prokoph}, {P{\"u}hlhofer}, {Punch}, {Quel}, {Quirrenbach}, {Rajda}, {Rando}, {Rataj}, {Raue}, {Reimann}, {Reimann}, {Reimer}, {Reimer}, {Renaud}, {Renner}, {Reymond}, {Rhode}, {Rib{\'o}}, {Ribordy}, {Rico}, {Rieger}, {Ringegni}, {Ripken}, {Ristori}, {Rivoire}, {Rob}, {Rodriguez}, {Roeser}, {Romano}, {Romero}, {Rosier-Lees}, {Rovero}, {Roy}, {Royer}, {Rudak}, {Rulten}, {Ruppel}, {Russo}, {Ryde}, {Sacco}, {Saggion}, {Sahakian}, {Saito}, {Saito}, {Sakaki}, {Salazar}, {Salini}, {S{\'a}nchez}, {S{\'a}nchez Conde}, {Santangelo}, {Santos}, {Sanuy}, {Sapozhnikov}, {Sarkar}, {Scalzotto}, {Scapin}, {Scarcioffolo}, {Schanz}, {Schlenstedt}, {Schlickeiser}, {Schmidt}, {Schmoll}, {Schroedter}, {Schultz}, {Schultze}, {Schulz}, {Schwanke}, {Schwarzburg}, {Schweizer},
  {Seiradakis}, {Selmane}, {Seweryn}, {Shayduk}, {Shellard}, {Shibata}, {Sikora}, {Silk}, {Sillanp{\"a}{\"a}}, {Sitarek}, {Skole}, {Smith}, {Sobczy{\'n}ska}, {Sofo Haro}, {Sol}, {Spanier}, {Spiga}, {Spyrou}, {Stamatescu}, {Stamerra}, {Starling}, {Stawarz}, {Steenkamp}, {Stegmann}, {Steiner}, {Stergioulas}, {Sternberger}, {Stinzing}, {Stodulski}, {Straumann}, {Su{\'a}rez}, {Suchenek}, {Sugawara}, {Sulanke}, {Sun}, {Supanitsky}, {Sutcliffe}, {Szanecki}, {Szepieniec}, {Szostek}, {Szymkowiak}, {Tagliaferri}, {Tajima}, {Takahashi}, {Takahashi}, {Takalo}, {Takami}, {Talbot}, {Tam}, {Tanaka}, {Tanimori}, {Tavani}, {Tavernet}, {Tchernin}, {Tejedor}, {Telezhinsky}, {Temnikov}, {Tenzer}, {Terada}, {Terrier}, {Teshima}, {Testa}, {Tibaldo}, {Tibolla}, {Tluczykont}, {Todero Peixoto}, {Tokanai}, {Tokarz}, {Toma}, {Torres}, {Tosti}, {Totani}, {Toussenel}, {Vallania}, {Vallejo}, {van der Walt}, {van Eldik}, {Vandenbroucke}, {Vankov}, {Vasileiadis}, {Vassiliev}, {Vegas}, {Venter}, {Vercellone}, {Veyssiere}, {Vialle},
  {Videla}, {Vincent}, {Vink}, {Vlahakis}, {Vlahos}, {Vogler}, {Vollhardt}, {Volpe}, {von Gunten}, {Vorobiov}, {Wagner}, {Wagner}, {Wagner}, {Wakely}, {Walter}, {Walter}, {Warwick}, {Wawer}, {Wawrzaszek}, {Webb}, {Wegner}, {Weinstein}, {Weitzel}, {Welsing}, {Wetteskind}, {White}, {Wierzcholska}, {Wilkinson}, {Williams}, {Winde}, {Wischnewski}, {Wi{\'s}niewski}, {Wolczko}, {Wood}, {Xiong}, {Yamamoto}, {Yamaoka}, {Yamazaki}, {Yanagita}, {Yoffo}, {Yonetani}, {Yoshida}, {Yoshida}, {Yoshikoshi}, {Zabalza}, {Zagda{\'n}ski}, {Zajczyk}, {Zdziarski}, {Zech}, {Zi{\c{e}}tara}, {Zi{\'o}{\l}kowski}, {Zitelli}, \& {Zychowski}}]{CTA2011}
{Actis}, M., {Agnetta}, G., {Aharonian}, F., {et~al.} 2011, Experimental Astronomy, 32, 193, \dodoi{10.1007/s10686-011-9247-0}

\bibitem[{{Adri{\'a}n-Mart{\'\i}nez} {et~al.}(2016){Adri{\'a}n-Mart{\'\i}nez}, {Ageron}, {Aharonian}, {Aiello}, {Albert}, {Ameli}, {Anassontzis}, {Andre}, {Androulakis}, {Anghinolfi}, {Anton}, {Ardid}, {Avgitas}, {Barbarino}, {Barbarito}, {Baret}, {Barrios-Mart{\'\i}}, {Belhorma}, {Belias}, {Berbee}, {van den Berg}, {Bertin}, {Beurthey}, {van Beveren}, {Beverini}, {Biagi}, {Biagioni}, {Billault}, {Bond{\`\i}}, {Bormuth}, {Bouhadef}, {Bourlis}, {Bourret}, {Boutonnet}, {Bouwhuis}, {Bozza}, {Bruijn}, {Brunner}, {Buis}, {Busto}, {Cacopardo}, {Caillat}, {Calamai}, {Calvo}, {Capone}, {Caramete}, {Cecchini}, {Celli}, {Champion}, {Cherkaoui El Moursli}, {Cherubini}, {Chiarusi}, {Circella}, {Classen}, {Cocimano}, {Coelho}, {Coleiro}, {Colonges}, {Coniglione}, {Cordelli}, {Cosquer}, {Coyle}, {Creusot}, {Cuttone}, {D'Amico}, {De Bonis}, {De Rosa}, {De Sio}, {Di Capua}, {Di Palma}, {D{\'\i}az Garc{\'\i}a}, {Distefano}, {Donzaud}, {Dornic}, {Dorosti-Hasankiadeh}, {Drakopoulou}, {Drouhin}, {Drury}, {Durocher}, {Eberl},
  {Eichie}, {van Eijk}, {El Bojaddaini}, {El Khayati}, {Elsaesser}, {Enzenh{\"o}fer}, {Fassi}, {Favali}, {Fermani}, {Ferrara}, {Filippidis}, {Frascadore}, {Fusco}, {Gal}, {Galat{\`a}}, {Garufi}, {Gay}, {Gebyehu}, {Giordano}, {Gizani}, {Gracia}, {Graf}, {Gr{\'e}goire}, {Grella}, {Habel}, {Hallmann}, {van Haren}, {Harissopulos}, {Heid}, {Heijboer}, {Heine}, {Henry}, {Hern{\'a}ndez-Rey}, {Hevinga}, {Hofest{\"a}dt}, {Hugon}, {Illuminati}, {James}, {Jansweijer}, {Jongen}, {de Jong}, {Kadler}, {Kalekin}, {Kappes}, {Katz}, {Keller}, {Kieft}, {Kie{\ss}ling}, {Koffeman}, {Kooijman}, {Kouchner}, {Kulikovskiy}, {Lahmann}, {Lamare}, {Leisos}, {Leonora}, {Clark}, {Liolios}, {Llorens Alvarez}, {Lo Presti}, {L{\"o}hner}, {Lonardo}, {Lotze}, {Loucatos}, {Maccioni}, {Mannheim}, {Margiotta}, {Marinelli}, {Mari{\c{s}}}, {Markou}, {Mart{\'\i}nez-Mora}, {Martini}, {Mele}, {Melis}, {Michael}, {Migliozzi}, {Migneco}, {Mijakowski}, {Miraglia}, {Mollo}, {Mongelli}, {Morganti}, {Moussa}, {Musico}, {Musumeci}, {Navas}, {Nicolau},
  {Olcina}, {Olivetto}, {Orlando}, {Papaikonomou}, {Papaleo}, {P{\u{a}}v{\u{a}}la{\c{s}}}, {Peek}, {Pellegrino}, {Perrina}, {Pfutzner}, {Piattelli}, {Pikounis}, {Poma}, {Popa}, {Pradier}, {Pratolongo}, {P{\"u}hlhofer}, {Pulvirenti}, {Quinn}, {Racca}, {Raffaelli}, {Randazzo}, {Rapidis}, {Razis}, {Real}, {Resvanis}, {Reubelt}, {Riccobene}, {Rossi}, {Rovelli}, {Salda{\~n}a}, {Salvadori}, {Samtleben}, {S{\'a}nchez Garc{\'\i}a}, {S{\'a}nchez Losa}, {Sanguineti}, {Santangelo}, {Santonocito}, {Sapienza}, {Schimmel}, {Schmelling}, {Sciacca}, {Sedita}, {Seitz}, {Sgura}, {Simeone}, {Siotis}, {Sipala}, {Spisso}, {Spurio}, {Stavropoulos}, {Steijger}, {Stellacci}, {Stransky}, {Taiuti}, {Tayalati}, {T{\'e}zier}, {Theraube}, {Thompson}, {Timmer}, {T{\"o}nnis}, {Trasatti}, {Trovato}, {Tsirigotis}, {Tzamarias}, {Tzamariudaki}, {Vallage}, {Van Elewyck}, {Vermeulen}, {Vicini}, {Viola}, {Vivolo}, {Volkert}, {Voulgaris}, {Wiggers}, {Wilms}, {de Wolf}, {Zachariadou}, {Zornoza}, \& {Z{\'u}{\~n}iga}}]{KM3NeT2016}
{Adri{\'a}n-Mart{\'\i}nez}, S., {Ageron}, M., {Aharonian}, F., {et~al.} 2016, Journal of Physics G Nuclear Physics, 43, 084001, \dodoi{10.1088/0954-3899/43/8/084001}

\bibitem[{{Adriani} {et~al.}(2022){Adriani}, {Akaike}, {Asano}, {Asaoka}, {Berti}, {Bigongiari}, {Binns}, {Bongi}, {Brogi}, {Bruno}, {Buckley}, {Cannady}, {Castellini}, {Checchia}, {Cherry}, {Collazuol}, {de Nolfo}, {Ebisawa}, {Ficklin}, {Fuke}, {Gonzi}, {Guzik}, {Hams}, {Hibino}, {Ichimura}, {Ioka}, {Ishizaki}, {Israel}, {Kasahara}, {Kataoka}, {Kataoka}, {Katayose}, {Kato}, {Kawanaka}, {Kawakubo}, {Kobayashi}, {Kohri}, {Krawczynski}, {Krizmanic}, {Maestro}, {Marrocchesi}, {Messineo}, {Mitchell}, {Miyake}, {Moiseev}, {Mori}, {Mori}, {Motz}, {Munakata}, {Nakahira}, {Nishimura}, {Okuno}, {Ormes}, {Ozawa}, {Pacini}, {Papini}, {Rauch}, {Ricciarini}, {Sakai}, {Sakamoto}, {Sasaki}, {Shimizu}, {Shiomi}, {Spillantini}, {Stolzi}, {Sugita}, {Sulaj}, {Takita}, {Tamura}, {Terasawa}, {Torii}, {Tsunesada}, {Uchihori}, {Vannuccini}, {Wefel}, {Yamaoka}, {Yanagita}, {Yoshida}, {Yoshida}, {Zober}, \& {Calet Collaboration}}]{Adriani+2022}
{Adriani}, O., {Akaike}, Y., {Asano}, K., {et~al.} 2022, \prl, 129, 251103, \dodoi{10.1103/PhysRevLett.129.251103}

\bibitem[{{Aguilar} {et~al.}(2016){Aguilar}, {Ali Cavasonza}, {Ambrosi}, {Arruda}, {Attig}, {Aupetit}, {Azzarello}, {Bachlechner}, {Barao}, {Barrau}, {Barrin}, {Bartoloni}, {Basara}, {Ba{\c{s}}e{\v{g}}mez-du Pree}, {Battarbee}, {Battiston}, {Becker}, {Behlmann}, {Beischer}, {Berdugo}, {Bertucci}, {Bindel}, {Bindi}, {Boella}, {de Boer}, {Bollweg}, {Bonnivard}, {Borgia}, {Boschini}, {Bourquin}, {Bueno}, {Burger}, {Cadoux}, {Cai}, {Capell}, {Caroff}, {Casaus}, {Castellini}, {Cervelli}, {Chae}, {Chang}, {Chen}, {Chen}, {Chen}, {Cheng}, {Chou}, {Choumilov}, {Choutko}, {Chung}, {Clark}, {Clavero}, {Coignet}, {Consolandi}, {Contin}, {Corti}, {Creus}, {Crispoltoni}, {Cui}, {Dai}, {Delgado}, {Della Torre}, {Demakov}, {Demirk{\"o}z}, {Derome}, {Di Falco}, {Dimiccoli}, {D{\'\i}az}, {von Doetinchem}, {Dong}, {Donnini}, {Duranti}, {D'Urso}, {Egorov}, {Eline}, {Eronen}, {Feng}, {Fiandrini}, {Finch}, {Fisher}, {Formato}, {Galaktionov}, {Gallucci}, {Garc{\'\i}a}, {Garc{\'\i}a-L{\'o}pez}, {Gargiulo}, {Gast}, {Gebauer},
  {Gervasi}, {Ghelfi}, {Giovacchini}, {Goglov}, {G{\'o}mez-Coral}, {Gong}, {Goy}, {Grabski}, {Grandi}, {Graziani}, {Guo}, {Haino}, {Han}, {He}, {Heil}, {Hoffman}, {Hsieh}, {Huang}, {Huang}, {Huh}, {Incagli}, {Ionica}, {Jang}, {Jinchi}, {Kang}, {Kanishev}, {Kim}, {Kim}, {Kirn}, {Konak}, {Kounina}, {Kounine}, {Koutsenko}, {Krafczyk}, {La Vacca}, {Laudi}, {Laurenti}, {Lazzizzera}, {Lebedev}, {Lee}, {Lee}, {Leluc}, {Li}, {Li}, {Li}, {Li}, {Li}, {Li}, {Li}, {Li}, {Li}, {Lim}, {Lin}, {Lipari}, {Lippert}, {Liu}, {Liu}, {Lordello}, {Lu}, {Lu}, {Luebelsmeyer}, {Luo}, {Luo}, {Lv}, {Machate}, {Majka}, {Ma{\~n}{\'a}}, {Mar{\'\i}n}, {Martin}, {Mart{\'\i}nez}, {Masi}, {Maurin}, {Menchaca-Rocha}, {Meng}, {Mikuni}, {Mo}, {Morescalchi}, {Mott}, {Nelson}, {Ni}, {Nikonov}, {Nozzoli}, {Oliva}, {Orcinha}, {Palmonari}, {Palomares}, {Paniccia}, {Pauluzzi}, {Pensotti}, {Pereira}, {Picot-Clemente}, {Pilo}, {Pizzolotto}, {Plyaskin}, {Pohl}, {Poireau}, {Putze}, {Quadrani}, {Qi}, {Qin}, {Qu}, {R{\"a}ih{\"a}}, {Rancoita}, {Rapin},
  {Ricol}, {Rosier-Lees}, {Rozhkov}, {Rozza}, {Sagdeev}, {Sandweiss}, {Saouter}, {Schael}, {Schmidt}, {Schulz von Dratzig}, {Schwering}, {Seo}, {Shan}, {Shi}, {Siedenburg}, {Son}, {Song}, {Sun}, {Tacconi}, {Tang}, {Tang}, {Tao}, {Tescaro}, {Ting}, {Ting}, {Tomassetti}, {Torsti}, {T{\"u}rko{\v{g}}lu}, {Urban}, {Vagelli}, {Valente}, {Vannini}, {Valtonen}, {V{\'a}zquez Acosta}, {Vecchi}, {Velasco}, {Vialle}, {Vitale}, {Vitillo}, {Wang}, {Wang}, {Wang}, {Wang}, {Wang}, {Wang}, {Wei}, {Weng}, {Whitman}, {Wienkenh{\"o}ver}, {Wu}, {Wu}, {Xia}, {Xiong}, {Xu}, {Yan}, {Yang}, {Yang}, {Yang}, {Yi}, {Yu}, {Yu}, {Zeissler}, {Zhang}, {Zhang}, {Zhang}, {Zhang}, {Zhang}, {Zhang}, {Zheng}, {Zhu}, {Zhuang}, {Zhukov}, {Zichichi}, {Zimmermann}, {Zuccon}, \& {AMS Collaboration}}]{Aguilar+2016}
{Aguilar}, M., {Ali Cavasonza}, L., {Ambrosi}, G., {et~al.} 2016, \prl, 117, 231102, \dodoi{10.1103/PhysRevLett.117.231102}

\bibitem[{{Albert} {et~al.}(2021){Albert}, {Alfaro}, {Alvarez}, {Angeles Camacho}, {Arteaga-Vel{\'a}zquez}, {Arunbabu}, {Avila Rojas}, {Ayala Solares}, {Baghmanyan}, {Belmont-Moreno}, {BenZvi}, {Brisbois}, {Caballero-Mora}, {Capistr{\'a}n}, {Carrami{\~n}ana}, {Casanova}, {Cotti}, {Cotzomi}, {De la Fuente}, {de Le{\'o}n}, {Diaz Hernandez}, {D{\'\i}az-V{\'e}lez}, {Dingus}, {Durocher}, {DuVernois}, {Ellsworth}, {Espinoza}, {Fan}, {Fang}, {Fraija}, {Galv{\'a}n-G{\'a}mez}, {Garc{\'\i}a-Gonz{\'a}lez}, {Garfias}, {Gonz{\'a}lez}, {Goodman}, {Harding}, {Hernandez}, {Hona}, {Huang}, {Hueyotl-Zahuantitla}, {H{\"u}ntemeyer}, {Iriarte}, {Jardin-Blicq}, {Joshi}, {Kieda}, {Lara}, {Lee}, {Lee}, {Le{\'o}n Vargas}, {Linnemann}, {Longinotti}, {Luis-Raya}, {Lundeen}, {Malone}, {Martinez}, {Mart{\'\i}nez-Castro}, {Matthews}, {Miranda-Romagnoli}, {Morales-Soto}, {Moreno}, {Mostaf{\'a}}, {Nayerhoda}, {Nellen}, {Newbold}, {Nisa}, {Noriega-Papaqui}, {Olivera-Nieto}, {Omodei}, {Peisker}, {P{\'e}rez Araujo}, {Rho}, {Roh},
  {Rosa-Gonz{\'a}lez}, {Salesa Greus}, {Sandoval}, {Schneider}, {Serna-Franco}, {Smith}, {Springer}, {Tollefson}, {Torres}, {Torres-Escobedo}, {Turner}, {Ure{\~n}a-Mena}, {Villase{\~n}or}, {Watson}, {Weisgarber}, {Willox}, \& {Zhou}}]{HAWC2021}
{Albert}, A., {Alfaro}, R., {Alvarez}, C., {et~al.} 2021, \apjl, 912, L4, \dodoi{10.3847/2041-8213/abf35a}

\bibitem[{Alfaro {et~al.}(2024{\natexlab{a}})}]{HAWC:2023wdq}
Alfaro, R., {et~al.} 2024{\natexlab{a}}, Astrophys. J., 961, 104, \dodoi{10.3847/1538-4357/ad00b6}

\bibitem[{Alfaro {et~al.}(2024{\natexlab{b}})}]{Alfaro:2024cjd}
---. 2024{\natexlab{b}}, Nature, 634, 557, \dodoi{10.1038/s41586-024-07995-9}

\bibitem[{{Amenomori} {et~al.}(2008){Amenomori}, {Bi}, {Chen}, {Cui}, {Danzengluobu}, {Ding}, {Ding}, {Fan}, {Feng}, {Feng}, {Feng}, {Gao}, {Geng}, {Guo}, {He}, {He}, {Hibino}, {Hotta}, {Hu}, {Hu}, {Huang}, {Huang}, {Jia}, {Kajino}, {Kasahara}, {Katayose}, {Kato}, {Kawata}, {Labaciren}, {Le}, {Li}, {Li}, {Lou}, {Lu}, {Lu}, {Meng}, {Mizutani}, {Mu}, {Munakata}, {Nagai}, {Nanjo}, {Nishizawa}, {Ohnishi}, {Ohta}, {Onuma}, {Ouchi}, {Ozawa}, {Ren}, {Saito}, {Saito}, {Sakata}, {Sako}, {Shibata}, {Shiomi}, {Shirai}, {Sugimoto}, {Takita}, {Tan}, {Tateyama}, {Torii}, {Tsuchiya}, {Udo}, {Wang}, {Wang}, {Wang}, {Wang}, {Wang}, {Wu}, {Xue}, {Yamamoto}, {Yan}, {Yang}, {Yasue}, {Ye}, {Yu}, {Yuan}, {Yuda}, {Zhang}, {Zhang}, {Zhang}, {Zhang}, {Zhang}, {Zhang}, {Zhaxisangzhu}, {Zhou}, \& {Tibet AS{\ensuremath{\gamma}} Collaboration}}]{Amenomori+2008_TibetIII}
{Amenomori}, M., {Bi}, X.~J., {Chen}, D., {et~al.} 2008, \apj, 678, 1165, \dodoi{10.1086/529514}

\bibitem[{{Amenomori} {et~al.}(2021){Amenomori}, {Bao}, {Bi}, {Chen}, {Chen}, {Chen}, {Chen}, {Chen}, {Cirennima}, {Danzengluobu}, {Fang}, {Fang}, {Feng}, {Feng}, {Feng}, {Gao}, {Gou}, {Guo}, {Guo}, {He}, {He}, {Hibino}, {Hotta}, {Hu}, {Hu}, {Huang}, {Jia}, {Jiang}, {Jin}, {Kasahara}, {Katayose}, {Kato}, {Kato}, {Kawata}, {Kihara}, {Ko}, {Kozai}, {Labaciren}, {Li}, {Li}, {Li}, {Lin}, {Liu}, {Liu}, {Liu}, {Liu}, {Liu}, {Lou}, {Lu}, {Meng}, {Munakata}, {Nakada}, {Nakamura}, {Nanjo}, {Nishizawa}, {Ohnishi}, {Ohura}, {Ozawa}, {Qian}, {Qu}, {Saito}, {Sakata}, {Sako}, {Shao}, {Shibata}, {Shiomi}, {Sugimoto}, {Takano}, {Takita}, {Tan}, {Tateyama}, {Torii}, {Tsuchiya}, {Udo}, {Wang}, {Wu}, {Xue}, {Yamamoto}, {Yang}, {Yokoe}, {Yuan}, {Zhai}, {Zhang}, {Zhang}, {Zhang}, {Zhang}, {Zhang}, {Zhang}, {Zhang}, {Zhao}, {Zhaxisangzhu}, \& {Tibet AS<SUB>{\ensuremath{\gamma}}</SUB> Collaboration}}]{TibetAsg2021}
{Amenomori}, M., {Bao}, Y.~W., {Bi}, X.~J., {et~al.} 2021, \prl, 126, 141101, \dodoi{10.1103/PhysRevLett.126.141101}

\bibitem[{{Apel} {et~al.}(2013){Apel}, {Arteaga-Vel{\'a}zquez}, {Bekk}, {Bertaina}, {Bl{\"u}mer}, {Bozdog}, {Brancus}, {Cantoni}, {Chiavassa}, {Cossavella}, {Daumiller}, {de Souza}, {Di Pierro}, {Doll}, {Engel}, {Engler}, {Finger}, {Fuchs}, {Fuhrmann}, {Gils}, {Glasstetter}, {Grupen}, {Haungs}, {Heck}, {H{\"o}randel}, {Huber}, {Huege}, {Kampert}, {Kang}, {Klages}, {Link}, {{\L}uczak}, {Ludwig}, {Mathes}, {Mayer}, {Melissas}, {Milke}, {Mitrica}, {Morello}, {Oehlschl{\"a}ger}, {Ostapchenko}, {Palmieri}, {Petcu}, {Pierog}, {Rebel}, {Roth}, {Schieler}, {Schoo}, {Schr{\"o}der}, {Sima}, {Toma}, {Trinchero}, {Ulrich}, {Weindl}, {Wochele}, {Wommer}, \& {Zabierowski}}]{Apel+2013}
{Apel}, W.~D., {Arteaga-Vel{\'a}zquez}, J.~C., {Bekk}, K., {et~al.} 2013, Astroparticle Physics, 47, 54, \dodoi{10.1016/j.astropartphys.2013.06.004}

\bibitem[{{Barnier} \& {Done}(2024)}]{Barner24}
{Barnier}, S., \& {Done}, C. 2024, arXiv e-prints, arXiv:2404.12815, \dodoi{10.48550/arXiv.2404.12815}

\bibitem[{{Belloni}(2010)}]{Belloni2010}
{Belloni}, T.~M. 2010, in Lecture Notes in Physics, Berlin Springer Verlag, ed. T.~{Belloni}, Vol. 794, 53, \dodoi{10.1007/978-3-540-76937-8_3}

\bibitem[{{Bisnovatyi-Kogan} \& {Ruzmaikin}(1974)}]{Bisnovatyi-Kogan1974}
{Bisnovatyi-Kogan}, G.~S., \& {Ruzmaikin}, A.~A. 1974, \apss, 28, 45, \dodoi{10.1007/BF00642237}

\bibitem[{{Borse} {et~al.}(2021){Borse}, {Acharya}, {Vaidya}, {Mukherjee}, {Bodo}, {Rossi}, \& {Mignone}}]{Borse+2021}
{Borse}, N., {Acharya}, S., {Vaidya}, B., {et~al.} 2021, \aap, 649, A150, \dodoi{10.1051/0004-6361/202140440}

\bibitem[{Buxton {et~al.}(2012)Buxton, Bailyn, Capelo, Chatterjee, Din{\c c}er, Kalemci, \& Tomsick}]{Buxton+2012}
Buxton, M.~M., Bailyn, C.~D., Capelo, H.~L., {et~al.} 2012, The Astronomical Journal, 143, 130, \dodoi{10.1088/0004-6256/143/6/130}

\bibitem[{{Cao}(2011)}]{Cao2011}
{Cao}, X. 2011, \apj, 737, 94, \dodoi{10.1088/0004-637X/737/2/94}

\bibitem[{{Cao} {et~al.}(2019){Cao}, {della Volpe}, {Liu}, {Editors}, {:}, {Bi}, {Chen}, {D'Ettorre Piazzoli}, {Feng}, {Jia}, {Li}, {Ma}, {Wang}, {Zhang}, {Referees}, {:}, {Qie}, {Hu}, {Referees}, {:}, {S{\'a}iz}, {Yang}, {Contributors}, {:}, {Addazi}, {Belotsky}, {Beylin}, {Bi}, {Che}, {Chen}, {Cheng}, {Chiavassa}, {Cirelli}, {Di Sciascio}, {Esmaili}, {Fang}, {Fornengo}, {Gou}, {Guo}, {Gan}, {Gong}, {Gu}, {He}, {He}, {Hou}, {Huang}, {Huang}, {Kachekriess}, {Khlopov}, {Korchagin}, {Korochkin}, {Kuksa}, {Ksenofontov}, {Liu}, {Liu}, {Liu}, {Marciano}, {Martineau-Huynh}, {Martraire}, {Ma}, {Neronov}, {Panci}, {Pasechnick}, {Ruffolo}, {Sakharov}, {Sala}, {Semikoz}, {Shchegolev}, {Serpico}, {Sheng}, {Stenkin}, {Tam}, {Vernetto}, {Vallania}, {Volchanskiy}, {Wang}, {Wang}, {Wang}, {Wu}, {Wu}, {Wu}, {Xiao}, {Yang}, {Yan}, {Yao}, {Yin}, {Yuan}, {Zhang}, {Zeng}, {Zhang}, {Zhang}, {Zhou}, {Zhu}, \& {Zuo}}]{LHAASO2019}
{Cao}, Z., {della Volpe}, D., {Liu}, S., {et~al.} 2019, arXiv e-prints, arXiv:1905.02773, \dodoi{10.48550/arXiv.1905.02773}

\bibitem[{{Cao} {et~al.}(2023){Cao}, {Aharonian}, {An}, {Axikegu}, {Bao}, {Bastieri}, {Bi}, {Bi}, {Cai}, {Cao}, {Cao}, {Cao}, {Chang}, {Chang}, {Chen}, {Chen}, {Chen}, {Chen}, {Chen}, {Chen}, {Chen}, {Chen}, {Chen}, {Chen}, {Chen}, {Chen}, {Cheng}, {Cheng}, {Cui}, {Cui}, {Cui}, {Cui}, {Dai}, {Dai}, {Dai}, {Danzengluobu}, {Dong}, {Duan}, {Fan}, {Fan}, {Fang}, {Fang}, {Feng}, {Feng}, {Feng}, {Feng}, {Feng}, {Gabici}, {Gao}, {Gao}, {Gao}, {Gao}, {Gao}, {Gao}, {Ge}, {Geng}, {Giacinti}, {Gong}, {Gou}, {Gu}, {Guo}, {Guo}, {Guo}, {Guo}, {Han}, {He}, {He}, {He}, {He}, {He}, {Heller}, {Hor}, {Hou}, {Hou}, {Hou}, {Hu}, {Hu}, {Hu}, {Huang}, {Huang}, {Huang}, {Huang}, {Huang}, {Huang}, {Huang}, {Ji}, {Jia}, {Jia}, {Jiang}, {Jiang}, {Jiang}, {Jin}, {Kang}, {Ke}, {Kuleshov}, {Kurinov}, {Li}, {Li}, {Li}, {Li}, {Li}, {Li}, {Li}, {Li}, {Li}, {Li}, {Li}, {Li}, {Li}, {Li}, {Li}, {Li}, {Li}, {Li}, {Li}, {Liang}, {Liang}, {Lin}, {Liu}, {Liu}, {Liu}, {Liu}, {Liu}, {Liu}, {Liu}, {Liu}, {Liu}, {Liu}, {Liu}, {Liu}, {Liu}, {Liu},
  {Lu}, {Luo}, {Lv}, {Ma}, {Ma}, {Ma}, {Mao}, {Min}, {Mitthumsiri}, {Mu}, {Nan}, {Neronov}, {Ou}, {Pang}, {Pattarakijwanich}, {Pei}, {Qi}, {Qi}, {Qiao}, {Qin}, {Ruffolo}, {S{\'a}iz}, {Semikoz}, {Shao}, {Shao}, {Shchegolev}, {Sheng}, {Shu}, {Song}, {Stenkin}, {Stepanov}, {Su}, {Sun}, {Sun}, {Sun}, {Tam}, {Tang}, {Tang}, {Tian}, {Wang}, {Wang}, {Wang}, {Wang}, {Wang}, {Wang}, {Wang}, {Wang}, {Wang}, {Wang}, {Wang}, {Wang}, {Wang}, {Wang}, {Wang}, {Wang}, {Wang}, {Wang}, {Wang}, {Wang}, {Wang}, {Wei}, {Wei}, {Wei}, {Wen}, {Wu}, {Wu}, {Wu}, {Wu}, {Wu}, {Xi}, {Xia}, {Xia}, {Xiang}, {Xiao}, {Xiao}, {Xin}, {Xin}, {Xing}, {Xiong}, {Xu}, {Xu}, {Xu}, {Xu}, {Xue}, {Yan}, {Yan}, {Yan}, {Yang}, {Yang}, {Yang}, {Yang}, {Yang}, {Yang}, {Yang}, {Yang}, {Yang}, {Yao}, {Yao}, {Ye}, {Yin}, {Yin}, {You}, {You}, {Yu}, {Yuan}, {Yue}, {Zeng}, {Zeng}, {Zeng}, {Zha}, {Zhang}, {Zhang}, {Zhang}, {Zhang}, {Zhang}, {Zhang}, {Zhang}, {Zhang}, {Zhang}, {Zhang}, {Zhang}, {Zhang}, {Zhang}, {Zhang}, {Zhang}, {Zhang}, {Zhang}, {Zhang}, {Zhao},
  {Zhao}, {Zhao}, {Zhao}, {Zhao}, {Zheng}, {Zhou}, {Zhou}, {Zhou}, {Zhou}, {Zhou}, {Zhou}, {Zhou}, {Zhu}, {Zhu}, {Zhu}, {Zhu}, {Zuo}, \& {Lhaaso Collaboration}}]{CaoDiffuse2023}
{Cao}, Z., {Aharonian}, F., {An}, Q., {et~al.} 2023, \prl, 131, 151001, \dodoi{10.1103/PhysRevLett.131.151001}

\bibitem[{{Cao} {et~al.}(2024){Cao}, {Aharonian}, {Axikegu}, {Bai}, {Bao}, {Bastieri}, {Bi}, {Bi}, {Bian}, {Bukevich}, {Cao}, {Cao}, {Cao}, {Chang}, {Chang}, {Chen}, {Chen}, {Chen}, {Chen}, {Chen}, {Chen}, {Chen}, {Chen}, {Chen}, {Chen}, {Chen}, {Chen}, {Chen}, {Chen}, {Cheng}, {Cheng}, {Cui}, {Cui}, {Cui}, {Cui}, {Dai}, {Dai}, {Dai}, {Danzengluobu}, {Dong}, {Duan}, {Fan}, {Fan}, {Fang}, {Fang}, {Fang}, {Feng}, {Feng}, {Feng}, {Feng}, {Feng}, {Feng}, {Feng}, {Gabici}, {Gao}, {Gao}, {Gao}, {Gao}, {Gao}, {Ge}, {Geng}, {Giacinti}, {Gong}, {Gou}, {Gu}, {Guo}, {Guo}, {Guo}, {Guo}, {Han}, {Hasan}, {He}, {He}, {He}, {He}, {Hor}, {Hou}, {Hou}, {Hou}, {Hu}, {Hu}, {Hu}, {Huang}, {Huang}, {Huang}, {Huang}, {Huang}, {Huang}, {Ji}, {Jia}, {Jia}, {Jiang}, {Jiang}, {Jiang}, {Jin}, {Kang}, {Karpikov}, {Kuleshov}, {Kurinov}, {Li}, {Li}, {Li}, {Li}, {Li}, {Li}, {Li}, {Li}, {Li}, {Li}, {Li}, {Li}, {Li}, {Li}, {Li}, {Li}, {Li}, {Li}, {Li}, {Liang}, {Liang}, {Lin}, {Liu}, {Liu}, {Liu}, {Liu}, {Liu}, {Liu}, {Liu}, {Liu}, {Liu},
  {Liu}, {Liu}, {Liu}, {Liu}, {Liu}, {Luo}, {Luo}, {Lv}, {Ma}, {Ma}, {Ma}, {Mao}, {Min}, {Mitthumsiri}, {Mu}, {Nan}, {Neronov}, {Ou}, {Pattarakijwanich}, {Pei}, {Qi}, {Qi}, {Qiao}, {Qin}, {Raza}, {Ruffolo}, {S{\'a}iz}, {Saeed}, {Semikoz}, {Shao}, {Shchegolev}, {Sheng}, {Shu}, {Song}, {Stenkin}, {Stepanov}, {Su}, {Sun}, {Sun}, {Sun}, {Sun}, {Takata}, {Tam}, {Tang}, {Tang}, {Tang}, {Tian}, {Wang}, {Wang}, {Wang}, {Wang}, {Wang}, {Wang}, {Wang}, {Wang}, {Wang}, {Wang}, {Wang}, {Wang}, {Wang}, {Wang}, {Wang}, {Wang}, {Wang}, {Wang}, {Wang}, {Wang}, {Wang}, {Wang}, {Wei}, {Wei}, {Wei}, {Wen}, {Wu}, {Wu}, {Wu}, {Wu}, {Wu}, {Wu}, {Xi}, {Xia}, {Xiang}, {Xiao}, {Xiao}, {Xin}, {Xing}, {Xiong}, {Xiong}, {Xu}, {Xu}, {Xu}, {Xu}, {Xue}, {Yan}, {Yan}, {Yan}, {Yang}, {Yang}, {Yang}, {Yang}, {Yang}, {Yang}, {Yang}, {Yang}, {Yao}, {Yao}, {Yin}, {Yin}, {You}, {You}, {Yu}, {Yuan}, {Yue}, {Zeng}, {Zeng}, {Zeng}, {Zha}, {Zhang}, {Zhang}, {Zhang}, {Zhang}, {Zhang}, {Zhang}, {Zhang}, {Zhang}, {Zhang}, {Zhang}, {Zhang}, {Zhang},
  {Zhang}, {Zhang}, {Zhang}, {Zhang}, {Zhang}, {Zhang}, {Zhao}, {Zhao}, {Zhao}, {Zhao}, {Zhao}, {Zhao}, {Zheng}, {Zhong}, {Zhou}, {Zhou}, {Zhou}, {Zhou}, {Zhou}, {Zhou}, {Zhou}, {Zhou}, {Zhu}, {Zhu}, {Zhu}, {Zhu}, {Zhu}, {Zou}, {Zuo}, \& {Lhaaso Collaboration}}]{LHAASO_CRspe_2024}
{Cao}, Z., {Aharonian}, F., {Axikegu}, {et~al.} 2024, \prl, 132, 131002, \dodoi{10.1103/PhysRevLett.132.131002}

\bibitem[{{Chatterjee} {et~al.}(2019){Chatterjee}, {Debnath}, {Jana}, \& {Chakrabarti}}]{Chatterjee+2019}
{Chatterjee}, D., {Debnath}, D., {Jana}, A., \& {Chakrabarti}, S.~K. 2019, \apss, 364, 14, \dodoi{10.1007/s10509-019-3495-2}

\bibitem[{{Chaty} {et~al.}(2003){Chaty}, {Haswell}, {Malzac}, {Hynes}, {Shrader}, \& {Cui}}]{Chaty+2003}
{Chaty}, S., {Haswell}, C.~A., {Malzac}, J., {et~al.} 2003, \mnras, 346, 689, \dodoi{10.1111/j.1365-2966.2003.07115.x}

\bibitem[{{Chen} {et~al.}(2023){Chen}, {Uzdensky}, \& {Dexter}}]{Chen+2022}
{Chen}, A.~Y., {Uzdensky}, D., \& {Dexter}, J. 2023, \apj, 944, 173, \dodoi{10.3847/1538-4357/acb68a}

\bibitem[{{Cooper} {et~al.}(2020){Cooper}, {Gaggero}, {Markoff}, \& {Zhang}}]{Cooper+2020}
{Cooper}, A.~J., {Gaggero}, D., {Markoff}, S., \& {Zhang}, S. 2020, \mnras, 493, 3212, \dodoi{10.1093/mnras/staa373}

\bibitem[{{Coppi} \& {Blandford}(1990)}]{Coppi1990}
{Coppi}, P.~S., \& {Blandford}, R.~D. 1990, \mnras, 245, 453

\bibitem[{{Corral-Santana} {et~al.}(2016){Corral-Santana}, {Casares}, {Mu{\~n}oz-Darias}, {Bauer}, {Mart{\'\i}nez-Pais}, \& {Russell}}]{Corral+2016}
{Corral-Santana}, J.~M., {Casares}, J., {Mu{\~n}oz-Darias}, T., {et~al.} 2016, \aap, 587, A61, \dodoi{10.1051/0004-6361/201527130}

\bibitem[{Dekker {et~al.}(2024)Dekker, Holst, Hooper, Leone, Simon, \& Xiao}]{Dekker:2023six}
Dekker, A., Holst, I., Hooper, D., {et~al.} 2024, Phys. Rev. D, 109, 083026, \dodoi{10.1103/PhysRevD.109.083026}

\bibitem[{{Done} {et~al.}(2007){Done}, {Gierli{\'n}ski}, \& {Kubota}}]{Done+2007}
{Done}, C., {Gierli{\'n}ski}, M., \& {Kubota}, A. 2007, \aapr, 15, 1, \dodoi{10.1007/s00159-007-0006-1}

\bibitem[{{Esin} {et~al.}(1997){Esin}, {McClintock}, \& {Narayan}}]{Esin+1997}
{Esin}, A.~A., {McClintock}, J.~E., \& {Narayan}, R. 1997, \apj, 489, 865, \dodoi{10.1086/304829}

\bibitem[{Fang {et~al.}(2024{\natexlab{a}})Fang, Gallagher, \& Halzen}]{Fang:2023azx}
Fang, K., Gallagher, J.~S., \& Halzen, F. 2024{\natexlab{a}}, Nature Astron., 8, 241, \dodoi{10.1038/s41550-023-02128-0}

\bibitem[{Fang {et~al.}(2024{\natexlab{b}})Fang, Halzen, Heinz, \& Gallagher}]{Fang:2024wmf}
Fang, K., Halzen, F., Heinz, S., \& Gallagher, J.~S. 2024{\natexlab{b}}, Astrophys. J. Lett., 975, L35, \dodoi{10.3847/2041-8213/ad887b}

\bibitem[{Fang {et~al.}(2022)Fang, Kerr, Blandford, Fleischhack, \& Charles}]{Fang:2022uge}
Fang, K., Kerr, M., Blandford, R., Fleischhack, H., \& Charles, E. 2022, Phys. Rev. Lett., 129, 071101, \dodoi{10.1103/PhysRevLett.129.071101}

\bibitem[{{Fang} \& {Murase}(2023)}]{Fang&Muraase2023}
{Fang}, K., \& {Murase}, K. 2023, \apjl, 957, L6, \dodoi{10.3847/2041-8213/ad012f}

\bibitem[{{Fernandez-Barral} {et~al.}(2017){Fernandez-Barral}, {Blanch}, {de O{\~n}a Wilhemi}, {Galindo}, {Herrera}, {Rib{\'o}}, {Rico}, {Stamerra}, {Aharonian}, {MAGIC Collaboration}, {Bosch-Ramon}, \& {Zanin}}]{MAGIC2017}
{Fernandez-Barral}, A., {Blanch}, O., {de O{\~n}a Wilhemi}, E., {et~al.} 2017, in International Cosmic Ray Conference, Vol. 301, 35th International Cosmic Ray Conference (ICRC2017), 734, \dodoi{10.22323/1.301.0734}

\bibitem[{{Fujita} {et~al.}(2017){Fujita}, {Murase}, \& {Kimura}}]{FujitaMuraseKimura2017}
{Fujita}, Y., {Murase}, K., \& {Kimura}, S.~S. 2017, \jcap, 2017, 037, \dodoi{10.1088/1475-7516/2017/04/037}

\bibitem[{{Galishnikova} {et~al.}(2023){Galishnikova}, {Philippov}, {Quataert}, {Bacchini}, {Parfrey}, \& {Ripperda}}]{Galishnikova+2023}
{Galishnikova}, A., {Philippov}, A., {Quataert}, E., {et~al.} 2023, \prl, 130, 115201, \dodoi{10.1103/PhysRevLett.130.115201}

\bibitem[{{Gandhi} {et~al.}(2010){Gandhi}, {Dhillon}, {Durant}, {Fabian}, {Kubota}, {Makishima}, {Malzac}, {Marsh}, {Miller}, {Shahbaz}, {Spruit}, \& {Casella}}]{Gandhi+2010}
{Gandhi}, P., {Dhillon}, V.~S., {Durant}, M., {et~al.} 2010, \mnras, 407, 2166, \dodoi{10.1111/j.1365-2966.2010.17083.x}

\bibitem[{{Gonz{\'a}lez Hern{\'a}ndez} {et~al.}(2012){Gonz{\'a}lez Hern{\'a}ndez}, {Rebolo}, \& {Casares}}]{Hernandez+2012}
{Gonz{\'a}lez Hern{\'a}ndez}, J.~I., {Rebolo}, R., \& {Casares}, J. 2012, \apjl, 744, L25, \dodoi{10.1088/2041-8205/744/2/L25}

\bibitem[{{Hakobyan} {et~al.}(2023){Hakobyan}, {Ripperda}, \& {Philippov}}]{Hakobyan2023}
{Hakobyan}, H., {Ripperda}, B., \& {Philippov}, A.~A. 2023, \apjl, 943, L29, \dodoi{10.3847/2041-8213/acb264}

\bibitem[{{Hayashi} {et~al.}(2024){Hayashi}, {Kiuchi}, {Kyutoku}, {Sekiguchi}, \& {Shibata}}]{Hayashi2024}
{Hayashi}, K., {Kiuchi}, K., {Kyutoku}, K., {Sekiguchi}, Y., \& {Shibata}, M. 2024, arXiv e-prints, arXiv:2410.10958, \dodoi{10.48550/arXiv.2410.10958}

\bibitem[{{Heida} {et~al.}(2017){Heida}, {Jonker}, {Torres}, \& {Chiavassa}}]{Heida2017}
{Heida}, M., {Jonker}, P.~G., {Torres}, M.~A.~P., \& {Chiavassa}, A. 2017, \apj, 846, 132, \dodoi{10.3847/1538-4357/aa85df}

\bibitem[{{Heinz} \& {Grimm}(2005)}]{Heinz&Grimm2005}
{Heinz}, S., \& {Grimm}, H.~J. 2005, \apj, 633, 384, \dodoi{10.1086/452624}

\bibitem[{{Helder} {et~al.}(2012){Helder}, {Vink}, {Bykov}, {Ohira}, {Raymond}, \& {Terrier}}]{Helder+2012}
{Helder}, E.~A., {Vink}, J., {Bykov}, A.~M., {et~al.} 2012, \ssr, 173, 369, \dodoi{10.1007/s11214-012-9919-8}

\bibitem[{{Hoshino}(2012)}]{HoshinoPhyLV2012}
{Hoshino}, M. 2012, \prl, 108, 135003, \dodoi{10.1103/PhysRevLett.108.135003}

\bibitem[{Huang {et~al.}(2023)Huang, Cao, Chen, Liu, Wang, You, \& Qi}]{HUNT2023}
Huang, T.-Q., Cao, Z., Chen, M., {et~al.} 2023, PoS, ICRC2023, 1080, \dodoi{10.22323/1.444.1080}

\bibitem[{{Huentemeyer} {et~al.}(2019){Huentemeyer}, {BenZvi}, {Dingus}, {Fleischhack}, {Schoorlemmer}, \& {Weisgarber}}]{SWGO2019}
{Huentemeyer}, P., {BenZvi}, S., {Dingus}, B., {et~al.} 2019, in Bulletin of the American Astronomical Society, Vol.~51, 109, \dodoi{10.48550/arXiv.1907.07737}

\bibitem[{{Icecube Collaboration} {et~al.}(2023){Icecube Collaboration}, {Abbasi}, {Ackermann}, {Adams}, {Aguilar}, {Ahlers}, {Ahrens}, {Alameddine}, {Alves}, {Amin}, {Andeen}, {Anderson}, {Anton}, {Arguelles}, {Ashida}, {Athanasiadou}, {Axani}, {Bai}, {Balagopal}, {Barwick}, {Basu}, {Baur}, {Bay}, {Beatty}, {Becker}, {Becker Tjus}, {Beise}, {Bellenghi}, {Benda}, {Benzvi}, {Berley}, {Bernardini}, {Besson}, {Binder}, {Bindig}, {Blaufuss}, {Blot}, {Boddenberg}, {Bontempo}, {Book}, {Borowka}, {Boser}, {Botner}, {Bottcher}, {Bourbeau}, {Bradascio}, {Braun}, {Brinson}, {Bron}, {Brostean-Kaiser}, {Burley}, {Busse}, {Campana}, {Carnie-Bronca}, {Chen}, {Chen}, {Chirkin}, {Choi}, {Clark}, {Clark}, {Classen}, {Coleman}, {Collin}, {Connolly}, {Conrad}, {Coppin}, {Correa}, {Cowen}, {Cross}, {Dappen}, {Dave}, {de Clercq}, {Delaunay}, {Delgado Lopez}, {Dembinski}, {Deoskar}, {Desai}, {Desiati}, {de Vries}, {de Wasseige}, {Deyoung}, {Diaz}, {Diaz-Velez}, {Dittmer}, {Dujmovic}, {Dunkman}, {Duvernois}, {Ehrhardt}, {Eller},
  {Engel}, {Erpenbeck}, {Evans}, {Evenson}, {Fan}, {Fazely}, {Fedynitch}, {Feigl}, {Fiedlschuster}, {Fienberg}, {Finley}, {Fischer}, {Fox}, {Franckowiak}, {Friedman}, {Fritz}, {Furst}, {Gaisser}, {Gallagher}, {Ganster}, {Garcia}, {Garrappa}, {Gerhardt}, {Ghadimi}, {Glaser}, {Glauch}, {Glusenkamp}, {Goehlke}, {Goldschmidt}, {Gonzalez}, {Goswami}, {Grant}, {Gregoire}, {Griswold}, {Gunther}, {Gutjahr}, {Haack}, {Hallgren}, {Halliday}, {Halve}, {Halzen}, {Ha}, {Hanson}, {Hardin}, {Harnisch}, {Haungs}, {Helbing}, {Henningsen}, {Hettinger}, {Hickford}, {Hignight}, {Hill}, {Hill}, {Hoffman}, {Hoshina}, {Hou}, {Huang}, {Huber}, {Huber}, {Hultqvist}, {Hunnefeld}, {Hussain}, {Hymon}, {in}, {Iovine}, {Ishihara}, {Jansson}, {Japaridze}, {Jeong}, {Jin}, {Jones}, {Kang}, {Kang}, {Kang}, {Kappes}, {Kappesser}, {Kardum}, {Karg}, {Karl}, {Karle}, {Katz}, {Kauer}, {Kellermann}, {Kelley}, {Kheirandish}, {Kin}, {Kiryluk}, {Klein}, {Kochocki}, {Koirala}, {Kolanoski}, {Kontrimas}, {Kopke}, {Kopper}, {Kopper}, {Koskinen},
  {Koundal}, {Kovacevich}, {Kowalski}, {Kozynets}, {Krupczak}, {Kun}, {Kurahashi}, {Lad}, {Lagunas Gualda}, {Lanfranchi}, {Larson}, {Lauber}, {Lazar}, {Lee}, {Leonard}, {Leszczynska}, {Li}, {Lincetto}, {Liu}, {Liubarska}, {Lohfink}, {Lozano Mariscal}, {Lu}, {Lucarelli}, {Ludwig}, {Luszczak}, {Lyu}, {Ma}, {Madsen}, {Mahn}, {Makino}, {Mancina}, {Maris}, {Martinez-Soler}, {Maruyama}, {McCarthy}, {McElroy}, {McNally}, {Mead}, {Meagher}, {Mechbal}, {Medina}, {Meier}, {Meighen-Berger}, {Merckx}, {Micallef}, {Mockler}, {Montaruli}, {Moore}, {Morik}, {Morse}, {Moulai}, {Mukherjee}, {Naab}, {Nagai}, {Nahnhauer}, {Naumann}, {Necker}, {Nguyen}, {Niederhausen}, {Nisa}, {Nowicki}, {Nygren}, {Obertacke Pollmann}, {Oehler}, {Oeyen}, {Olivas}, {O'Sullivan}, {Pandya}, {Pankova}, {Park}, {Parker}, {Paudel}, {Paul}, {Perez de Los Heros}, {Peters}, {Peterson}, {Philippen}, {Pieper}, {Pizzuto}, {Plum}, {Popovych}, {Porcelli}, {Rodriguez}, {Pries}, {Przybylski}, {Raab}, {Rack-Helleis}, {Raissi}, {Rameez}, {Rawlins}, {Rea},
  {Rechav}, {Rehman}, {Reichherzer}, {Reimann}, {Renzi}, {Resconi}, {Reusch}, {Rhode}, {Richman}, {Riedel}, {Roberts}, {Robertson}, {Roellinghoff}, {Rongen}, {Rott}, {Ruhe}, {Ryckbosch}, {Rysewyk Cantu}, {Safa}, {Saffer}, {Salazar-Gallegos}, {Sampathkumar}, {Herrera}, {Sandrock}, {Santander}, {Sarkar}, {Sarkar}, {Satalecka}, {Schaufel}, {Schieler}, {Schindler}, {Schmidt}, {Schneider}, {Schneider}, {Schroder}, {Schumacher}, {Schwefer}, {Sclafani}, {Seckel}, {Seunarine}, {Sharma}, {Shefali}, {Shimizu}, {Silva}, {Skrzypek}, {Smithers}, {Snihur}, {Soedingrekso}, {Sogaard}, {Soldin}, {Spannfellner}, {Spiczak}, {Spiering}, {Stamatikos}, {Stanev}, {Stein}, {Stettner}, {Stezelberger}, {Stokstad}, {Sturwald}, {Stuttard}, {Sullivan}, {Taboada}, {Ter-Antonyan}, {Thwaites}, {Tilav}, {Tischbein}, {Tollefson}, {Tonnis}, {Toscano}, {Tosi}, {Trettin}, {Tselengidou}, {Tung}, {Turcati}, {Turcotte}, {Turley}, {Twagirayezu}, {Ty}, {Elorrieta}, {Valtonen-Mattila}, {Vandenbroucke}, {van Eijndhoven}, {Vannerom}, {van Santen},
  {Veitch-Michaelis}, {Verpoest}, {Walck}, {Wang}, {Watson}, {Weaver}, {Weigel}, {Weindl}, {Weiss}, {Weldert}, {Wendt}, {Werthebach}, {Weyrauch}, {Whitehorn}, {Wiebusch}, {Willey}, {Williams}, {Wolf}, {Wrede}, {Wulff}, {Xu}, {Yanez}, {Yildizci}, {Yoshida}, {Yu}, {Yuan}, {Zhang}, \& {Zhelnin}}]{IceCubeDiffuse2023}
{Icecube Collaboration}, {Abbasi}, R., {Ackermann}, M., {et~al.} 2023, Science, 380, 1338, \dodoi{10.1126/science.adc9818}

\bibitem[{{Kantzas} {et~al.}(2023){Kantzas}, {Markoff}, {Cooper}, {Gaggero}, {Petropoulou}, \& {De La Torre Luque}}]{Kantzas+2023}
{Kantzas}, D., {Markoff}, S., {Cooper}, A.~J., {et~al.} 2023, \mnras, 524, 1326, \dodoi{10.1093/mnras/stad1909}

\bibitem[{{Kantzas} {et~al.}(2021){Kantzas}, {Markoff}, {Beuchert}, {Lucchini}, {Chhotray}, {Ceccobello}, {Tetarenko}, {Miller-Jones}, {Bremer}, {Garcia}, {Grinberg}, {Uttley}, \& {Wilms}}]{Kantzas+2021}
{Kantzas}, D., {Markoff}, S., {Beuchert}, T., {et~al.} 2021, \mnras, 500, 2112, \dodoi{10.1093/mnras/staa3349}

\bibitem[{{Kawashima} \& {The ALPACA Collaboration}(2022)}]{ALPACA2022}
{Kawashima}, T., \& {The ALPACA Collaboration}. 2022, arXiv e-prints, arXiv:2208.14659, \dodoi{10.48550/arXiv.2208.14659}

\bibitem[{{Kawazura} {et~al.}(2020){Kawazura}, {Schekochihin}, {Barnes}, {TenBarge}, {Tong}, {Klein}, \& {Dorland}}]{Kawazura2020}
{Kawazura}, Y., {Schekochihin}, A.~A., {Barnes}, M., {et~al.} 2020, Physical Review X, 10, 041050, \dodoi{10.1103/PhysRevX.10.041050}

\bibitem[{{Kimura} {et~al.}(2018){Kimura}, {Murase}, \& {M{\'e}sz{\'a}ros}}]{KimuraMurase2018}
{Kimura}, S.~S., {Murase}, K., \& {M{\'e}sz{\'a}ros}, P. 2018, \apj, 866, 51, \dodoi{10.3847/1538-4357/aadc0a}

\bibitem[{{Kimura} {et~al.}(2019{\natexlab{a}}){Kimura}, {Murase}, \& {M{\'e}sz{\'a}ros}}]{KimuraMurase2019}
---. 2019{\natexlab{a}}, \prd, 100, 083014, \dodoi{10.1103/PhysRevD.100.083014}

\bibitem[{{Kimura} {et~al.}(2015){Kimura}, {Murase}, \& {Toma}}]{Kimura2015}
{Kimura}, S.~S., {Murase}, K., \& {Toma}, K. 2015, \apj, 806, 159, \dodoi{10.1088/0004-637X/806/2/159}

\bibitem[{{Kimura} {et~al.}(2021){Kimura}, {Sudoh}, {Kashiyama}, \& {Kawanaka}}]{KimuraSudoh2021}
{Kimura}, S.~S., {Sudoh}, T., {Kashiyama}, K., \& {Kawanaka}, N. 2021, \apj, 915, 31, \dodoi{10.3847/1538-4357/abff58}

\bibitem[{{Kimura} \& {Toma}(2020)}]{Kimura2020}
{Kimura}, S.~S., \& {Toma}, K. 2020, \apj, 905, 178, \dodoi{10.3847/1538-4357/abc343}

\bibitem[{{Kimura} {et~al.}(2022){Kimura}, {Toma}, {Noda}, \& {Hada}}]{Kimura+2022}
{Kimura}, S.~S., {Toma}, K., {Noda}, H., \& {Hada}, K. 2022, \apjl, 937, L34, \dodoi{10.3847/2041-8213/ac8d5a}

\bibitem[{{Kimura} {et~al.}(2016){Kimura}, {Toma}, {Suzuki}, \& {Inutsuka}}]{Kimura2016}
{Kimura}, S.~S., {Toma}, K., {Suzuki}, T.~K., \& {Inutsuka}, S.-i. 2016, \apj, 822, 88, \dodoi{10.3847/0004-637X/822/2/88}

\bibitem[{{Kimura} {et~al.}(2019{\natexlab{b}}){Kimura}, {Tomida}, \& {Murase}}]{KimuraTomida2019}
{Kimura}, S.~S., {Tomida}, K., \& {Murase}, K. 2019{\natexlab{b}}, \mnras, 485, 163, \dodoi{10.1093/mnras/stz329}

\bibitem[{{Koljonen} {et~al.}(2023){Koljonen}, {Long}, {Matthews}, \& {Knigge}}]{Koljonen+2023}
{Koljonen}, K.~I.~I., {Long}, K.~S., {Matthews}, J.~H., \& {Knigge}, C. 2023, \mnras, 521, 4190, \dodoi{10.1093/mnras/stad809}

\bibitem[{{Kosenkov} {et~al.}(2020){Kosenkov}, {Veledina}, {Suleimanov}, \& {Poutanen}}]{Kosenkov+2020}
{Kosenkov}, I.~A., {Veledina}, A., {Suleimanov}, V.~F., \& {Poutanen}, J. 2020, \aap, 638, A127, \dodoi{10.1051/0004-6361/201936143}

\bibitem[{{Krawczynski} {et~al.}(2022){Krawczynski}, {Muleri}, {Dov{\v{c}}iak}, {Veledina}, {Rodriguez Cavero}, {Svoboda}, {Ingram}, {Matt}, {Garcia}, {Loktev}, {Negro}, {Poutanen}, {Kitaguchi}, {Podgorn{\'y}}, {Rankin}, {Zhang}, {Berdyugin}, {Berdyugina}, {Bianchi}, {Blinov}, {Capitanio}, {Di Lalla}, {Draghis}, {Fabiani}, {Kagitani}, {Kravtsov}, {Kiehlmann}, {Latronico}, {Lutovinov}, {Mandarakas}, {Marin}, {Marinucci}, {Miller}, {Mizuno}, {Molkov}, {Omodei}, {Petrucci}, {Ratheesh}, {Sakanoi}, {Semena}, {Skalidis}, {Soffitta}, {Tennant}, {Thalhammer}, {Tombesi}, {Weisskopf}, {Wilms}, {Zhang}, {Agudo}, {Antonelli}, {Bachetti}, {Baldini}, {Baumgartner}, {Bellazzini}, {Bongiorno}, {Bonino}, {Brez}, {Bucciantini}, {Castellano}, {Cavazzuti}, {Ciprini}, {Costa}, {De Rosa}, {Del Monte}, {Di Gesu}, {Di Marco}, {Donnarumma}, {Doroshenko}, {Ehlert}, {Enoto}, {Evangelista}, {Ferrazzoli}, {Gunji}, {Hayashida}, {Heyl}, {Iwakiri}, {Jorstad}, {Karas}, {Kolodziejczak}, {La Monaca}, {Liodakis}, {Maldera}, {Manfreda},
  {Marscher}, {Marshall}, {Mitsuishi}, {Ng}, {O{\textquoteright}Dell}, {Oppedisano}, {Papitto}, {Pavlov}, {Peirson}, {Perri}, {Pesce-Rollins}, {Pilia}, {Possenti}, {Puccetti}, {Ramsey}, {Romani}, {Sgr{\`o}}, {Slane}, {Spandre}, {Tamagawa}, {Tavecchio}, {Taverna}, {Tawara}, {Thomas}, {Trois}, {Tsygankov}, {Turolla}, {Vink}, {Wu}, {Xie}, \& {Zane}}]{IXPE2022}
{Krawczynski}, H., {Muleri}, F., {Dov{\v{c}}iak}, M., {et~al.} 2022, Science, 378, 650, \dodoi{10.1126/science.add5399}

\bibitem[{{Kuze} {et~al.}(2022){Kuze}, {Kimura}, \& {Toma}}]{Kuze+2022}
{Kuze}, R., {Kimura}, S.~S., \& {Toma}, K. 2022, \apj, 935, 159, \dodoi{10.3847/1538-4357/ac7ec1}

\bibitem[{{Kuze} {et~al.}(2024){Kuze}, {Kimura}, \& {Toma}}]{RK+2024}
---. 2024, \apj, 977, 22, \dodoi{10.3847/1538-4357/ad88f4}

\bibitem[{{LHAASO Collaboration}(2024)}]{LHAASO2024_mq}
{LHAASO Collaboration}. 2024, arXiv e-prints, arXiv:2410.08988, \dodoi{10.48550/arXiv.2410.08988}

\bibitem[{Lipari \& Vernetto(2024)}]{Lipari:2024pzo}
Lipari, P., \& Vernetto, S. 2024.
\newblock \doarXiv{2412.08861}

\bibitem[{{Liska} {et~al.}(2022){Liska}, {Musoke}, {Tchekhovskoy}, {Porth}, \& {Beloborodov}}]{Liska+2022}
{Liska}, M.~T.~P., {Musoke}, G., {Tchekhovskoy}, A., {Porth}, O., \& {Beloborodov}, A.~M. 2022, \apjl, 935, L1, \dodoi{10.3847/2041-8213/ac84db}

\bibitem[{{Liu} {et~al.}(2007){Liu}, {van Paradijs}, \& {van den Heuvel}}]{Liu+2007_GRO}
{Liu}, Q.~Z., {van Paradijs}, J., \& {van den Heuvel}, E.~P.~J. 2007, \aap, 469, 807, \dodoi{10.1051/0004-6361:20077303}

\bibitem[{{Lynn} {et~al.}(2014){Lynn}, {Quataert}, {Chandran}, \& {Parrish}}]{Lynn+2014}
{Lynn}, J.~W., {Quataert}, E., {Chandran}, B. D.~G., \& {Parrish}, I.~J. 2014, \apj, 791, 71, \dodoi{10.1088/0004-637X/791/1/71}

\bibitem[{{Makishima} {et~al.}(2008){Makishima}, {Takahashi}, {Yamada}, {Done}, {Kubota}, {Dotani}, {Ebisawa}, {Itoh}, {Kitamoto}, {Negoro}, {Ueda}, \& {Yamaoka}}]{Suzaku2008}
{Makishima}, K., {Takahashi}, H., {Yamada}, S., {et~al.} 2008, \pasj, 60, 585, \dodoi{10.1093/pasj/60.3.585}

\bibitem[{{Malzac} {et~al.}(2018){Malzac}, {Kalamkar}, {Vincentelli}, {Vue}, {Drappeau}, {Belmont}, {Casella}, {Clavel}, {Corbel}, {Coriat}, {Dornic}, {Ferreira}, {Henri}, {Maccarone}, {Marcowith}, {O'Brien}, {P{\'e}ault}, {Petrucci}, {Rodriguez}, {Russell}, \& {Uttley}}]{Malzac+2018}
{Malzac}, J., {Kalamkar}, M., {Vincentelli}, F., {et~al.} 2018, \mnras, 480, 2054, \dodoi{10.1093/mnras/sty2006}

\bibitem[{{Markoff} {et~al.}(2001){Markoff}, {Falcke}, \& {Fender}}]{Markoff+2001}
{Markoff}, S., {Falcke}, H., \& {Fender}, R. 2001, \aap, 372, L25, \dodoi{10.1051/0004-6361:20010420}

\bibitem[{{Marshall} {et~al.}(2018){Marshall}, {Avara}, \& {McKinney}}]{Marshall+2018}
{Marshall}, M.~D., {Avara}, M.~J., \& {McKinney}, J.~C. 2018, \mnras, 478, 1837, \dodoi{10.1093/mnras/sty1184}

\bibitem[{{McClintock} {et~al.}(2001){McClintock}, {Garcia}, {Caldwell}, {Falco}, {Garnavich}, \& {Zhao}}]{McClintock+2001}
{McClintock}, J.~E., {Garcia}, M.~R., {Caldwell}, N., {et~al.} 2001, \apjl, 551, L147, \dodoi{10.1086/320030}

\bibitem[{{McConnell} {et~al.}(2002){McConnell}, {Zdziarski}, {Bennett}, {Bloemen}, {Collmar}, {Hermsen}, {Kuiper}, {Paciesas}, {Phlips}, {Poutanen}, {Ryan}, {Sch{\"o}nfelder}, {Steinle}, \& {Strong}}]{McConnell+2002}
{McConnell}, M.~L., {Zdziarski}, A.~A., {Bennett}, K., {et~al.} 2002, \apj, 572, 984, \dodoi{10.1086/340436}

\bibitem[{{McKinney} {et~al.}(2012){McKinney}, {Tchekhovskoy}, \& {Blandford}}]{Mckinney2012}
{McKinney}, J.~C., {Tchekhovskoy}, A., \& {Blandford}, R.~D. 2012, \mnras, 423, 3083, \dodoi{10.1111/j.1365-2966.2012.21074.x}

\bibitem[{{Migliari} {et~al.}(2007){Migliari}, {Tomsick}, {Markoff}, {Kalemci}, {Bailyn}, {Buxton}, {Corbel}, {Fender}, \& {Kaaret}}]{Migliari+2007}
{Migliari}, S., {Tomsick}, J.~A., {Markoff}, S., {et~al.} 2007, \apj, 670, 610, \dodoi{10.1086/522023}

\bibitem[{{Miller-Jones} {et~al.}(2021){Miller-Jones}, {Bahramian}, {Orosz}, {Mandel}, {Gou}, {Maccarone}, {Neijssel}, {Zhao}, {Zi{\'o}{\l}kowski}, {Reid}, {Uttley}, {Zheng}, {Byun}, {Dodson}, {Grinberg}, {Jung}, {Kim}, {Marcote}, {Markoff}, {Rioja}, {Rushton}, {Russell}, {Sivakoff}, {Tetarenko}, {Tudose}, \& {Wilms}}]{MillerJones+2021}
{Miller-Jones}, J. C.~A., {Bahramian}, A., {Orosz}, J.~A., {et~al.} 2021, Science, 371, 1046, \dodoi{10.1126/science.abb3363}

\bibitem[{{Moscibrodzka}(2024)}]{Moscibrodzka2024}
{Moscibrodzka}, M. 2024, \apss, 369, 68, \dodoi{10.1007/s10509-024-04333-3}

\bibitem[{{Motta} {et~al.}(2021){Motta}, {Rodriguez}, {Jourdain}, {Del Santo}, {Belanger}, {Cangemi}, {Grinberg}, {Kajava}, {Kuulkers}, {Malzac}, {Pottschmidt}, {Roques}, {S{\'a}nchez-Fern{\'a}ndez}, \& {Wilms}}]{Motta+21}
{Motta}, S.~E., {Rodriguez}, J., {Jourdain}, E., {et~al.} 2021, \nar, 93, 101618, \dodoi{10.1016/j.newar.2021.101618}

\bibitem[{Murase \& Fukugita(2019)}]{MuraseFukugita2018}
Murase, K., \& Fukugita, M. 2019, Phys. Rev. D, 99, 063012, \dodoi{10.1103/PhysRevD.99.063012}

\bibitem[{{Nakanishi} \& {Sofue}(2016)}]{Nakanishi&Sofue2016}
{Nakanishi}, H., \& {Sofue}, Y. 2016, \pasj, 68, 5, \dodoi{10.1093/pasj/psv108}

\bibitem[{{Narayan} {et~al.}(2012){Narayan}, {S{\"A} dowski}, {Penna}, \& {Kulkarni}}]{Narayan2012}
{Narayan}, R., {S{\"A} dowski}, A., {Penna}, R.~F., \& {Kulkarni}, A.~K. 2012, \mnras, 426, 3241, \dodoi{10.1111/j.1365-2966.2012.22002.x}

\bibitem[{{Narayan} \& {Yi}(1994)}]{Narayah&Yi1994}
{Narayan}, R., \& {Yi}, I. 1994, \apjl, 428, L13, \dodoi{10.1086/187381}

\bibitem[{Nishikawa {et~al.}(2016)Nishikawa, Frederiksen, Nordlund, Mizuno, Hardee, Niemiec, Gómez, Pe’er, Duţan, Meli, Sol, Pohl, \& Hartmann}]{Nishikawa_2016}
Nishikawa, K.-I., Frederiksen, J.~T., Nordlund, {\AA}., {et~al.} 2016, The Astrophysical Journal, 820, 94, \dodoi{10.3847/0004-637X/820/2/94}

\bibitem[{{{\"O}zbey Arabac{\i}} {et~al.}(2022){{\"O}zbey Arabac{\i}}, {Kalemci}, {Din{\c{c}}er}, {Bailyn}, {Altamirano}, \& {Ak}}]{Arabaci+2022}
{{\"O}zbey Arabac{\i}}, M., {Kalemci}, E., {Din{\c{c}}er}, T., {et~al.} 2022, \mnras, 514, 3894, \dodoi{10.1093/mnras/stac1574}

\bibitem[{{Pepe} {et~al.}(2015){Pepe}, {Vila}, \& {Romero}}]{Pepe+2015}
{Pepe}, C., {Vila}, G.~S., \& {Romero}, G.~E. 2015, \aap, 584, A95, \dodoi{10.1051/0004-6361/201527156}

\bibitem[{{Ripperda} {et~al.}(2020){Ripperda}, {Bacchini}, \& {Philippov}}]{Ripperda2020}
{Ripperda}, B., {Bacchini}, F., \& {Philippov}, A.~A. 2020, \apj, 900, 100, \dodoi{10.3847/1538-4357/ababab}

\bibitem[{{Ripperda} {et~al.}(2022){Ripperda}, {Liska}, {Chatterjee}, {Musoke}, {Philippov}, {Markoff}, {Tchekhovskoy}, \& {Younsi}}]{Ripperda2022}
{Ripperda}, B., {Liska}, M., {Chatterjee}, K., {et~al.} 2022, \apjl, 924, L32, \dodoi{10.3847/2041-8213/ac46a1}

\bibitem[{{Rodrigues} {et~al.}(2018){Rodrigues}, {Biehl}, {Boncioli}, \& {Taylor}}]{RBB18a}
{Rodrigues}, X., {Biehl}, D., {Boncioli}, D., \& {Taylor}, A.~M. 2018, ArXiv e-prints.
\newblock \doarXiv{1806.01624}

\bibitem[{{Saikia} {et~al.}(2019){Saikia}, {Russell}, {Bramich}, {Miller-Jones}, {Baglio}, \& {Degenaar}}]{Saikia+2019}
{Saikia}, P., {Russell}, D.~M., {Bramich}, D.~M., {et~al.} 2019, \apj, 887, 21, \dodoi{10.3847/1538-4357/ab4a09}

\bibitem[{{Salvesen} {et~al.}(2016){Salvesen}, {Simon}, {Armitage}, \& {Begelman}}]{Salvesen2016}
{Salvesen}, G., {Simon}, J.~B., {Armitage}, P.~J., \& {Begelman}, M.~C. 2016, \mnras, 457, 857, \dodoi{10.1093/mnras/stw029}

\bibitem[{{Shakura} \& {Sunyaev}(1973)}]{Shakura1973}
{Shakura}, N.~I., \& {Sunyaev}, R.~A. 1973, \aap, 500, 33

\bibitem[{{Shao} \& {Li}(2020)}]{Shao&Li2020}
{Shao}, Y., \& {Li}, X.-D. 2020, \apj, 898, 143, \dodoi{10.3847/1538-4357/aba118}

\bibitem[{{Shidatsu} {et~al.}(2016){Shidatsu}, {Done}, \& {Ueda}}]{Shidatsu+2016}
{Shidatsu}, M., {Done}, C., \& {Ueda}, Y. 2016, \apj, 823, 159, \dodoi{10.3847/0004-637X/823/2/159}

\bibitem[{{Sironi} {et~al.}(2021){Sironi}, {Rowan}, \& {Narayan}}]{Sironi+2021}
{Sironi}, L., {Rowan}, M.~E., \& {Narayan}, R. 2021, \apjl, 907, L44, \dodoi{10.3847/2041-8213/abd9bc}

\bibitem[{{Stawarz} \& {Petrosian}(2008)}]{Stawarz2008}
{Stawarz}, {\L}., \& {Petrosian}, V. 2008, \apj, 681, 1725, \dodoi{10.1086/588813}

\bibitem[{{Sun} \& {Bai}(2021)}]{SunBai2021}
{Sun}, X., \& {Bai}, X.-N. 2021, \mnras, 506, 1128, \dodoi{10.1093/mnras/stab1643}

\bibitem[{{Takahara} \& {Kusunose}(1985)}]{Takahara&Kusunose1985}
{Takahara}, F., \& {Kusunose}, M. 1985, Progress of Theoretical Physics, 73, 1390, \dodoi{10.1143/PTP.73.1390}

\bibitem[{{Tchekhovskoy} {et~al.}(2011){Tchekhovskoy}, {Narayan}, \& {McKinney}}]{Tchekhovskoy2011}
{Tchekhovskoy}, A., {Narayan}, R., \& {McKinney}, J.~C. 2011, \mnras, 418, L79, \dodoi{10.1111/j.1745-3933.2011.01147.x}

\bibitem[{{Tetarenko} {et~al.}(2019){Tetarenko}, {Casella}, {Miller-Jones}, {Sivakoff}, {Tetarenko}, {Maccarone}, {Gandhi}, \& {Eikenberry}}]{Tetarenko+2019}
{Tetarenko}, A.~J., {Casella}, P., {Miller-Jones}, J.~C.~A., {et~al.} 2019, \mnras, 484, 2987, \dodoi{10.1093/mnras/stz165}

\bibitem[{{Tetarenko} {et~al.}(2016){Tetarenko}, {Sivakoff}, {Heinke}, \& {Gladstone}}]{Tetarenko+2016}
{Tetarenko}, B.~E., {Sivakoff}, G.~R., {Heinke}, C.~O., \& {Gladstone}, J.~C. 2016, \apjs, 222, 15, \dodoi{10.3847/0067-0049/222/2/15}

\bibitem[{Twagirayezu {et~al.}(2023)Twagirayezu, Niederhausen, Sclafani, Whitehorn, Nisa, Yu, \& Halliday}]{Twagirayezu:2023Sd}
Twagirayezu, J.~P., Niederhausen, H., Sclafani, S., {et~al.} 2023, PoS, ICRC2023, 1175, \dodoi{10.22323/1.444.1175}

\bibitem[{Vecchiotti {et~al.}(2022)Vecchiotti, Zuccarini, Villante, \& Pagliaroli}]{Vecchiotti_2022}
Vecchiotti, V., Zuccarini, F., Villante, F.~L., \& Pagliaroli, G. 2022, The Astrophysical Journal, 928, 19, \dodoi{10.3847/1538-4357/ac4df4}

\bibitem[{{Vos} {et~al.}(2024){Vos}, {Cerutti}, {Moscibrodzka}, \& {Parfrey}}]{Vos+2024}
{Vos}, J., {Cerutti}, B., {Moscibrodzka}, M., \& {Parfrey}, K. 2024, arXiv e-prints, arXiv:2410.19061, \dodoi{10.48550/arXiv.2410.19061}

\bibitem[{{Wang} {et~al.}(2024){Wang}, {Liu}, {Qiao}, \& {Cheng}}]{Wang+2024}
{Wang}, Y., {Liu}, B.~F., {Qiao}, E., \& {Cheng}, H. 2024, \mnras, 527, 1333, \dodoi{10.1093/mnras/stad3224}

\bibitem[{White {et~al.}(2019)White, Stone, \& Quataert}]{White2019}
White, C.~J., Stone, J.~M., \& Quataert, E. 2019, The Astrophysical Journal, 874, 168, \dodoi{10.3847/1538-4357/ab0c0c}

\bibitem[{{Wood} {et~al.}(2021){Wood}, {Miller-Jones}, {Homan}, {Bright}, {Motta}, {Fender}, {Markoff}, {Belloni}, {K{\"o}rding}, {Maitra}, {Migliari}, {Russell}, {Russell}, {Sarazin}, {Soria}, {Tetarenko}, \& {Tudose}}]{Wood+2021}
{Wood}, C.~M., {Miller-Jones}, J.~C.~A., {Homan}, J., {et~al.} 2021, \mnras, 505, 3393, \dodoi{10.1093/mnras/stab1479}

\bibitem[{{Xie} \& {Zdziarski}(2019)}]{XieZdziarski2019}
{Xie}, F.-G., \& {Zdziarski}, A.~A. 2019, \apj, 887, 167, \dodoi{10.3847/1538-4357/ab5848}

\bibitem[{{Xu} \& {Lazarian}(2023)}]{Xu2023}
{Xu}, S., \& {Lazarian}, A. 2023, \apj, 942, 21, \dodoi{10.3847/1538-4357/aca32c}

\bibitem[{{Yan} {et~al.}(2024){Yan}, {Liu}, {Zhang}, {Li}, {Yuan}, \& {Wang}}]{Yan+2024}
{Yan}, K., {Liu}, R.-Y., {Zhang}, R., {et~al.} 2024, Nature Astronomy, 8, 628, \dodoi{10.1038/s41550-024-02221-y}

\bibitem[{{Yang} {et~al.}(2024){Yang}, {Yuan}, {Li}, {Mizuno}, {Guo}, {Lu}, {Ho}, {Lin}, {Zdziarski}, \& {Wang}}]{Yang+2024}
{Yang}, H., {Yuan}, F., {Li}, H., {et~al.} 2024, Science Advances, 10, eadn3544, \dodoi{10.1126/sciadv.adn3544}

\bibitem[{{Ye} {et~al.}(2022){Ye}, {Hu}, {Tian}, {Chang}, {Chang}, {Cheng}, {Gao}, {Ge}, {Gong}, {Guo}, {Guo}, {He}, {Huang}, {Jiang}, {Jiang}, {Jing}, {Li}, {Li}, {Li}, {Li}, {Li}, {Liao}, {Lin}, {Liu}, {Liu}, {Liu}, {Miao}, {Mo}, {Morton-Blake}, {Peng}, {Sun}, {Tang}, {Tang}, {Tao}, {Tian}, {Wang}, {Wang}, {Wang}, {Wei}, {Wei}, {Wu}, {Xian}, {Xiang}, {Xu}, {Xue}, {Yang}, {Yang}, {Yu}, {Zeng}, {Zhang}, {Zhang}, {Zhang}, {Zhang}, {Zhi}, {Zhong}, {Zhou}, {Zhu}, \& {Zhuang}}]{TRIDENT2022}
{Ye}, Z.~P., {Hu}, F., {Tian}, W., {et~al.} 2022, arXiv e-prints, arXiv:2207.04519, \dodoi{10.48550/arXiv.2207.04519}

\bibitem[{{Yoshitake} {et~al.}(2024){Yoshitake}, {Shidatsu}, {Ueda}, {Nogami}, {Murata}, {Higuchi}, {Isogai}, {Maehara}, {Mineshige}, {Negoro}, {Kawai}, {Yatsu}, {Sasada}, {Takahashi}, {Niwano}, {Saito}, {Takayama}, {Oasa}, {Takarada}, {Shigeyoshi}, \& {Oister Collaboration}}]{Yoshitake+2024}
{Yoshitake}, T., {Shidatsu}, M., {Ueda}, Y., {et~al.} 2024, \pasj, 76, 251, \dodoi{10.1093/pasj/psae005}

\bibitem[{{You} {et~al.}(2023){You}, {Cao}, {Yan}, {Hameury}, {Czerny}, {Wu}, {Xia}, {Sikora}, {Zhang}, {Du}, \& {Zycki}}]{You+2023}
{You}, B., {Cao}, X., {Yan}, Z., {et~al.} 2023, Science, 381, 961, \dodoi{10.1126/science.abo4504}

\bibitem[{{Yuan} \& {Narayan}(2014)}]{Yuan&Narayan2014}
{Yuan}, F., \& {Narayan}, R. 2014, \araa, 52, 529, \dodoi{10.1146/annurev-astro-082812-141003}

\bibitem[{{Yuan} {et~al.}(2003){Yuan}, {Quataert}, \& {Narayan}}]{Yuan+2003}
{Yuan}, F., {Quataert}, E., \& {Narayan}, R. 2003, \apj, 598, 301, \dodoi{10.1086/378716}

\bibitem[{{Yungelson} {et~al.}(2006){Yungelson}, {Lasota}, {Nelemans}, {Dubus}, {van den Heuvel}, {Dewi}, \& {Portegies Zwart}}]{Yungelson+2006}
{Yungelson}, L.~R., {Lasota}, J.~P., {Nelemans}, G., {et~al.} 2006, \aap, 454, 559, \dodoi{10.1051/0004-6361:20064984}

\bibitem[{{Zdziarski} {et~al.}(2017){Zdziarski}, {Malyshev}, {Chernyakova}, \& {Pooley}}]{Zdziarski+2017}
{Zdziarski}, A.~A., {Malyshev}, D., {Chernyakova}, M., \& {Pooley}, G.~G. 2017, \mnras, 471, 3657, \dodoi{10.1093/mnras/stx1846}

\bibitem[{{Zdziarski} {et~al.}(2009){Zdziarski}, {Malzac}, \& {Bednarek}}]{Zdziarski+2009}
{Zdziarski}, A.~A., {Malzac}, J., \& {Bednarek}, W. 2009, \mnras, 394, L41, \dodoi{10.1111/j.1745-3933.2008.00605.x}

\bibitem[{{Zhdankin} {et~al.}(2018){Zhdankin}, {Uzdensky}, {Werner}, \& {Begelman}}]{Zhdankin2018}
{Zhdankin}, V., {Uzdensky}, D.~A., {Werner}, G.~R., \& {Begelman}, M.~C. 2018, \apjl, 867, L18, \dodoi{10.3847/2041-8213/aae88c}

\bibitem[{{Zhdankin} {et~al.}(2019){Zhdankin}, {Uzdensky}, {Werner}, \& {Begelman}}]{Zhdankin+2019}
---. 2019, \prl, 122, 055101, \dodoi{10.1103/PhysRevLett.122.055101}

\end{thebibliography}
\bibliographystyle{aasjournal}



\end{document}